\documentclass{report}
\usepackage{chngpage}
\usepackage[utf8]{inputenc}
\usepackage[T1]{fontenc}
\usepackage[english]{babel}
\usepackage{amsmath}
\usepackage{amssymb,amsfonts,textcomp}
\usepackage{color}
\usepackage{array}
\usepackage{supertabular}
\usepackage{hhline}
\usepackage{hyperref}
\usepackage{graphicx}
\usepackage{floatrow}
\usepackage{pdfpages}
\usepackage{longtable}
\usepackage{subfig}
\usepackage{multirow}
\setboolean{@twoside}{false}
\usepackage{titletoc}% http://ctan.org/pkg/titletoc
\usepackage{lipsum}
\usepackage{authblk}

\hypersetup{pdftex,   colorlinks=false, linkcolor=blue, citecolor=blue, filecolor=false, urlcolor=false, pdftitle=, pdfauthor=Tao Chen, pdfsubject=, pdfkeywords=}
\makeatletter
\newcommand\arraybslash{\let\\\@arraycr}
\newcommand\textsubscript[1]{\ensuremath{{}_{\text{#1}}}}
\newcommand\textstyleDefaultParagraphFont[1]{#1}
\makeatother
\makeatletter
\newcommand\ps@Standard{
  \renewcommand\@oddhead{\footnotesize{Chen et al. \emph{The Handbook of Engineering Self-Aware and Self-Expressive Systems}}}
  \renewcommand\@evenhead{\footnotesize\emph{Chen et al. -- The Handbook of Engineering Self-Aware and Self-Expressive Systems}}
  \renewcommand\@oddfoot{}
  \renewcommand\@evenfoot{}
  \renewcommand\thepage{\arabic{page}}
}
\makeatother
\title{The Handbook of Engineering Self-Aware and Self-Expressive Systems}
\date{5 September 2014}
\author[1]{Tao Chen}
\author[1]{Funmilade Faniyi}
\author[1]{Rami Bahsoon}
\author[2]{Peter R. Lewis}
\author[1]{Xin Yao}
\author[1]{Leandro L. Minku}
\author[3]{Lukas Esterle}
\affil[1]{University of Birmigham, UK}
\affil[2]{Aston University, Birmigham, UK}
\affil[3]{Alpen-Adria Universität Klagenfurt, Austria}

\titlecontents*{chapter}% <section-type>
  [0pt]% <left>
  {}% <above-code>
  {\bfseries\chaptername\ \thecontentslabel\quad}% <numbered-entry-format>
  {}% <numberless-entry-format>
  {\bfseries\hfill\contentspage}% <filler-page-format>

\begin{document}

\maketitle

\chapter*{Acknowledgement}

This work was partially supported by the European Union Seventh Framework Programme under grant agreement 257906 (EPiCS).

\begin{center}
\includegraphics[width=5cm]{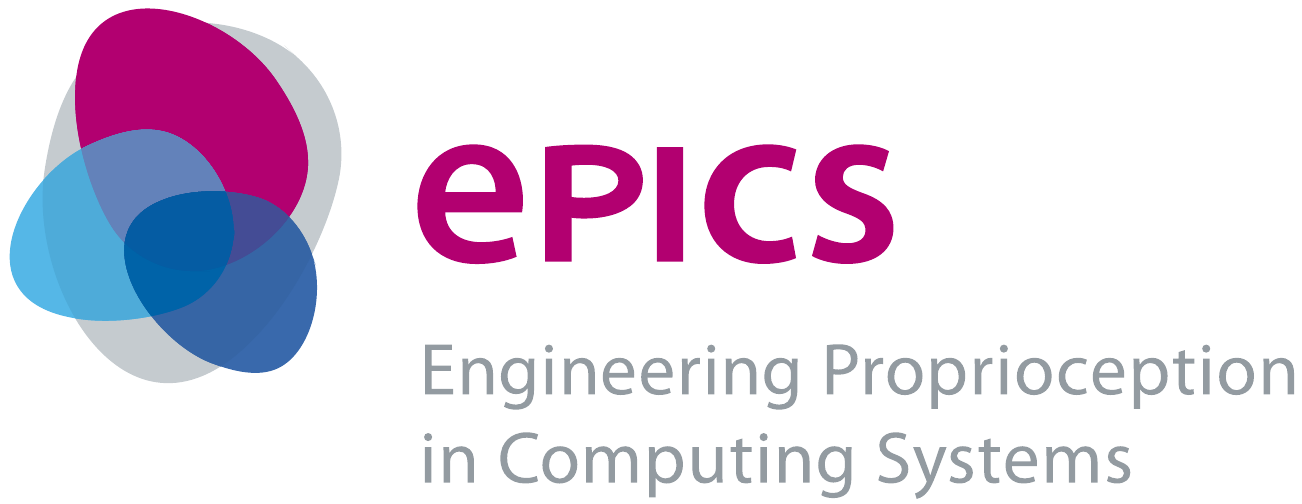}
\end{center}

\tableofcontents

\chapter{Patterns for Self-aware Architecture Style}

During previous Task 2.1, we have developed the notions of self-expression and different levels of computational self-awareness, inspired by corresponding psychological levels. In the context of architecture, we refer to the self-expression and different levels of computational self-awareness as \textbf{capability} of the systems to obtain and react upon certain knowledge. In this report, we study the categorization of different capabilities from the architecture perspective; this could create the possibility to ensure that, when designing self-aware systems, only relevant capabilities are included, and their inclusion justified by identified benefits. There is no need for a system to become unnecessarily complex, learning and maintaining capabilities which do nothing to advance the outcomes for that system, generating only overhead. We have codified the knowledge about how to architecture self-aware applications in the form of architecture patterns, each contains different capabilities. In this task, an architecture pattern refers to an architectural problem-solution pair using the capabilities in a given context. We have elicited some patterns, where each pattern is decentralized by design. That is, structurally our self-aware patterns resemble a peer-to-peer network of inter- connecting self-aware nodes, varying only in the number of the capabilities and the type of interconnection between them.

Until recently, architecture patterns for self-adaptive systems have received little attention \cite{Weyns2013}. Many existing patterns target specific application domains \cite{Menasce2010}, limiting their reuse outside the domains where they were originally conceived. Weyns et al. \cite{Weyns2013} argued that UML notations are limited in their ability to characterize self-adaptive architecture patterns, hence they proposed a simple, generic notation for describing patterns for Monitor-Analyze-Plan-Execute (MAPE) architecture style. Our patterns are distinct in focus from Weyns' in the sense that while we model self-aware capability and knowledge concerns in the architecture, their attention was about MAPE component interaction.

\begin{figure}[h!]
\centering
\includegraphics[width=4in]{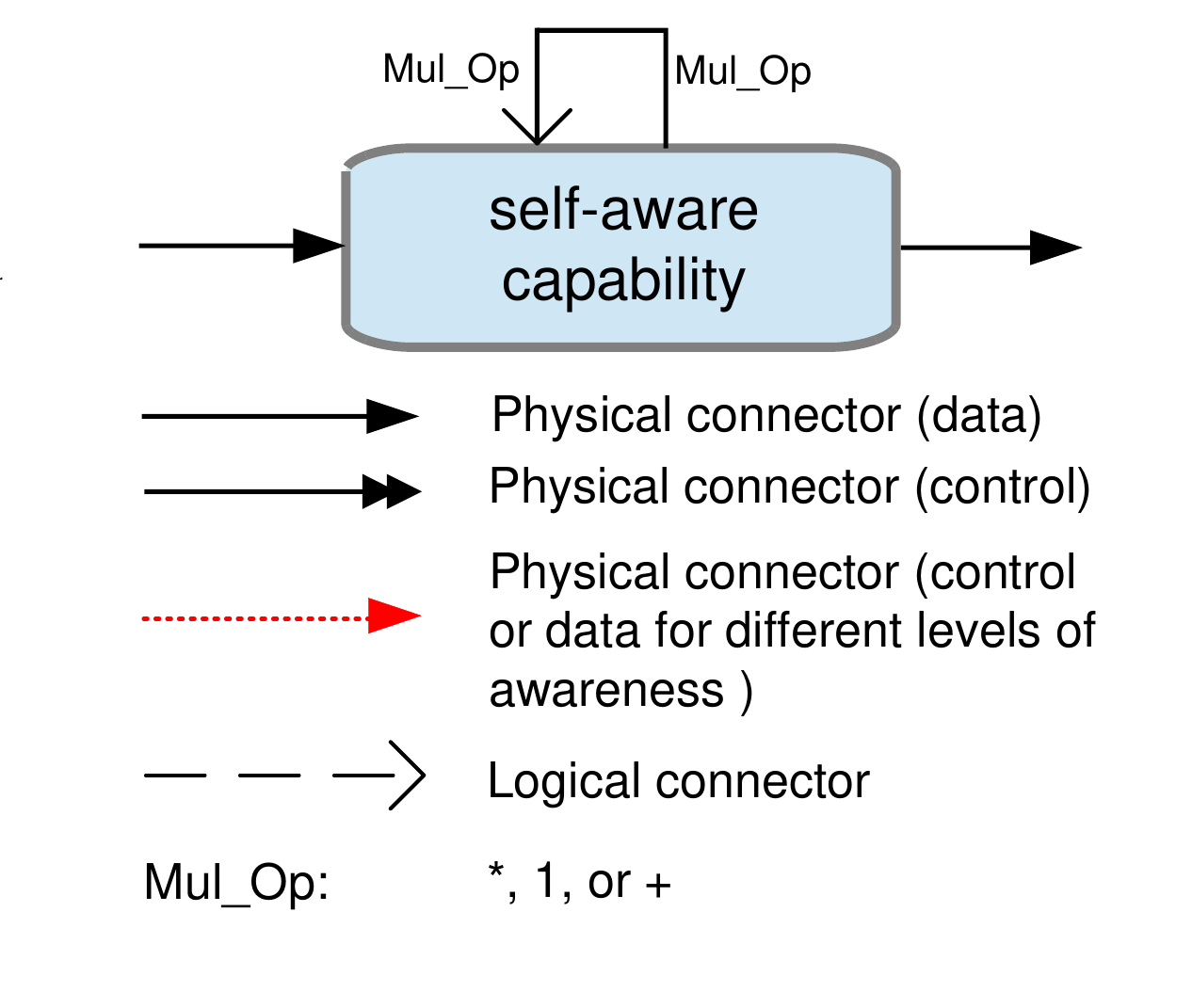}
\caption{Notation for Describing Self-aware Architecture Pattern}
\label{fig:pattern-notation}
\end{figure}

We adopt a pattern notation, similar to the one in \cite{Weyns2013} for describing our self-aware patterns. Firstly, Weyns's notation \cite{Weyns2013} is simple and easy to comprehend. Secondly, we believe describing our self-aware patterns using existing notation in the self-adaptive community makes our work accessible to other researchers and paves the way for others to build on our work. Existing work on architecture patterns focus on modeling the \textbf{components} and \textbf{connectors} of architecture; in such context, components are specializations of modules in the architecture and therefore have attributes and operations, but are also associated with the \textit{provide} and \textit{required} interfaces; and connectors could be the assembly that connects the \textit{required} interface of one component to the \textit{provided} interface of the second, or they could be the delegation that links the ports of a component to its internal parts. In our self-aware patterns, instead of modeling components, we model the capabilities of self-awareness and self-expression (e.g., stimulus awareness) in the architecture. In this way, our patterns preserve the flexibility for the concrete architectural implementation; since whether two or more capabilities are combined and realized in one component; or one capability is implemented in separate components could be based on the context. On the other hand, the connectors in our patterns are based on the standard definition but they are associated with capability rather than component. Although the capabilities of patterns are designed in a flexible manner,  it is important that the interactions amongst these capabilities should not be violated when realizing the pattern. For instance, one should not realize a direct interaction between a sensor and an actuator if it is not presented in a pattern. The pattern notation is depicted in figure~\ref{fig:pattern-notation}.

Two types of connectors are used to express the logical and physical interactions. physical connector means there is a direct interaction between two or more capabilities (from the same or different node), and each capability is required to directly interact with the others. Notably, physical connector (between different levels of awareness ), or the red arrow, particularly refers to the interactions for the self-awareness of different types (e.g., goal and time awareness); in contrast, the other black solid arrows represent the interactions for the self-awareness of the same type (e.g., the interaction-awareness from different nodes).  On the other hand, the logical connector does not require direct interaction, but rather the data or control in the interaction is sent/received through the other capabilities (e.g., Sensors and Actuators), which have the physical connector. For instance, self-expression might be logically required to reach consensus amongst different nodes, but such interaction is physically realized through Sensors and Actuators. The benefit of additionally introducing the logical connector is that, when design a self-aware capability where the communication protocol (e.g., local/remote function call, multi-cast and broadcast etc) is not needed, the pattern can still show that such capability needs to interact with the others. Thus, this provides the designers with a more precise view about he architecture.

We have used multiplicity operator to represents how many capabilities and their components (a capability can be realized in one or more components),  including those from different nodes, are involved in the interaction.There are three types of multiplicity operators (mul\_op):

\begin{itemize}
\item \textbf{+} expresses that the number of capability of the same type in the interaction is restricted to at least one.  
\item \textbf{1} indicates that one and only one capability of the same type is permitted. 

\item \textbf{*} indicates that zero, one or many of the type specified is permitted in the interaction. 
\end{itemize}

\noindent It is worth noting that when the operator is *, it means that the associated interaction may not exist but does not represent that the corresponding capability can be eliminated. In case a capability is interact with itself, e.g., a + on both sides of the intra-capability arrow of a capability means that it can interact with the same capability implemented in other nodes. To better clarify the operators, suppose that there is a physical interaction between stimulus awareness and external sensors where the stimulus awareness is associated with 1 whereas the external sensors is associated +. This means that within the interaction, the stimulus awareness can only have one whereas the number of external sensors presented in the interaction needs to be one or many. Other multiplicity arrangements can be similarly interpreted. We document our patterns using standard pattern template \cite{Buschmann2007} as follows.
%Other multiplicity arrangements can be similarly interpreted.
\begin{itemize}
\item Problem/Motivation: A scenario where the pattern is applicable
\item Solution: A representation of the said pattern in a graphical form
\item Consequences: A narration of the outcome of applying the pattern
\item Example: Instance of the pattern in real applications or systems
\end{itemize}

\noindent Next, we present the definition of different self-awareness and self-expression capabilities.

\section{Definition of Self-awareness}
\subsection{Private and Public Self-awareness}
The sources of the relevant
knowledge (i.e. internal or external sensors) for a node, underlie the notion
of private and private self-awareness.

\begin{enumerate}
  \item \textbf{Private self-awareness}: This concerns with a node possessing
  knowledge of and/or based on phenomena that are internal to itself.
  \item \textbf{Public self-awareness}: This concerns with a node possessing
  knowledge of phenomena external to itself. Such knowledge grounds meaning
  (e.g. in the form of models) to the node's own perspective on phenomena
  external to itself, i.e. it is subjective to the node. This subjectivity
  is what underlies the notion of self.
\end{enumerate}

\subsection{Levels of Self-awareness}
Described below are the levels of self-awareness,
along with their relevance to either public or private self-awareness or both.

\begin{enumerate}

  \item \textbf{Stimulus-aware}\\
    A node is stimulus-aware if it has knowledge of stimuli. The node is not
    able to distinguish between the sources of stimuli. It does not have
    knowledge of past/future stimuli. It enables the ability in a node to
    respond to events. It is a prerequisite for all other levels of self-awareness.
    Since stimuli may originate both internally and externally,
    stimulus-awareness can either be \textbf{private}, \textbf{public} or
    \textbf{both}.

    \bigskip
  \item \textbf{Interaction-aware}\\
    A node is interaction-aware if it has knowledge that stimuli and its own
    actions form part of interactions with other nodes and the environment. It
    has knowledge via feedback loops that its actions can provoke, generate or
    cause specific reactions from the social or physical environment. It enables
    a node to distinguish between other nodes and environments. Simple
    interaction-awareness may just enable a node to reason about individual
    interactions. More advanced interaction-awareness may involve the node
    possessing knowledge of social structures such as communities or network
    topology. Interaction-awareness is typically based on external phenomena,
    whereby it is therefore a form of \textbf{public} self-awareness, however
    one can also envisage a system which learns about the effects of internal
    interactions with itself, which would constitute a form of \textbf{private}
    self-awareness.

    \bigskip
  \item \textbf{Time-aware}\\
    A node is time-aware if it has knowledge of historical and/or likely future
    phenomena. Implementing time-awareness may involve the node possessing an
    explicit memory, capabilities of time series modeling and/or anticipation.
    Since time-awareness can apply to both internal and external phenomena, it
    can either be \textbf{private}, \textbf{public} or \textbf{both}.

    \bigskip
  \item \textbf{Goal-aware}\\
    A node is goal-aware if it has knowledge of current goals, objectives,
    preferences and constraints. It is important to note that there is a
    difference between a goal existing implicitly in the design of a node, and
    the node having knowledge of that goal in such a way that it can reason
    about it. The former does not describe goal-awareness; the latter does.
    Example implementations of such knowledge in a node include state based
    goals (i.e. knowing what is a goal state and what is not) and utility based
    goals (i.e. having a utility or objective function). Goal-awareness permits
    acknowledgment of and adaptation to changes in goals. When coupled with
    interaction-awareness or time-awareness, goal-awareness permits the ability
    to reason about goals in relation to other nodes, or about likely future
    goals, respectively.  Since goals may exist privately to the node, or
    collectively as a shared or externally imposed goal, goal-awareness can
    either be \textbf{private}, \textbf{public} or \textbf{both}.

    \bigskip
  \item \textbf{Meta-self-aware}\\
    A node is meta-self-aware if it has knowledge of its own level(s) of
    awareness and the degree of complexity with which the level(s) are
    exercised. Such awareness permits a node to reason about the benefits and
    costs of maintaining a certain level of awareness (and degree of complexity
    with which it exercises this level). It further allows the node to adapt the
    way in which the level(s) of self-awareness are realized (e.g. by changing
    algorithms realizing the level(s), thus changing the degree of complexity of
    realization of the level(s)). As an example, this awareness may involve a
    node being able to dynamically select a particular technique out of a set of
    possibilities for realizing one or more levels, in order to meet or manage
    trade-off between its goals or objectives. Since meta-self-awareness is
    concerned only with knowledge of internal processes, it is a form of
    \textbf{private} self-awareness.

\end{enumerate}

\section{Definition of Self-expression}
The following ideas underpin what we mean by self-expression within a
computing node.
\begin{itemize}

    \item \emph{A node exhibits self-expression if it is able to assert its
        behavior upon either itself or other nodes.}

    \item \emph{This behavior is based upon the node's state, context, goals,
        values, objectives and constraints.}

\end{itemize}

\noindent Next, we present the eight self-aware patterns using the template described above. For our purposes, the state of the node comprises the self-aware capability and knowledge captured in its self-awareness processes. Thus, self-expression can be thought of as behavior based on self-awareness.

\section{Basic Pattern} 

\noindent \textbf{Problem/Motivation.}  In some cases, a system may need to 
trigger some actions in order to cope with emergent events and stimuli. Such capacity could 
greatly help to manage system at runtime. As a result, there is an increasing 
need for system to react upon stimuli, based on either static or dynamic 
rules.

\begin{figure}[h!]
\centering
\includegraphics[width=5in]{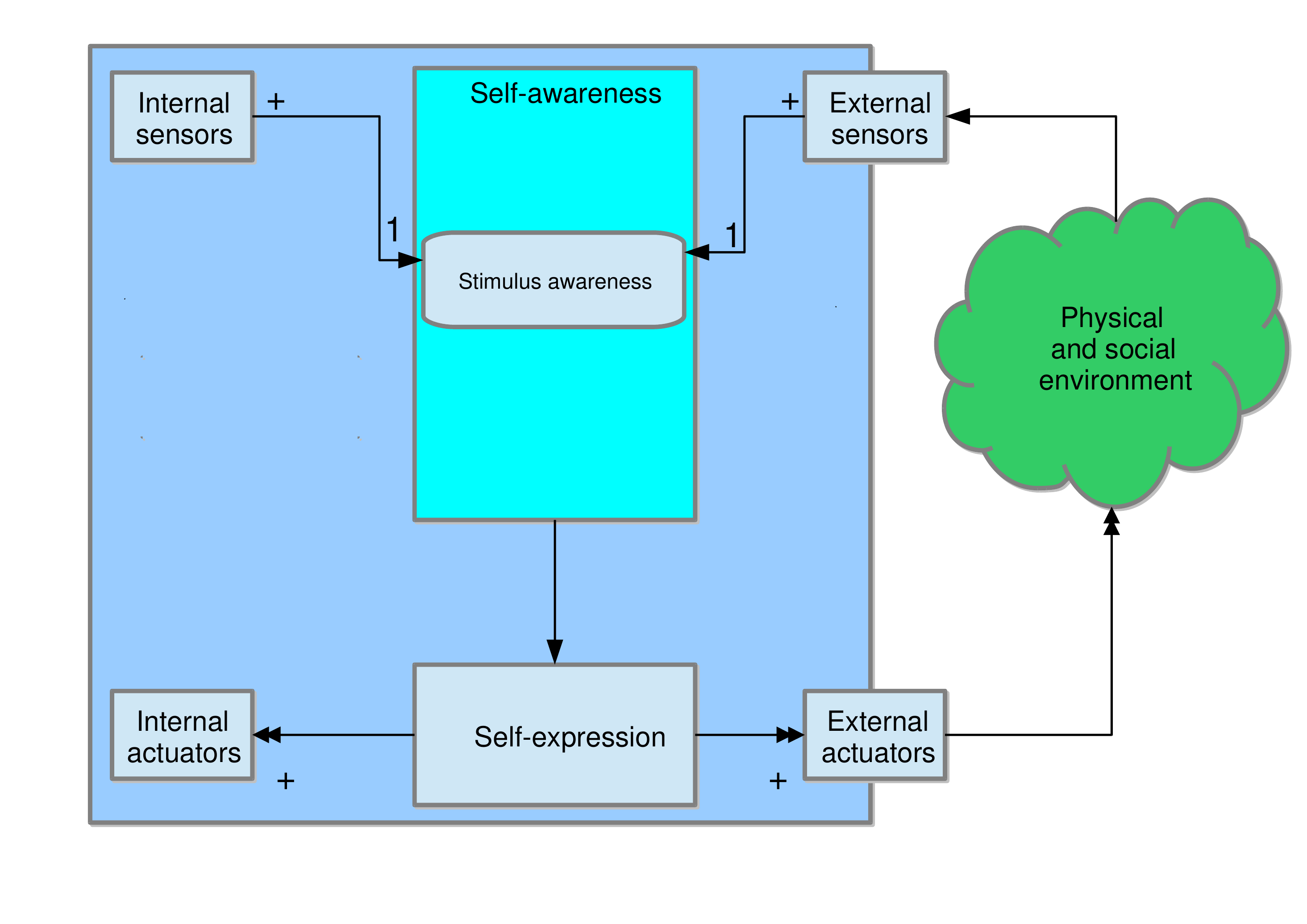}
\caption{Basic Pattern}
\label{fig:basic-8-pattern}
\end{figure}

\begin{figure}[h!]
\centering
\includegraphics[width=6in]{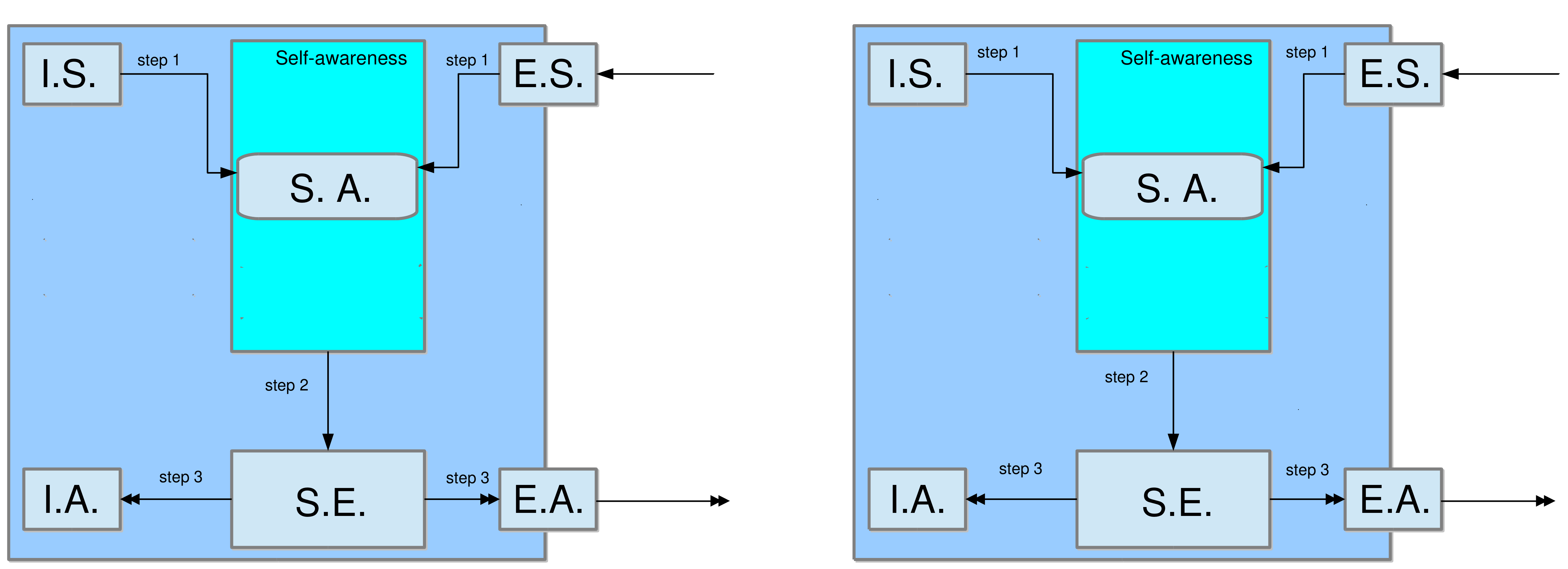}
\caption{Concrete Instance of the Basic Pattern}
\label{fig:basic-pattern-8-concrete-example1}
\end{figure}

\noindent \textbf{Solution.}  The simplest pattern to enable self-aware node is 
what we call Basic Pattern, as shown in figure~\ref{fig:basic-8-pattern}.  This 
pattern contains only stimulus awareness, which receiving data flow from sensors 
and actuators. Proper actions of self-expression could be triggered based on the type of stimulus 
detected. A concrete example has been shown in figure~\ref{fig:basic-pattern-8-concrete-example1} 
where each node only aware of its own stimuli.

\noindent  \textbf{Consequences.} A major limitation of this pattern is no 
information is shared amongst nodes, therefore the node is not aware of the 
environment and the other node. This could become a major problem in some cases (e.g., the smart camera case study) 
where there are intensive interaction and/or interference amongst nodes.

\noindent  \textbf{Examples.} Consider the case of server farm or private cloud where the 
numbers of deployed applications/services are limited. The 
basic pattern could be realized in such context by defining \textit{if-condition-then-action rules}, in 
which case the conditions could be various stimuli (e.g., QoS is low and utilization is 
low); the action could be changing software configuration and/or resource 
provisioning.

\section{Basic Information Sharing Pattern}

\noindent \textbf{Problem/Motivation.} Sometimes one computing node may not be sufficient to cope with the complexity of an application or to meet the demands of users as they scale. To manage application complexity, functionalities could be divided among several self-aware nodes, where each node is specialized in a few functionalities, collaborating to provide the application's service. More self-aware nodes may also be introduced to meet the scalability requirement of the system. In each case, at the basic level, there is a need to provide a means for the nodes to interact with one another to carry out their respective roles.

\begin{figure}[h!]
\centering
\includegraphics[width=5in]{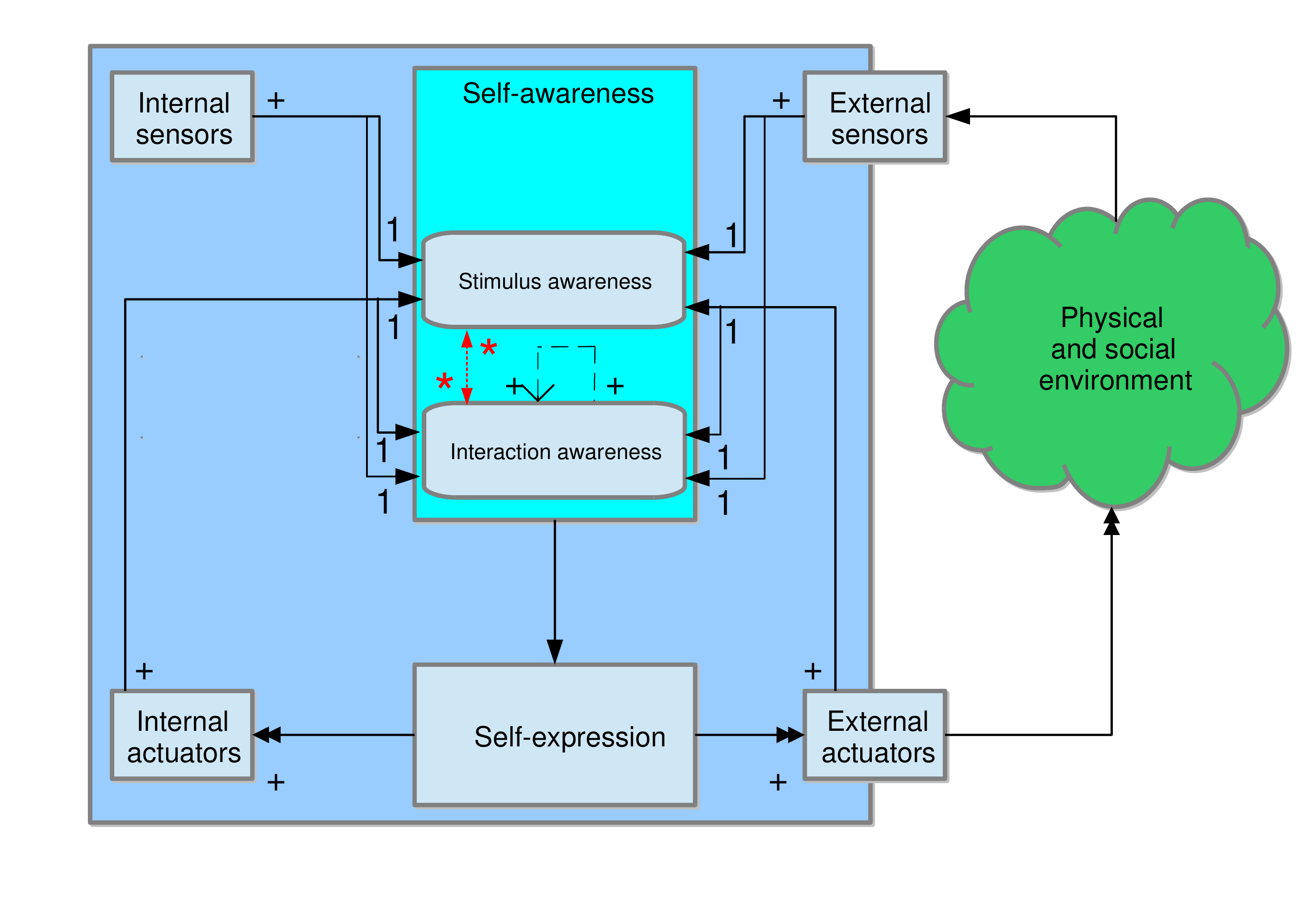}
\caption{Basic Information Sharing Pattern}
\label{fig:basic-pattern}
\end{figure}

\noindent \textbf{Solution.} The simplest pattern for interacting self-aware nodes is the basic information sharing pattern. In this pattern, a self-aware node contains only the interaction-awareness capability other than the stimulus-awareness. Interaction-awareness can be connected to one or more self-aware nodes as shown in figure~\ref{fig:basic-pattern}. Each self-aware node may have one or more sensors (internal/external) and actuators (internal/external). The underlying characteristic of this pattern is that peers are linked only at the level of interaction-awareness. It is important to note that nodes can not only interact with neighbors but also with their environment. For example, in the financial modeling application, interaction is all about communication between nodes and the market rather than amongst nodes themselves.

\begin{figure}[h!]
\centering
\includegraphics[width=6in]{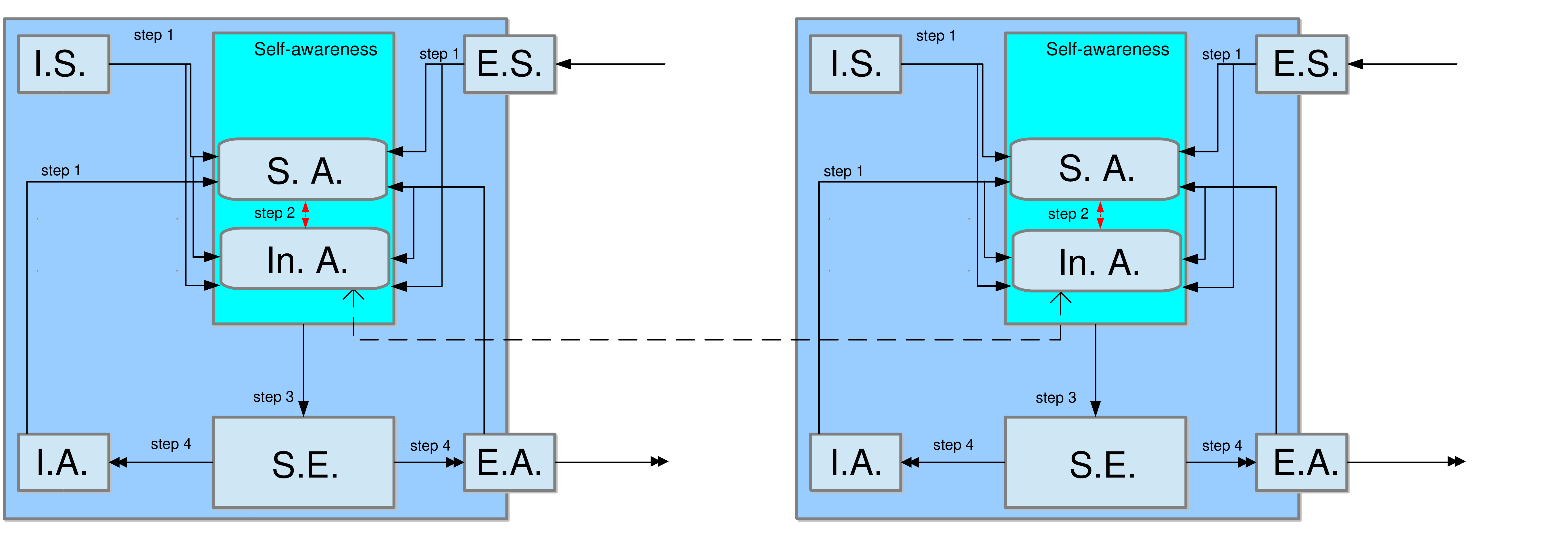}
\caption{Concrete Instance of the Basic Information Sharing Pattern}
\label{fig:basic-pattern-concrete-example1}
\end{figure}

An example of the basic pattern where two nodes are connected via their interaction-awareness capabilities is shown in figure~\ref{fig:basic-pattern-concrete-example1}. Although only two nodes are shown in figure~\ref{fig:basic-pattern-concrete-example1}, the number of connected nodes is not limited to two. The number of nodes is limited by the scalability of the interaction mechanism. For instance, a broadcast mechanism may limit the number of interconnected nodes when compared to a gossip protocol. In practice, a node may be connected to either all or a subset of nodes in the systems depending on its role in the system.

\noindent  \textbf{Consequences.} Self-aware nodes could use the interconnection between them to negotiate the protocol to use for communicating in a network. As observed in the smart camera case study, this pattern can be used to facilitate sharing information among nodes about neighbourhood relation in a network of smart camera. 

Crucially, in this pattern each self-aware node maintains its autonomy about how to make adaptation decisions via its self-expression capability. This means that each node is responsible for its interpretation and reaction to the information shared via interaction-awareness. Therefore, this pattern is not suitable for cooperative problem-solving scenarios, where nodes need to reach an agreement among themselves about the best course of action for the problem. This limitation is addressed in the \textit{coordinated decision-making pattern} (see next section). The basic information sharing pattern assumes the system's goal is pr-econfigured at design time, consequently, constraining the system's adaptation.

\noindent  \textbf{Examples.} Federated datacenters and clouds, owned by distinct entities, are good candidate applications of the basic information sharing pattern. The owners of such clouds or datacenters may choose only to share status information about availability of resources or current load and not cooperate beyond this level. Thus, each cloud provider maintains autonomy over its resources while collaborating with other cloud providers in a limited way to facilitate outsourcing of resources, if required. Participants in a grid computing set-up utilize similar communication model and rely on incentive-based mechanisms to facilitate resource sharing \cite{Xiao2005}.

\section{Coordinated Decision-making Pattern}

\noindent \textbf{Problem/Motivation.} Decisions made by individual self-aware nodes in a group may be suboptimal due to their limited view of the system and its operating environment. As noted in the basic information sharing pattern, individual self-aware nodes do not cooperate when making decisions. In applications requiring near-optimal and consistent global decision making in a cooperative setting, a more advanced architectural pattern may be required. In particular, such a pattern should make it possible for nodes to synchronize their self-expressive actions.

\begin{figure}[h!]
\centering
\includegraphics[width=5in]{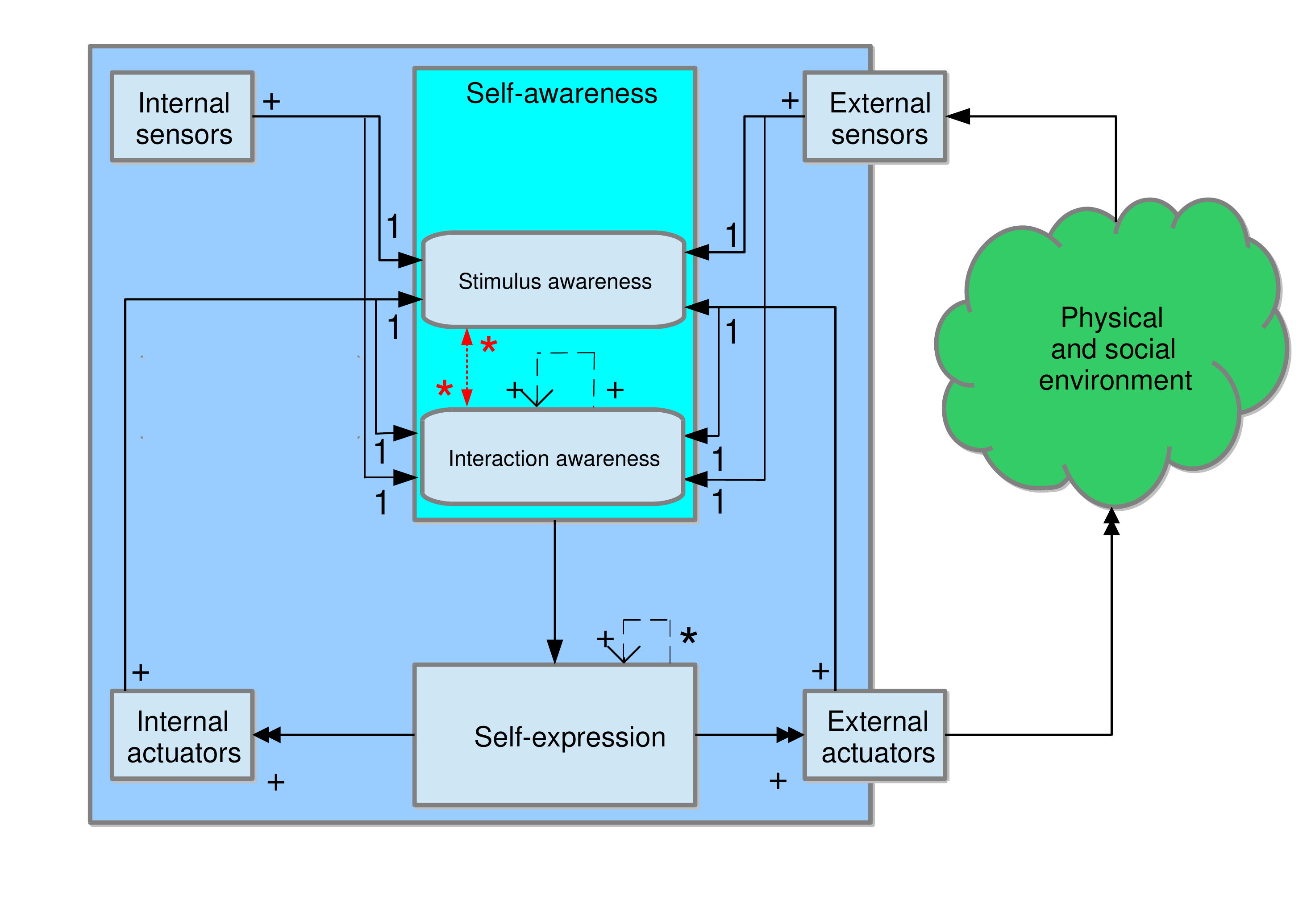}
\caption{Coordinated Decision-making Pattern}
\label{fig:coordinate-decision-making-pattern}
\end{figure}

\noindent \textbf{Solution.} The coordinated decision-making pattern provides a means of coordinating actions of multiple, interconnected self-aware nodes. Figure~\ref{fig:coordinate-decision-making-pattern} shows this pattern. It differs from the basic pattern in that self-expressive nodes are linked to one another, such that they are able to agree on \emph{what} action to take. It is clear to see that the coordinated decision-making pattern is a related pattern to the basic information sharing pattern as they only differ on the self-expression capability. However, they are designed to aim for different problems and forces, therefore such separation of concepts paves a better way in pattern selection.

\begin{figure}[h!]
\centering
\includegraphics[width=6in]{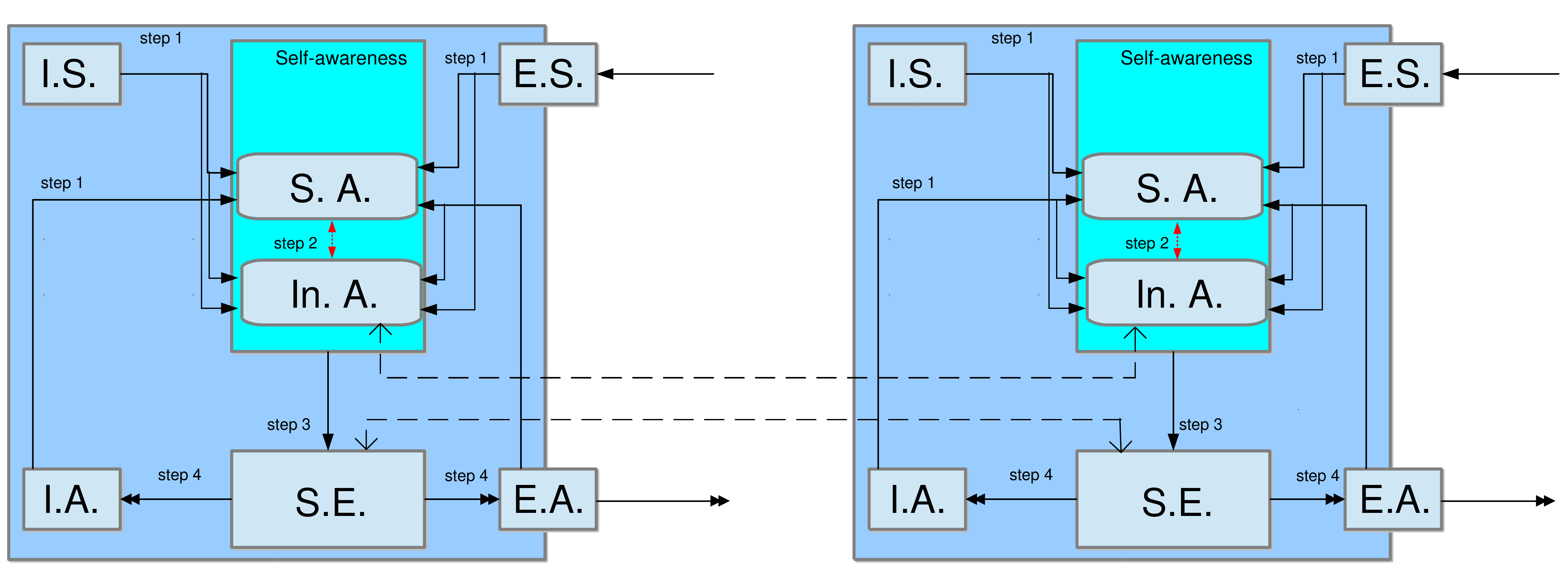}
\caption{Concrete Instance of Coordinated Decision-making Pattern}
\label{fig:coordinated-decision-making-example}
\end{figure}

\noindent  \textbf{Consequences.} Unlike the basic pattern, given the * to 0 multiplicity on the self-expression capability in figure~\ref{fig:coordinate-decision-making-pattern}, it is not mandatory for nodes to link their self-expression capabilities to each other. This makes it possible for nodes to form clusters, where nodes in a cluster cooperate to solve problems in one part of a system, while nodes in other clusters cooperate to solve problems in other parts. Figure~\ref{fig:coordinated-decision-making-example} shows an example where two self-aware nodes instantiate this pattern. As argued in the case of the basic pattern, using two nodes to illustrate the pattern as shown in Figure~\ref{fig:coordinated-decision-making-example} does not limit the number of nodes that can realize the pattern in a real system.

The downside of this pattern is that although nodes are able to form clusters and cooperate on \emph{what} action to take, they are unable to decide the timing of such actions, i.e. \emph{when} to act. This notion of time insensitivity is addressed in the \emph{Temporal Knowledge Sharing Pattern} (see next section). The temporal knowledge sharing pattern incorporates time-awareness capabilities into the coordinated decision making pattern.

\noindent  \textbf{Examples.} Large-scale cloud federations where providers agree to implement unified resource allocation policies, irrespective of how such policies are enforced at individual cloud levels, are a candidate application of this pattern. In such federated clouds, policy changes are negotiated via interaction-awareness capabilities, upon agreement the self-expression capability of each cloud enforce the agreed policy within its (local) cloud.

\section{Temporal Knowledge Sharing Pattern}

\noindent \textbf{Problem/Motivation.} As stated in the previous section, coordinated decision-making pattern does not provide a means of coordinating the \emph{timing} of actions agreed upon by cooperating nodes. This limitation may not be tolerated in applications where timing of actions has an impact on the integrity of the application. Also historic knowledge may be required to forecast future actions, in order to improve the accuracy of adaptive actions. %when compared time insensitive patterns.

\begin{figure}[h!]
\centering
\includegraphics[width=5in]{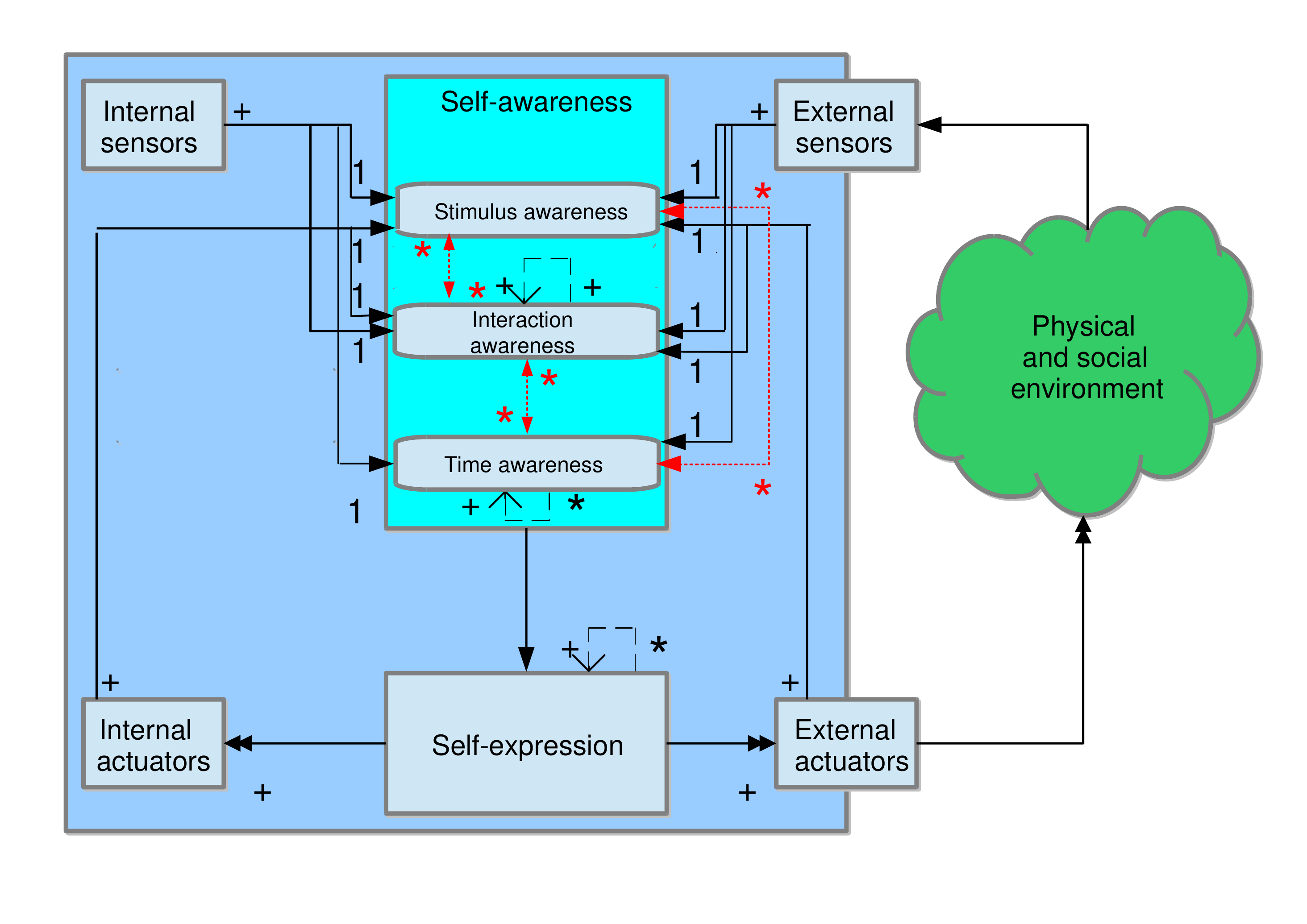}
\caption{Temporal Knowledge Sharing Pattern}
\label{fig:temporal-knowledge-sharing-pattern}
\end{figure}

\begin{figure}[h!]
\centering
\includegraphics[width=6in]{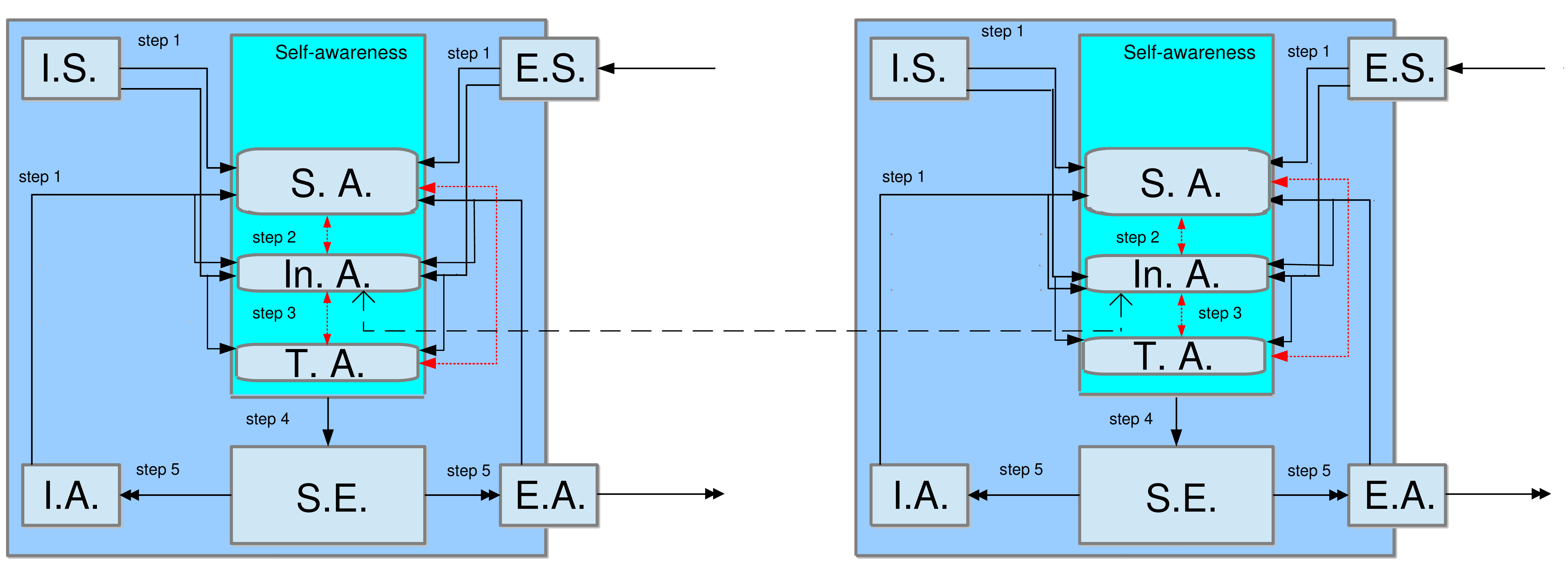}
\caption{Concrete Instance of Temporal Knowledge Sharing Pattern}
\label{fig:temporal-knowledge-sharing-pattern-example}
\end{figure}

\noindent \textbf{Solution.} The temporal knowledge sharing pattern solves this problem by incorporating time-awareness capabilities into the coordinated decision-making pattern. As shown in figure~\ref{fig:temporal-knowledge-sharing-pattern}, each self-aware node has a time-aware capability which is, optionally (as denoted by its multiplicity), linked to other self-aware nodes to represent timing information. An example where two nodes are connected using this pattern is shown in figure~\ref{fig:temporal-knowledge-sharing-pattern-example}. This timing information can be exploited by the self-expression capability to manage the timing of adaptation actions across multiple nodes.

\noindent  \textbf{Consequences.} The knowledge of timing information provides a rich basis to enrich the power of the adaptation action that is possible. However, there are a lot of design considerations left to the application designer who instantiates the style. For example, how often should timing information be recorded? In storage constrained systems, how long should acquired knowledge be stored for before forgetting (removing) them? Should the forgetting process be total, i.e. delete all knowledge acquired within a period at once, or selective? Depending on the concerns of the application at hand, these questions will have different answers.

\noindent  \textbf{Examples.} Clusters in cloud datacenters, where the servers in the cluster cooperate to execute tasks assigned to the cluster head, are able to exploit this pattern. For example, a parallel scientific  application assigned to the cluster, requiring coordination across different time-steps of the application could utilize the pattern to ensure actions taken in each time-step are coordinated to avoid compromising the integrity of the result.

\section{Temporal Knowledge Aware Pattern} 

\noindent \textbf{Problem/Motivation.}  The knowledge of timing enables the 
capability of proactive adaptation and potentially, better adaptation quality. Within the 
previously mentioned pattern, Temporal Knowledge Sharing pattern is the only one 
that applies time awareness capability. However, a drawback of such pattern is that the 
interaction awareness capability might not be a unnecessarity, therefore it could affect the self-aware system as it is suffering unnecessary overhead. 

\begin{figure}[h!]
\centering
\includegraphics[width=5in]{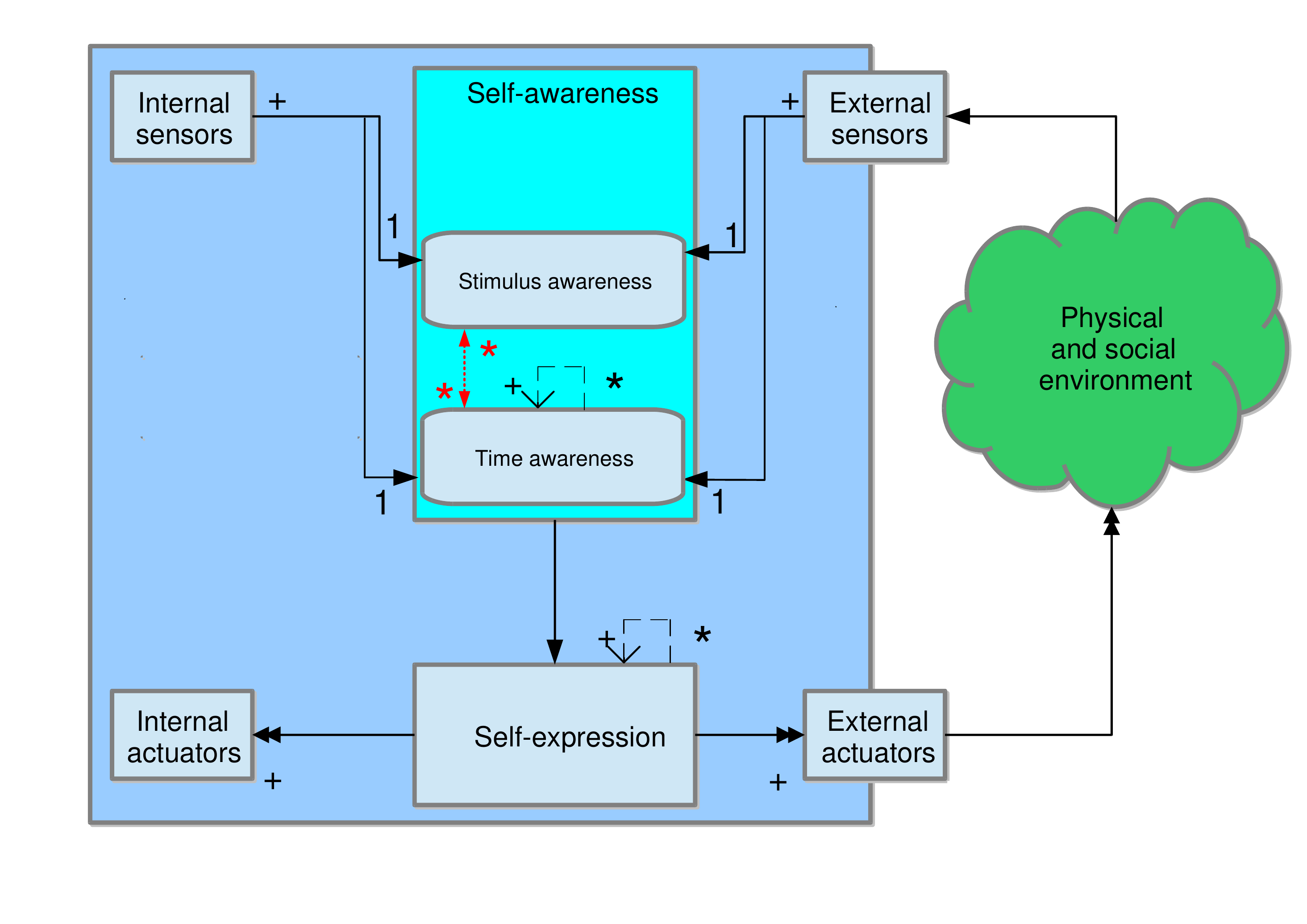}
\caption{Temporal Knowledge Aware Pattern}
\label{fig:temporal-knowledge-aware-pattern}
\end{figure}

\begin{figure}[h!]
\centering
\includegraphics[width=6in]{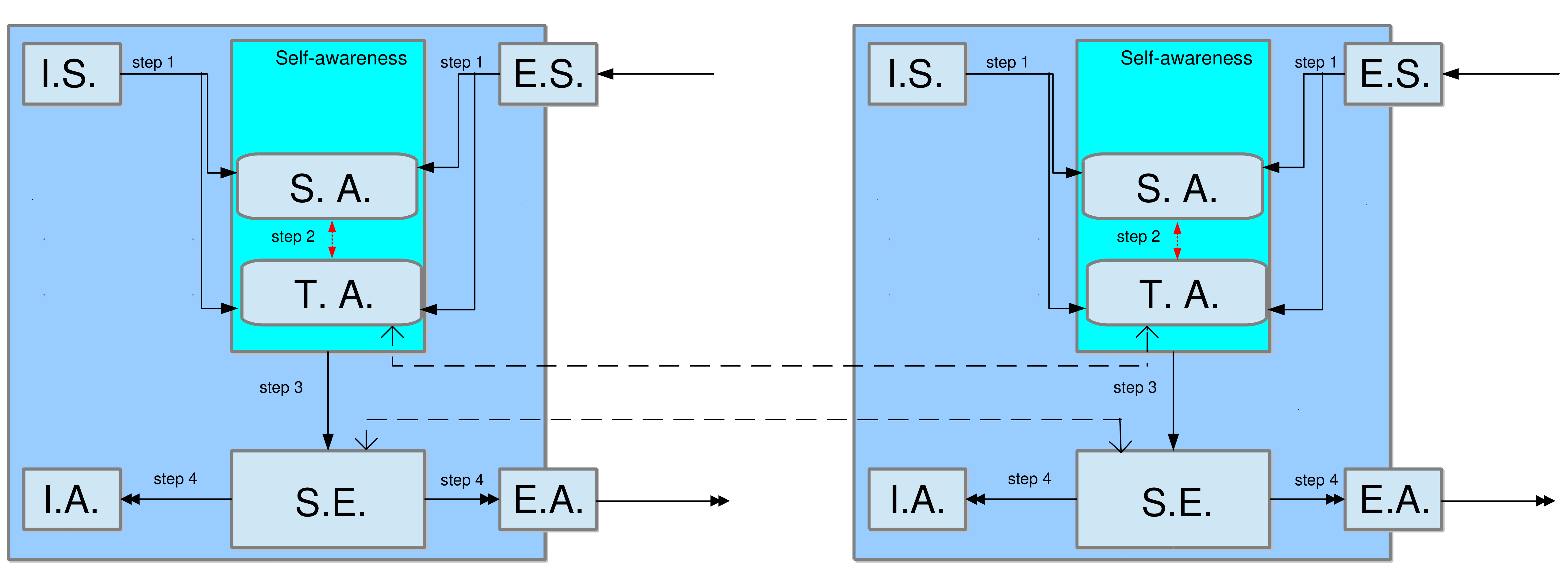}
\caption{Concrete Instance of Temporal Knowledge Aware Pattern}
\label{fig:temporal-knowledge-aware-pattern-example}
\end{figure}

\noindent \textbf{Solution.}  As shown in figure~\ref{fig:temporal-knowledge-aware-pattern}, the temporal knowledge aware pattern solves this 
problem by incorporating only time awareness working in conjunction with stimulus 
awareness. Again, the time awareness capabilities of different node is logically 
linked together (optionally). This pattern allows the knowledge of timing to 
assist the self-expression capability and overhead the extra overhead produced 
by unneeded level of awareness. A concrete example has been shown in figure~\ref{fig:temporal-knowledge-aware-pattern-example}.

\noindent  \textbf{Consequences.} There are scenarios where the software designer is uncertain about whether the lack of environmental information and 
information could affect the modeling of timing knowledge. This is highly depend on the concrete time-series prediction technique in the time awareness capability. An inappropriate use of certain 
time-series prediction technique could result in low accuracy, which eventually 
affect the quality of adaptation. As a result, the decision of which time-series prediction technique to be used is critical and  the designers are recommended to consult 
experts of time-series modeling when applying this pattern.

It should be noted that up till now, all the patterns discussed do not cater to changing goals. That is, they assume the goal of the self-adaptive system is known at design-time and statically encoded in the system, without opportunity to modify it at run-time. The pattern discussed in the next section - \emph{Goal Sharing Pattern} - will address the challenge of modifying or changing goal at run-time.

\noindent  \textbf{Examples.}  Cloud is an environment where resource is sharing 
via Virtual Machine (VM) on each node. In this context, by leveraging the historical usage of resources, time-series prediction would be able to predict the demand of VMs on a node for the nearly future, which assists 
proactive provisioning of resource and potentially, prevents SLA violation and/or resource 
exhaustion on a node.

\section{Goal Sharing Pattern} \noindent \textbf{Problem/Motivation.} User preferences are mostly dynamic, i.e. users want different things at different times. As an example, a user who is pleased with operating a computing system using a touch screen at one time may prefer a voice interaction mood at another time. These changes in user preferences may range from simple changes, such as mood of user-interaction, to more advanced ones. Furthermore, a computing system may itself decide to change its goal, depending on the amount of resources available to it. In the smart camera case \cite{esterle_et_al_2014}, a camera running low on battery may choose to bid for only the most valued objects within its field of view instead of aiming to track all objects in its vicinity. A specialized pattern that allows explicit representation of run-time goals and facilitate changes to these goals, as the system evolves, is needed .

\begin{figure}[h!]
\centering
\includegraphics[width=5in]{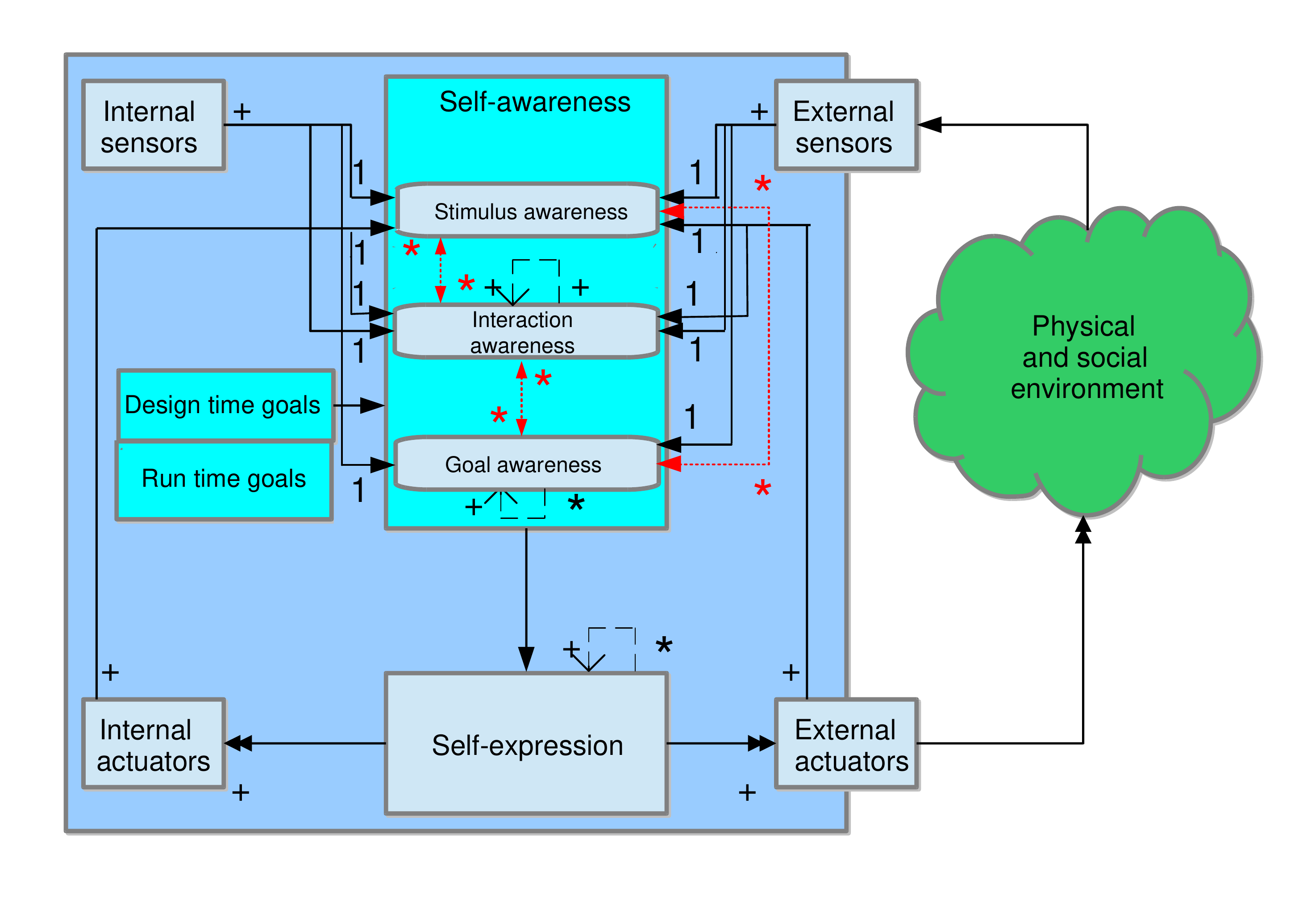}
\caption{Goal Sharing Pattern}
\label{fig:goal-sharing-pattern}
\end{figure}

\noindent \textbf{Solution.} Figure~\ref{fig:goal-sharing-pattern} shows the goal sharing pattern that address the concern of representing run-time goal. A goal-awareness capability represents knowledge about run-time goals, which can be changed as the system evolves. The goal-awareness capability in a self-aware node can, optionally, share its state information with goal-awareness capabilities in other self-aware nodes.

\begin{figure}[h!]
\centering
\includegraphics[width=5in]{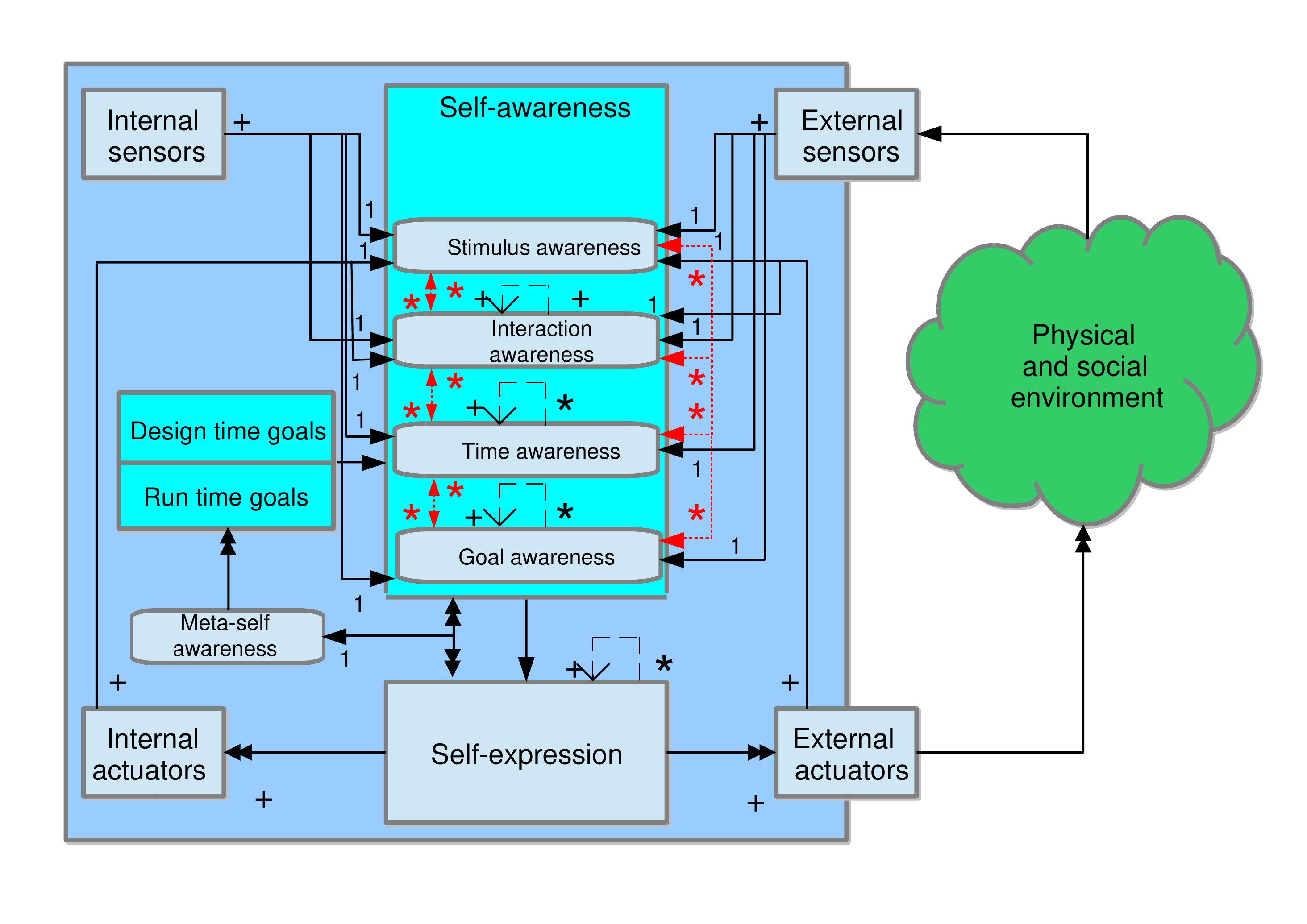}
\caption{Goal Sharing Pattern (with time-awareness capability)}
\label{fig:fully-coordinated-pattern}
\end{figure}

As with previous patterns, goal information sharing is not necessarily globally shared with all nodes. Hence, a subset of nodes in a system could share their goal state, while their goal information is disjoint from other nodes. It is important to note that sharing goal information is not equivalent to unifying goal state across nodes. It is possible for nodes to share goal information, while each pursues its distinct goal. The reverse scenario, where goal information are unified across nodes, is also possible.

\noindent  \textbf{Consequences.} As can be observed from figure~\ref{fig:goal-sharing-pattern}, a time-awareness capability is not included in this pattern. This implies that time-awareness is not a necessary perquisite for goal-awareness. While each node is able to change its goal at run-time, it does not represent temporal information to realize the capabilities of the temporal knowledge sharing pattern. For the sake of completeness, we include a different pattern that addresses this limitation (see figure~\ref{fig:fully-coordinated-pattern}). The pattern in figure~\ref{fig:fully-coordinated-pattern} makes the inclusion of temporal knowledge explicit, making it suitable for application domains where changing goals and forecasting are required.

\begin{figure}[h!]
\centering
\includegraphics[width=6in]{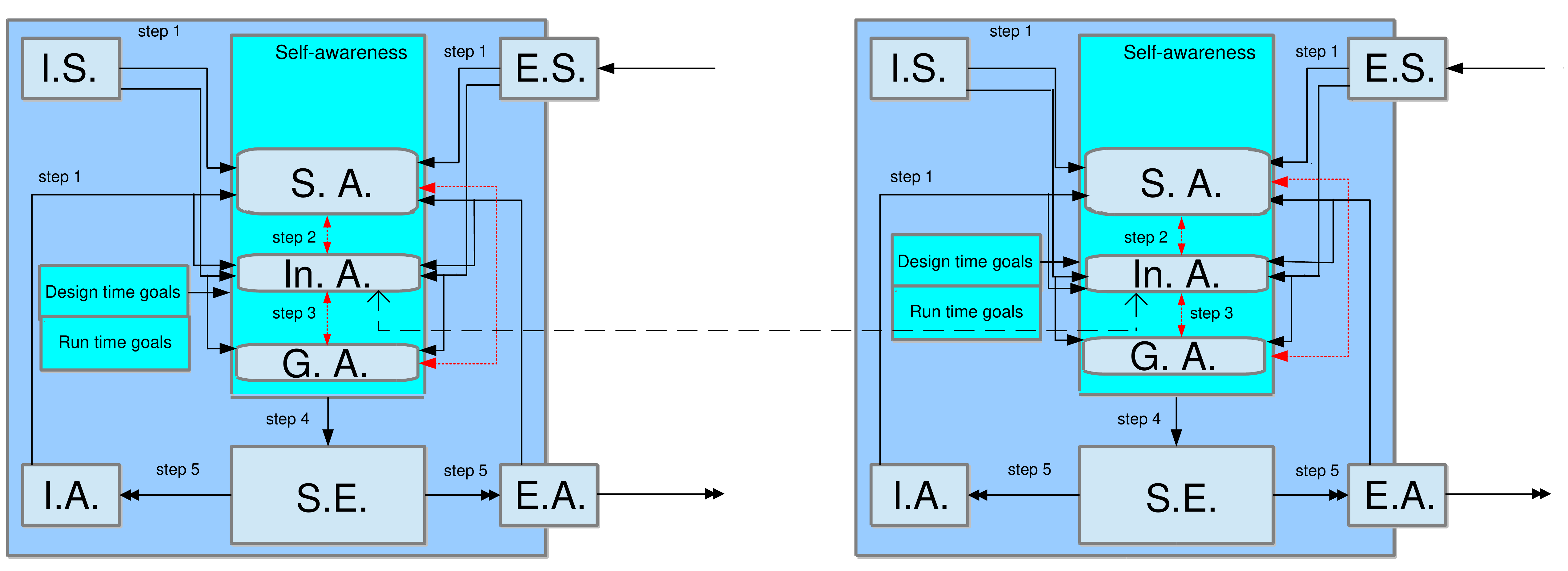}
\caption{Concrete Instance of Goal Sharing Pattern (without time-awareness capability)}
\label{fig:goal-sharing-pattern-example1}
\end{figure}

\begin{figure}[h!]
\centering
\includegraphics[width=6in]{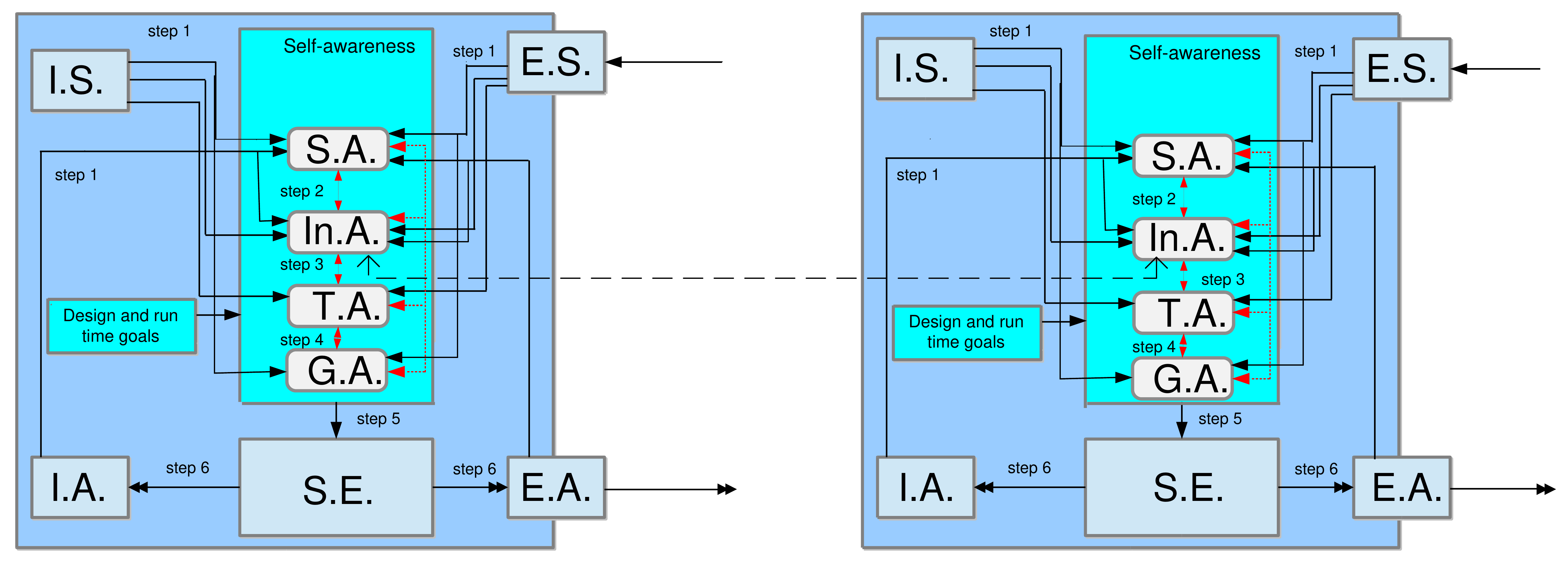}
\caption{Concrete Instance of Goal Sharing Pattern (with time-awareness capability)}
\label{fig:goal-sharing-pattern-example2}
\end{figure}

In both  patterns, self-expression capability makes use of the goal-awareness capability to make strategic decisions in line with the system's current goal. Figure~\ref{fig:goal-sharing-pattern-example1} shows an instance of the pattern (without time-awareness), while figure~\ref{fig:goal-sharing-pattern-example2} an instance (with time-awareness).

\noindent  \textbf{Examples.} Service-based applications operating in dynamic, open cloud environment are possible candidates of this pattern. Here, applications are composed from cloud services which are selected based on QoS and cost considerations. A service that is highly performed at one time may degrade in quality at later times due to overloading of the service. Each application has service level agreement (SLAs), to which it must adhere. Application goals encoded in SLAs may themselves change as users demand different levels of service from time to time.

Using the goal-sharing pattern with time-awareness capability (see figure~\ref{fig:fully-coordinated-pattern}) in this scenario has the benefit of making each application capable of representing temporal knowledge about service performance and forecasting which service(s) are likely to be more dependable and long-lasting. Also, the goal-awareness capability makes it possible to represent SLA terms of users and adapt such goals as they change. Lastly, by sharing temporal knowledge, applications can cascade knowledge of service performance among themselves. It should be noted that this introduces opportunities to falsely badmouth or inflate performance of services. Considerations for filtering out good knowledge are left to the computational models used to implement time- and goal-awareness.

\section{Temporal Goal Aware Pattern} 

\noindent \textbf{Problem/Motivation.}  The knowledge of goals and time might not necessarily to be shared amongst nodes,  especially in cases where the optimization of local goals could lead to acceptable global optimum. As a result, the presence of interaction awareness capability could cause extra overhead on the system.

\begin{figure}[h!]
\centering
\includegraphics[width=5in]{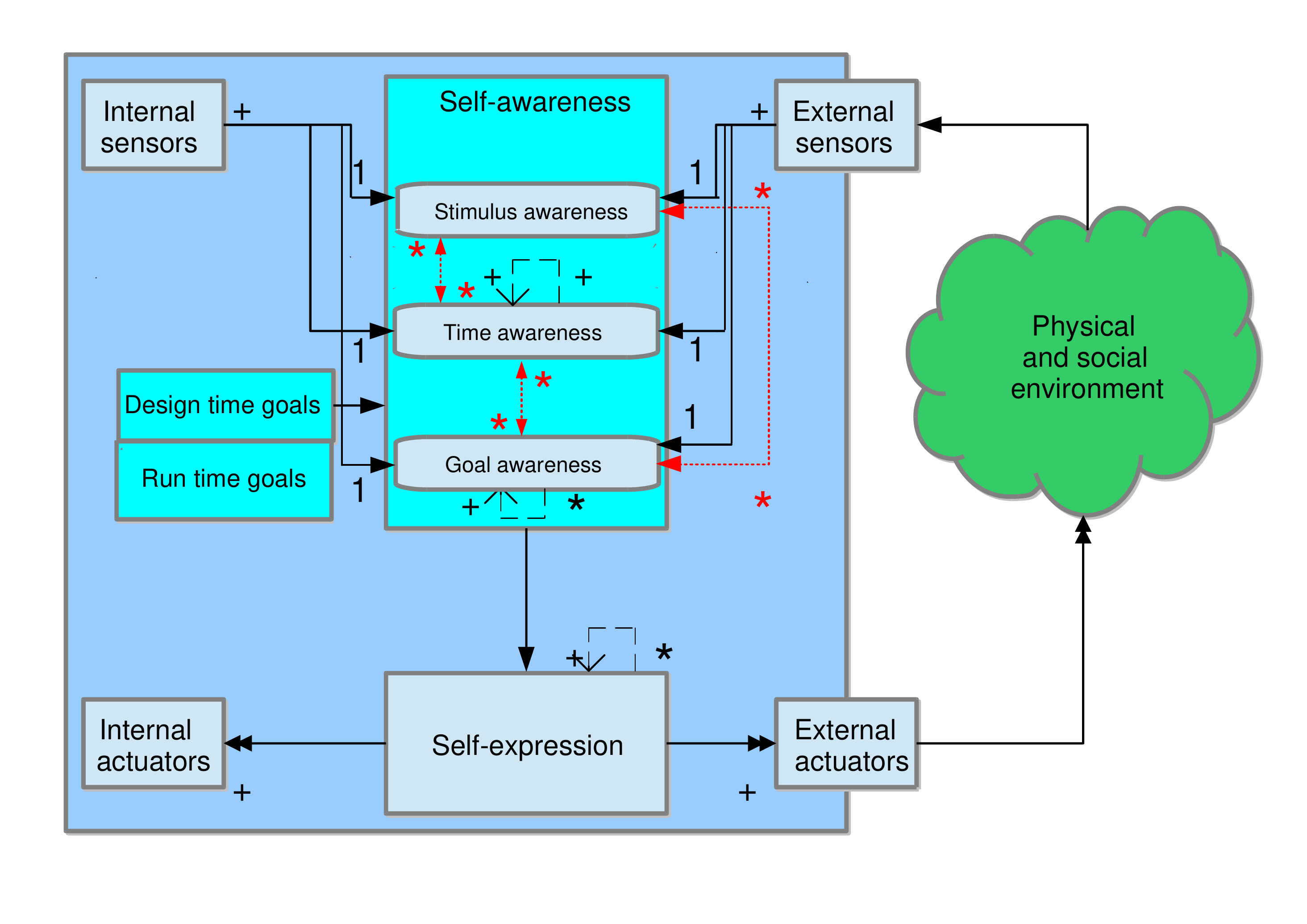}
\caption{Temporal Goal Aware Pattern}
\label{fig:temporal-goal-aware-pattern}
\end{figure}

\begin{figure}[h!]
\centering
\includegraphics[width=6in]{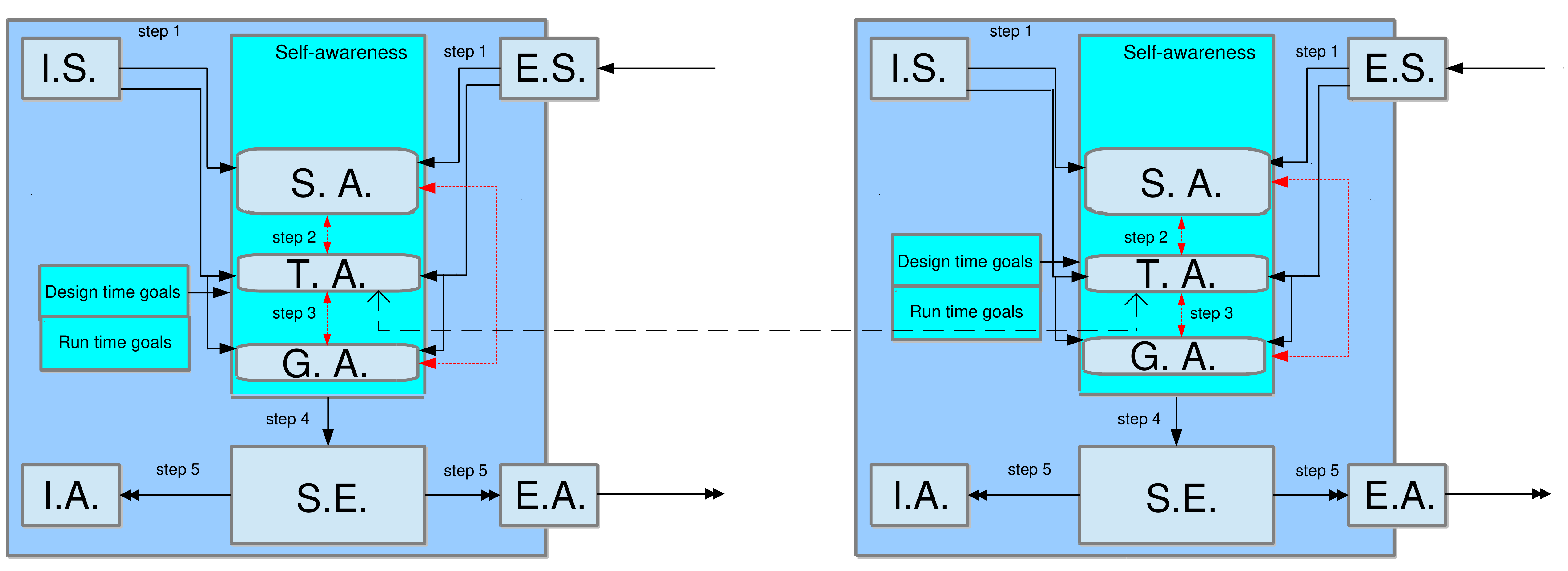}
\caption{Concrete Instance of Temporal Goal Aware Pattern}
\label{fig:temporal-goal-aware-pattern-example}
\end{figure}

\noindent \textbf{Solution.}  As shown in figure~\ref{fig:temporal-goal-aware-pattern}, the temporal goal aware pattern solves this 
problem by removing the interaction
awareness capability. In this pattern, there is no notion of 'sharing' as the nodes are not aware of any interactions and therefore not aware of the presence of the other nodes. It is worth noting that the absence of interaction awareness does not mean there is no interaction - nodes and the environment could still interact with each other, but the nodes are not aware of it. A concrete example has been shown in figure~\ref{fig:temporal-goal-aware-pattern-example}.

\noindent  \textbf{Consequences.} The removal of interaction awareness implies that the nodes could be in inconsistent state. The designer should carefully verify that such situation would not result in violations of system requirements. In addition, the self-expression capability could not use any information from other nodes when making decisions.

\noindent  \textbf{Examples.}  Orchestrate fully decentralized harmonic synchronization amongst different mobile devices requires each node to aware of stimulus, time and goal but not necessarily interaction. In such case, each node receives phase and frequency updates from the other nodes or the environment, and reacts upon based on its own time and goal information. This is a typical example where there are occurrences of interaction, but no occurrences of interaction awareness; because the nodes only aware of the incoming phase and frequency updates but it has no knowledge of where they come from; the sources could be other nodes, the environment or even some unexpected noise.

\section{Meta-self-awareness and Self-aware Patterns}

Meta-self-awareness is useful for managing the trade-off between various levels of self-awareness and for modifying goals at run-time. Since reasoning at the meta level is considered an advanced form of awareness, which may be beneficial or necessary in some contexts and not beneficial in others. This section specially treats the relation between meta-self-awareness and the patterns discussed in previous sections.

We reiterate that one of the distinct benefits of the self-aware style is to reduce the complexity of modeling adaptive behavior when compared to non-self-aware approaches. For the sake of illustration, consider the problem of modeling and tuning an online learning algorithm, e.g. neural network, for deciding actions of an application in different scenarios. It is known that this task is time-consuming and requires expertise mathematical skills, which may not be readily available \cite{Elkhodary2010}. Additionally, in some use cases, small changes in the application scenarios may render the solution proffered by the algorithm invalid or incorrect - another cycle of algorithm tuning may be needed to cater to these changes. An alternative approach is to provide families of algorithms for different contexts and dynamically select the appropriate algorithm at run-time using online learning capabilities of the meta-self-aware capability.

\begin{figure}[h!]
\centering
\includegraphics[width=6in]{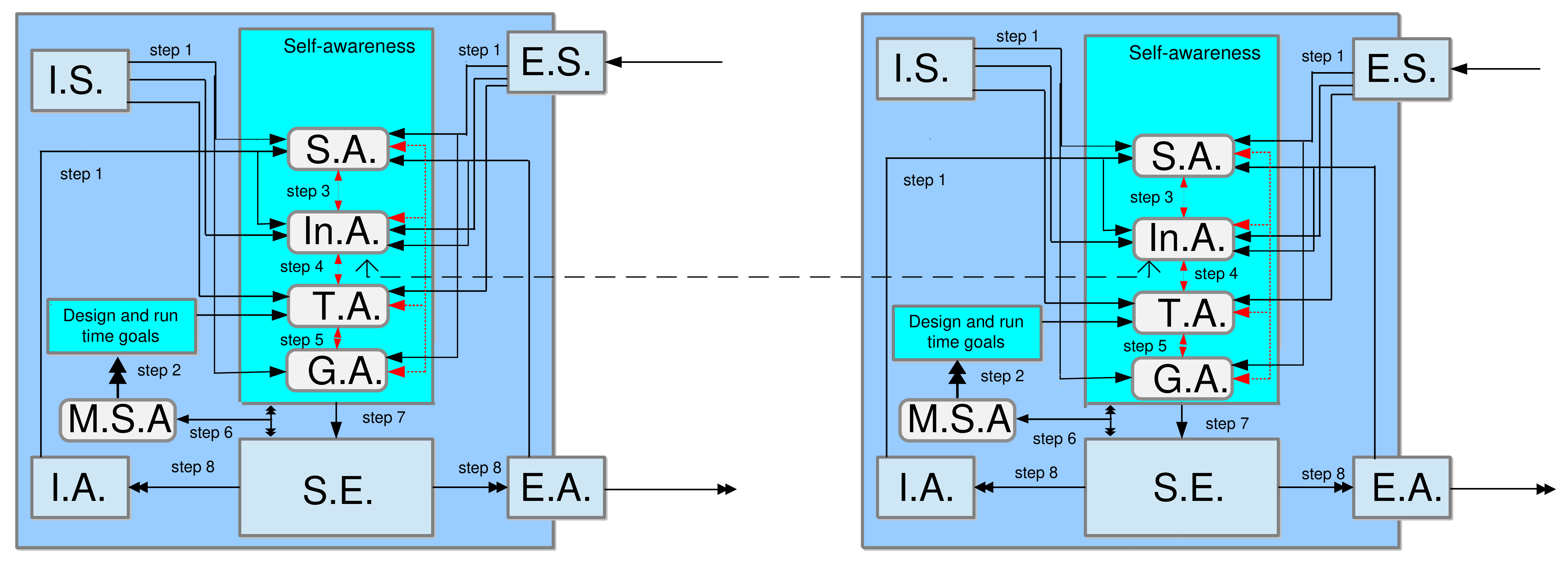}
\caption{Concrete Instance of Goal Sharing Pattern (including meta-self-awareness capability)}
\label{fig:goal-sharing-pattern-example3}
\end{figure}

While the first approach offers faster adaptation, if application scenarios are relatively stable, the second approach is able to better cope with complexity in highly perturbed environments, where one algorithm is insufficient to cover the scope of adaptive behavior. Accordingly, we recommend that every pattern can optionally incorporate the meta-self-aware capability depending on the complexity to be managed and expertise of the application designer. Figure~\ref{fig:goal-sharing-pattern-example3} shows the goal-sharing pattern with time-awareness capability where a meta-self-aware pattern is present to manage trade-off between goal-, interaction-, and time-awareness capabilities. Presumably, the presence of meta-self-awareness capability could help to switch between different pattern at runtime, which could be a very interesting direction for future work. 

There are also other examples of meta-self-awareness capability in EPiCS (e.g., \cite{2013_happe_seus}, \cite{2012_keller_ancs}, \cite{2011_becker_reconfig}). Specifically, in the smart camera demonstrator \cite{esterle_et_al_2014}, the meta-self-awareness is used to switch between different behavioral strategies in the interaction awareness and self-expression capabilities; In the hyper music demonstrator \cite{2012_chandra_cmmr}, the meta-self-awareness can help to control the degree of stimulus awareness based on confidence measure.

\section*{Questions \& Answers of the Patterns}

\begin{itemize}
\item Q: Can the nodes be heterogeneous (i.e., different nodes implementing different patterns)? 
\\A: Yes, different pattern can be chosen to realize different node if it is required.
\item Q: Can we realize different capabilities using the same technique? 
\\A: Yes, the notions of capability itself is flexible. 
\item Q: Can we have logical connections between different capabilities (e.g., goal and time awareness)? 
\\A: Yes, logical connections is not restricted as physical ones. 
\end{itemize}

\chapter{Architectural Primitives for Self-aware Systems}

These architectural patterns provide proven solution to recurring design problems that arise in a context of self-aware and self-expressive system. However, the pattern itself is an abstract form of the architecture. Often, the abstract patterns can be used as the first step for architecting and engineering self-aware systems by mapping them with an existing/candidate architecture instance, which is application specific. We have organized a workshop that allows the partners to present their results in using the patterns.  After the workshop, we have observed that the partners could use the same pattern in many different ways depending on the problem context and their concrete architecture instance (e.g., whether to use meta-self-awareness; whether to use a particular technique to realize a capability; whether to realize two capabilities in the same component); however, the concrete applications of patterns in different contexts still share many underlying concepts. These shared concepts lead to the notion of \textbf{architectural primitive}, which refers to the common concepts that can be constructively used to form a concrete architecture instance of a pattern. The architecture instances of a pattern differs in terms of the candidate techniques and/or attributes (i.e., a particular form of an architectural primitive) that used to realized each architectural primitive. These differences amongst architecture instances of a pattern are referred to as \textbf{pattern variability}; similarly, the alternative architecture instances of a pattern are the \textbf{variants} of this pattern.  

Architectural primitives and pattern variability are long-studied problems \cite{mehtaM03}. However, unlike traditional work that aim for generic system architectures, we are specifically interested in how these concepts can be used in self-aware systems. In particular, we aim to link these well-studied concepts to the self-awareness principles \cite{2012_becker_cse}, which are all about different levels of knowledge awareness. In this report, we have documented the architectural primitives in four categories (i.e., Capability, Behavior, Interaction and Topology), each of which cover an aspect of a self-aware system. We anticipate that a benefit from this will be a reduction in the chances to introduce faults, and easier fault detection, when using the proposed patterns during design.  We show the potential dependency between these primitives.  These architecture primitives and candidate techniques and/or attributes have been used in EPiCS to form different variants of a pattern.

\section{Architectural Primitives and Candidate Techniques}

To better understand the orthogonal aspects of self-aware systems, we have used four categories. This is motivated by our investigation about the critical aspects that could affect the functional and/or non-functional requirements of a self-aware system. These categories are shown as below:

\begin{itemize}
\item \textbf{Capability}:  The capability of the system to obtain certain knowledge or to react based upon the knowledge. e.g., the levels of awareness and expression.
\item \textbf{Behavior}: The process of a capability regarding how input data is consumed and output data is produced.
\item \textbf{Interaction}: The relationship between the capabilities or of a capability itself as expressed by the multiplicity operators.
\item \textbf{Topology}: The deployment about how capabilities are distributed to the components in the architecture instance.
\end{itemize}

We have structured the architectural primitives to express: (i) the characteristic of a components or connector; and (ii) the function of a components or connector. The architectural primitives and their attributes with respect to the four categories are shown as below:

\begin{itemize}
\item \textbf{Capability}: 
     \begin{itemize}
            \item \textit{stimulus-awareness}: the capability to obtain knowledge of stimuli
            \item \textit{interaction-awareness}: the capability to obtain knowledge of stimuli and its own
    actions form part of interactions with other nodes and the environment
            \item \textit{time-awareness}: the capability to obtain knowledge of historical and/or likely future phenomena
            \item \textit{goal-awareness}: the capability to obtain knowledge of current goals, objectives,
    preferences and constraints
            \item \textit{meta-self-awareness}: the capability to obtain knowledge of its own level(s) of
    awareness and the degree of complexity with which the level(s) are exercised
            \item \textit{self-expression}: the capability to react based upon the node's state, context, goals,  values, objectives and constraints
     \end{itemize}
\item \textbf{Behaviour}: 
      \begin{itemize} 
            \item \textit{send}: the output process of a sender (e.g., capability, sensors and actuators), it has two attributes
                    \begin{itemize} 
                             \item \textit{synchronous}: the scenario where after the sender sends a request, it needs to wait for a reply from the receiver before transit to the next actions
                             \item \textit{asynchronous}: the case that the sender does not require such blocked communication
                    \end{itemize}
             \item \textit{handle}: the input process of a receiver (e.g., capability, sensors and actuators), it has two attributes
                    \begin{itemize} 
                             \item \textit{sequential}: the cases where incoming data are processed by specific sequence using a queue
                             \item \textit{parallel}: the scenario that the incoming data is simply process in parallel upon its arrival
                    \end{itemize}
            \item \textit{state}: the behavior of a capability, it depends on the candidate techniques that realize such capability. Mainly, it has two attributes
                    \begin{itemize} 
                             \item \textit{proactive}: the behavior that predicts and reasons about when the change is going to occur, it then act upon
                             \item \textit{reactive}: the capability responds when a change has already happened
                    \end{itemize}
           \item \textit{transit}: a specific behavior of meta-self-awareness capability; it aims to reason about and switch on/off one or more other capabilities
      \end{itemize}
\item \textbf{Interaction}:
      \begin{itemize} 
            \item \textit{link}: the physical and logical relationship between the capabilities, it has four attributes
                    \begin{itemize} 
                             \item \textit{one-to-many}: the capability of a type interacts with many sensors or actuators
                             \item \textit{many-to-many}:  the capability of the same type from different nodes interact with each others
                             \item \textit{one-to-one}: one capability of a type interacts with one capability of another type
                             \item \textit{none}: there is no interactive relationship
                    \end{itemize}
      \end{itemize}
\item \textbf{Topology}:
      \begin{itemize} 
            \item \textit{structure}: how the capabilities are mapped to the components of an architectural instance, it has three attributes
                    \begin{itemize} 
                             \item \textit{combine}: the case where a capability is realized in conjunction with other capabilities within a single component
                             \item \textit{separate}: a single capability is realized using separate components
                             \item \textit{compact}: a capability is realized using exactly one component
                    \end{itemize}
             \item \textit{existence}: the existence of an optional capability of the pattern, it has two attributes
                    \begin{itemize} 
                             \item \textit{exist}: the optional capability is included
                             \item \textit{non-exist}: the optional capability is not included
                    \end{itemize}
      \end{itemize} 
\end{itemize}

The use of these primitives is application specific and depends on whether they would affect the functional and/or non-functional requirement of the self-aware systems. In the following, we explain those primitives and their attributes in great details.

\subsection{Capability}
The capable primitives consists of the 4 level of knowledge in self-awareness principles, which can be realized in one or more components in an architecture instance. The \textit{self-expression} and \textit{meta-self-awareness} capabilities are also belong to this category. These primitives belong to capability can be realized by using different candidate techniques. 

\subsection{Behavior}
In the behavioral primitives, \textit{send} is used to describe the output process of a sender. In particular, it has two attributes: \textit{synchronous} and \textit{asynchronous}. \textit{Synchronous} refer to the scenarios where after the sender sends a request, it needs to wait for a reply from the receiver before transit to the next actions. On the other hand,  \textit{asynchronous} refer to the case that the sender does not require such blocked communication. \textit{Handle} is used to express how the receiver process inputs. \textit{Sequential} handler refer to the cases where incoming data are processed by specific sequence using a queue. On the other hand, \textit{parallel} handler simply process all incoming data in parallel upon its arrival. Both \textit{send} and \textit{handle} primitives can be realized using different candidate techniques. \textit{State} primitive is used to express the behavior of a capability, it depends on the candidate techniques that realize such capability. In particular, In the \textit{reactive} state, the capability responds when a change has already happened, while in the \textit{proactive} state, the capability predicts and reasons about when the change is going to occur, it then act upon \cite{Lemos13}. It is also possible that a capability is realized using both \textit{proactive} and \textit{reactive} attributes. \textit{Transit} is a specific behavior of meta-self-awareness capability; it aims to reason about and switch on/off one or more other capabilities. Similarity, it can be realized using different candidate techniques. The main differences between the candidate techniques of \textit{meta-self-awareness} and those of \textit{transit} is that the former one focus on how to improve and optimize other capabilities; whereas the later on focus on the question of whether we should keep or shutdown certain capabilities? However, we have not seen any application of techniques belong to \textit{transit} primitive in EPiCS.

\subsection{Interaction}
Interactive primitive, \textit{link}, is used to describe the physical and logical relationship between the capabilities as expressed by the multiplicity operators. It has four attributes: \textit{one-to-many}, \textit{many-to-many}, \textit{one-to-one} or \textit{none}. The use of this primitive is constrained by the selected pattern.

\subsection{Topology}
Finally, topological primitives are design for expressing how the capabilities are mapped to the components of an architectural instance. \textit{Structure} primitive has three attributes: \textit{separate} or \textit{compact}. \textit{Combine} is used to describe the cases where a capability is realized in conjunction with other capabilities within a single component; conversely, the \textit{separate} refers to a single capability is realized using separate components; \textit{compact} refers to a capability is realized using exactly one component. In particular, it is possible to have both \textit{separate}  and \textit{compact}  for a capability. \textit{Existence} primitive is simply used to show whether a capability is needed, as certain pattern allows optional capabilities.

A detailed summary of the architectural primitives and a list of possible candidate techniques surveyed from the EPiCS projects has been shown in Table 2.1. Please note that the list of candidate techniques here is not exhaustive, therefore more exampled candidate techniques can be added when appropriate. We omitted some primitives (e.g., the \textit{link} ) here as they do not associated with any concrete techniques. 



Table 2.1: Architectural primitives and their candidate technique

| Architectural Primitives | Candidate Techniques/Options | | | |
|---|---|---|---|---|
| | Technique | Demonstrator | Approach Details | Nature |
| stimulus-awareness | Simple Update | Financial Application [6] | 1. Read from either human inputs or files | Stimulus: market data (price, interest rate, trade), parallelism, precision, IO bandwidth |
| | Auction Invitation/Bidding | Smart Camera [15] | 1. Perform background subtraction (private) 2. Bound boxes candidates (public) 3. Perform valuation of bids based on tracking results (private) | Stimulus: images, auctions, bids, handover, and object tracking results. |
| | Background Substitution | | | |
| | Bounding Boxes Generation | | | |
| | Threshold-based Algorithm | Dynamic Protocol Stack [18] | 1. Simple mapping: map most-used functional block to HW 2. Smart mapping: considers processing time per block | Stimulus: Workload, core utilization, performance, Network condition, link quality, signal-to-noise ratio, bandwidth |
| | Conditions-Actions Rule | HW/SW Platform [16] | 1. Trigger performance management / thermal management | Stimulus: performance counter, local temperature, input data, thermal diode |
| | Simple Update Function | Rhythmic Music Application [8] | 1. Trigger phase and frequency updates upon input from other nodes. | Stimulus: Sound input from other nodes |
| interaction-awareness | Market-based Mechanism | Smart Camera [15] | 1. Create model of objects to be tracked (public) 2. Hand-over process using broadcast 3. React to auctions, bids and handover (public) 4. Define neighbourhood based on auctions (private) | Interaction: Participation in auction & vision graph, Model of objects |
| | Object modelling and matching | | | |
| | Threshold-based Algorithm | Dynamic Protocol Stack [18] | 1. Packet processing engine (stack builder) 2. Source node proposes a set of protocol stacks 3. Meet application / node goals, network conditions 4. Destination node selects one and informs source 5. Both nodes can suggest run-time adaptations | Interaction: Source and destination nodes negotiate protocol stack. |
| time-awareness | Ant Inspired Artificial Pheromones | Smart Camera [15] | 1. Calculate pheromones in vision graph (private) | Time: Evaporation of pheromones |
| | Partial Autocorrelation | Dynamic Protocol Stack [18] | 1. Perform proactive reconfiguration of HW/SW mapping 2. Create fitness function for HW/SW mapping | Time: Historical traffic patterns |
| | 2-Layer Resistor-Capacitor (RC) Network | HW/SW Platform [16] | 1. Learn thermal model of chip | Time: Historical temperature data of the chip |
| | Randomized Hill Climbing | | | |
| | Median Filter with Moving Average | Rhythmic Music Application [8] | 1. Filter the errors in frequency | Time: Historical frequency errors. |
| | Hawkes Point Process Estimation | Financial Application [6] | 1. Perform point process 2. Correlate time series | Time: Historical financial trends |
| | HAC estimation and verification | | | |
| goal-awareness | Error/Confidence Measure | Rhythmic Music Application [8] | 1. Learn the tempo change and how it affects the goals. | Goal: minimizing error of producing music |
| self-expression | Hand-over Mechanism | Smart Camera [15] | 1. Send out auction invitations and bids (public) | Expression: Pan-tilt-zoom (internal), decision to participate in |

| | | | | auction (internal), communication channel – send bids/initiate auctions (external) |
|---|---|---|---|---|
| | | | 2. Automate load balancing | |
| | Static Mapping | Dynamic Protocol Stack [18] | 1. Update the protocol stack mapping<br><br>2. Minimize CPU utilization | Expression: Send normal packets, send messages for negotiation, setup protocol stacks, migrate functional blocks between HW/SW |
| | Brute Force Optimization | HW/SW Platform [16] | 1. Find the optimal thermal mappings | Expression: Start/stop/migrate threads, generate heat on the chip |
| | Simple Update | HW/SW Platform [16] | 1. A confident node will make sound at a steady tempo.<br><br>2. An insecure node will constantly try to adapt its tempo to the other nodes. | Expression: Send tones and make sound. |
| | Gaussian Process<br>Support Vector Machine<br>Reinforcement Learning | Financial Application [6] | 1. Optimize decision values and decisions for financial trading.<br><br>2. Improve design fitness by exploring / tuning several design parameters | Expression: Design configurations, trend prediction decisions |
| meta-self-awareness | Hand-over Mechanism<br>Bandit Solver | Smart Camera [15] | 1. Perform hand-over process using vision graph information (private/public)<br><br>2. Bandit solvers introduce very simple meta-self-awareness (private) | Meta-self-awareness: choose the cameras for the auction, start auction, hand-over, update vision graph |
| | Quality- driven and Threshold-based Algorithm | HW/SW Platform [16] | 1. Select best strategy to meet goals in current situation | Meta-self-awareness: Switch between different scheduling strategies for thermal management |
| | | Rhythmic Music Application [8] | 1. Change scaling of frequency updates based on confidence, i.e. managing the degree of stimulus awareness | Meta-self-awareness: Change the degree of adapting frequency, which is directly controlled by confidence measure |
| send | synchronous function call | Dynamic Protocol Stack [18] | Applied to every capability except self-expression | |
| | | HW/SW Platform [16] | Applied to every capability except self-expression | |
| | | Rhythmic Music Application [8] | Applied to every capability except self-expression | |
| | | Financial Application [6] | Applied to every capability except time-awareness | |
| | synchronous multicast | HW/SW Platform [16] | Applied to self-expression capability | |
| | asynchronous function call | Financial Application [6] | Applied to time-awareness capability | |
| | asynchronous multicast | Dynamic Protocol Stack [18] | Applied to self-expression capability | |
| | | HW/SW Platform [16] | Applied to self-expression capability | |
| | | Smart Camera [15] | Applied to every capability | |
| | asynchronous broadcast | HW/SW Platform [16] | Applied to self-expression capability | |
| | asynchronous stigmergy | Rhythmic Music Application [8] | Applied to self-expression capability | |
| handle | First-Come-First-Serve (sequential) | Dynamic Protocol Stack [18] | Applied to every capability except self-awareness to handle data from external sensors | |
| | | HW/SW Platform [16] | Applied to every capability except self-awareness to handle data from external sensors | |
| | multi-threading (parallel) | Smart Camera [15] | Applied to every capability | |
| | | Rhythmic Music Application [8] | Applied to every capability | |
| | | Financial Application [6] | Applied to every capability | |
| | | Dynamic Protocol Stack [18] | Applied to self-awareness to handle data from external sensors | |
| | | HW/SW Platform [16] | Applied to self-awareness to handle data from external sensors | |



It is worth noting that it is possible to select more than one candidate techniques for a primitives, each work on a particular aspect of a capability. In addition, we do not provide candidate techniques for sensor and actuator as they are highly application specific and in some cases, they are uncontrollable.

\section{The Dependency}

\begin{figure}[h!]
\centering
\includegraphics[width=4in]{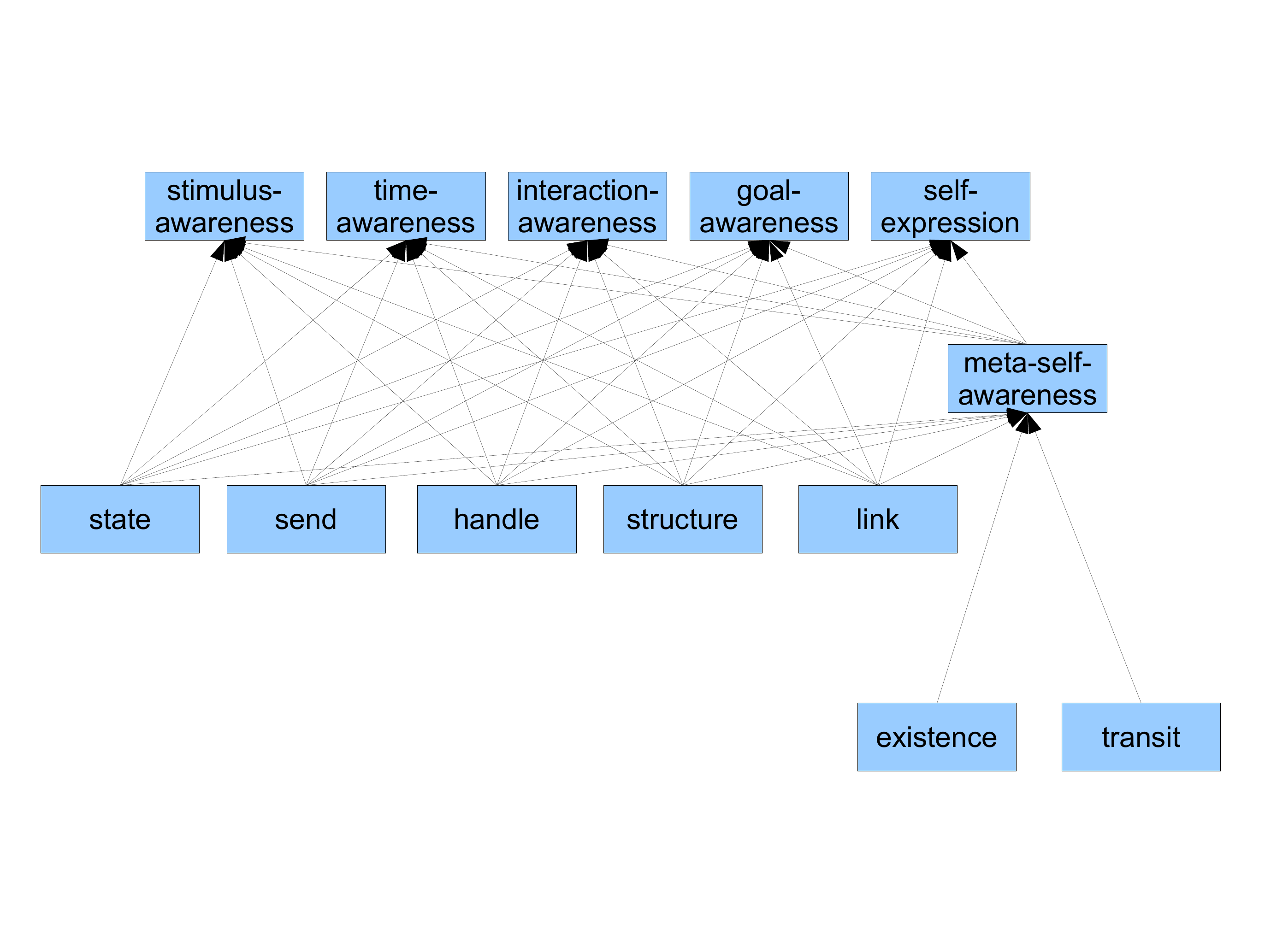}
\caption{Dependency amongst primitives}
\label{fig:dependency}
\end{figure}

It can be clearly seen that dependency exist amongst some of the aforementioned architectural primitives, that is, certain primitives can not be used if another primitive or attribute has not be considered. The dependency has been illustrated in Figure~\ref{fig:dependency}. Note that the arrow here means 'or' relationship, e.g.,  \textit{state} primitive can be considered only if any capability primitive exists.

\chapter{Pattern Driven Methodology for Engineering Self-aware and Self-expressive Systems}

Engineering self-aware and self-expressive systems is a widely important and complex activity. This is because the process involves many decision makings on the possible alternatives (e.g., what technique/attribute  should one apply in order to realize certain level of awareness?)  even in the early stage of development. In addition, it is difficult to thoroughly reason about the consequences of different design alternatives to the functional and non-functional requirements of the systems. Research for engineering methodology has been widely conducted on the area of conventional software architecture \cite{mehtaM03, Kazman2002, Kazmam1998, Al-Naeem2005}, however a systematic approach to the design and engineering of self-aware and self-expressive systems is still in its infancy. In this report, we present a pattern driven methodology to this engineering problem by leveraging on previously proposed patterns and architectural primitives. The methodology contains detailed guidance to make decisions with respect to the possible design alternatives. We evaluate the approach in two aspects: (i) a qualitative evaluation using two case studies: the smart camera networking problem within EPiCS \cite{2013_Esterle_SASOW, esterle_et_al_2014} and the elastic cloud autoscaling problem \cite{tao2013, tao2014online, tao2014}, which is an example outside EPiCS; and (ii) a quantitative assessment by comparing the resulted self-aware and self-expressive system to a conventional and non-self-aware system.

\begin{figure}[h!]
\centering
\includegraphics[width=5in]{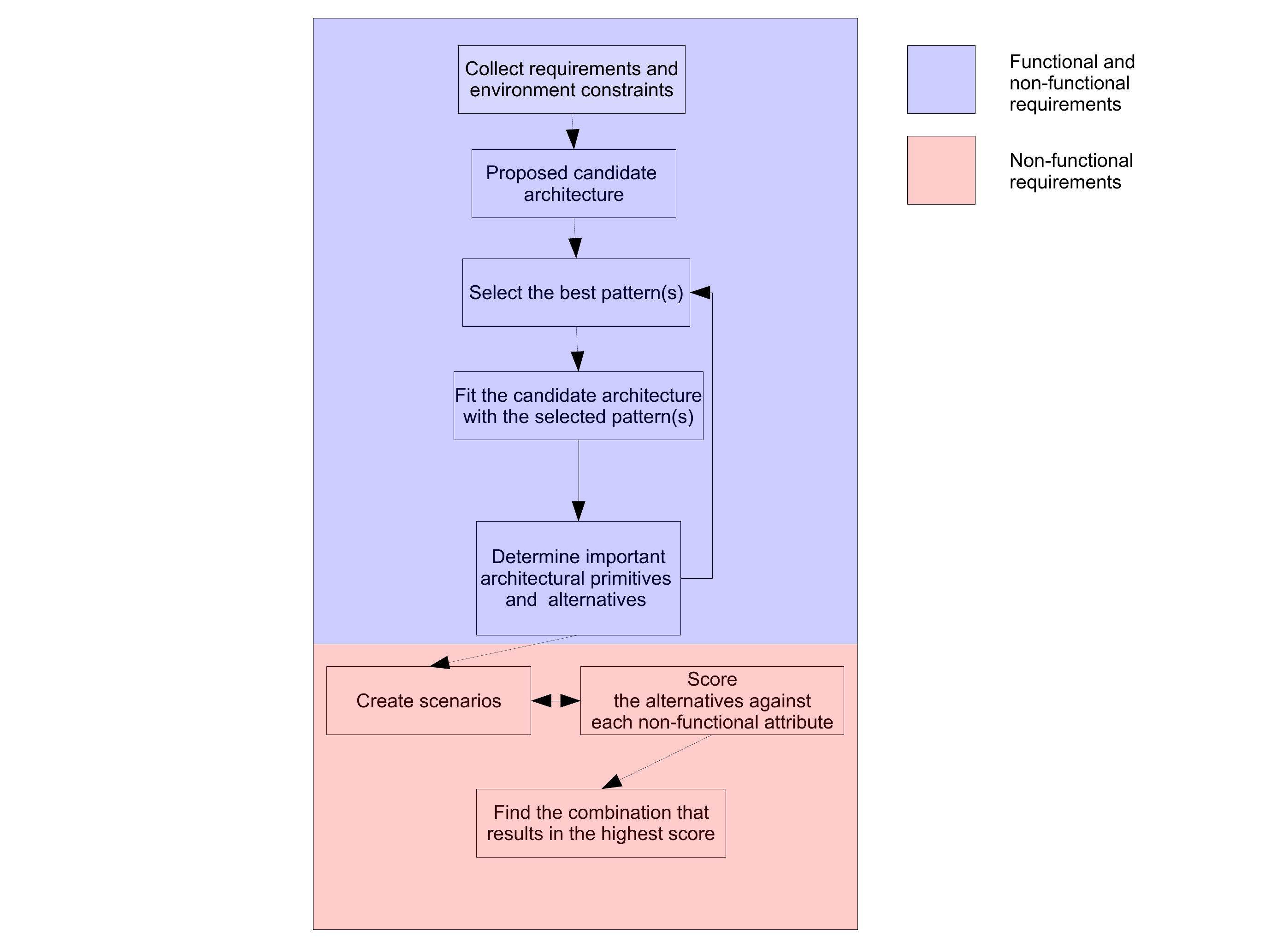}
\caption{Overview of the pattern driven methodology}
\label{fig:methodology}
\end{figure}

\section{The Methodology Overview}

To facilitate a systematic way of building self-aware and self-expressive systems, we proposed a pattern driven methodology leverage on the 8 proposed patterns and the defined architectural primitives. This methodology is a variation of ATAM \cite{Kazman2002} extended by applying patterns and quality-values analysis in relation to the self-awareness principles. In particular, the aim of such methodology is two-folds: (i) select the right pattern(s); and (ii) select the right variation of the chosen pattern(s). As shown in Figure ~\ref{fig:methodology}, we can see that the initial four steps, responsible for selecting the right pattern, are tightly related to the functional and possibly the non-functional requirements of systems. After the selection of pattern, there is an intermediate stage (step 5) aims to determine the important design decisions for selecting variant with respect to the selected pattern. The last three steps, on the other hand, are used to select the right variation of the chosen pattern; and they are associated with the non-functional aspects of the systems. We argue that the separate consideration of requirements for selection of pattern and pattern variation promote a concise and precise design of self-aware and self-expressive systems. In the following, we will see each of the steps in details.

\subsection{Step 1 - Collect Requirements and Constraints}

The first step in this method involves collecting requirements and constraints from the stakeholders and environment for engineering self-aware and self-expressive systems. Similar to the step in ATAM \cite{Kazman2002}, the purpose of this step is to gain depth knowledge about the problem context; to operationalise both functional and non-functional requirements, to facilitate communication between stakeholders, and to develop a common vision of the important activities the system should support.

The requirements could be either functional or non-functional, for instance, \textit{the system should be able to record historical data for analyzing its behavior} is clearly a functional requirement. On the other hand, \textit{the adaptation delay in the system should be no more than 1 minute} is an example of non-functional requirement. Constraints are also another important factor to be considered, in particular, they could come from the stakeholders, e.g., \textit{the hourly cost of infrastructure for running the system should not be more than 10 dollars}; or from the environment, e.g., \textit{the topology of nodes in the context is dynamic as nodes can join/leave on demand}. 

\subsection{Step 2 - Propose Candidate Architecture}

Our methodology assumes that there is an existing candidate architecture of the system.  This is because our self-aware patterns are generic and do not rely on any assumptions about the application domain. Therefore, in order to apply the pattern in practice, it is essential to obtain certain knowledge about the architecture related to the context of given application domain. Often, this knowledge is represented as a collection of components and connectors, which we call them existing candidate architecture. By doing so, we can obtain two major benefits: (i) it provides a clear view about what is really needed by the problem domain from an architecture perspective, and thus assist the engineers to reason about and select the right pattern; (ii) it is possible to refine the existing candidate architecture when mapping the components and connectors to the capabilities of chosen pattern(s). We believe this is a rational assumption as once the important requirements and constraints have been determined, the engineer should be able to propose a candidate architecture based on the obtained information. In addition, design almost never starts from a clean slate: legacy systems, interoperability, and the successes/failures of previous projects all constrain the space of architectures.

The candidate architecture must be described in terms of architectural components/elements. In particular, the architecture should express the module/component view \cite{Kazmam1998} of the system, which is usually used to reason about work assignments and information hiding. In this work, we do not assume any prerequisite of the modeling notations, therefore the engineer could use either the standard ones (e.g., UML)  or create his/her own notations.

\subsection{Step 3 - Select the Best Pattern(s)}

This phase is concerned with selecting the suitable pattern(s) based on the functional and non-functional requirements as well as the constraints. By doing so, the pattern helps the engineer to rethink the domain specific problem in a self-aware computing sense. The patterns differ mainly in terms of the self-awareness capabilities, which play an integral role in satisfying the functional requirements, therefore the selection of pattern is equivalent to the selection of the right set of self-awareness capabilities. 

To promote a systematic approach for pattern selection, we design certain questions and ask the software engineer to consider these questions. The suitable pattern(s) could be determined based on these answers. Specifically, one should answer the following questions for each self-awareness and self-expression capability:

\begin{itemize}
\item What does the capability mean in your problem context?
\item What are the functional requirements that affected by this capability?
\item What are the non-functional requirements that affected by this capability?
\item What are the constraints that could affect this capability?
\item Whether this capability is necessary or beneficial?
\end{itemize}

These questions could lead to the conclusion about whether a capability should be included in the design. If the problem domain requires two patterns, then there can be two sets of answers for each self-aware capability. Finally, Table 3.1 could be used to select the suitable pattern(s).

\begin{table}[H]
\begin{adjustwidth}{-1in}{-1in} 
\begin{center}

\caption{The patterns with respect to the principle of self-awareness. (y means the capability is included in the pattern; o denotes that this capability is optional)}

\tablefirsthead{}
\tablehead{}
\tabletail{}
\tablelasttail{}
\begin{supertabular}{|m{1.8199999cm}|m{1.0999999cm}|m{2.109cm}|m{1.938cm}|m{1.888cm}|m{1.9029999cm}|m{1.326cm}|m{1.4699999cm}|m{1.6489999cm}|}
\hline
~
 &
\centering Basic Pattern &
\centering Basic Information Sharing Pattern &
\centering Coordinated Decision Making Pattern &
\centering Temporal Knowledges Sharing Pattern &
\centering Temporal Knowledges Aware Pattern &
\centering Goal Sharing Pattern &
\centering Goal Sharing Pattern (with time) &
\centering\arraybslash Temporal Goal Aware Pattern\\\hline
\centering Stimulus-awareness &
\centering y &
\centering y &
\centering y &
\centering y &
\centering y &
\centering y &
\centering y &
\centering\arraybslash y\\\hline
\centering Time-awareness &
~
 &
~
 &
~
 &
\centering y &
\centering y &
~
 &
\centering y &
\centering\arraybslash y\\\hline
\centering Interaction-awareness &
~
 &
\centering y &
\centering y &
\centering y &
~
 &
\centering y &
\centering y &
~
\\\hline
\centering Goal-awareness &
~
 &
~
 &
~
 &
~
 &
~
 &
\centering y &
\centering y &
\centering\arraybslash y\\\hline
\centering Meta-self-awareness &
\centering o &
\centering o &
\centering o &
\centering o &
\centering o &
\centering o &
\centering o &
\centering\arraybslash o\\\hline
\end{supertabular}

\end{center}
 \end{adjustwidth}
\end{table}

\subsection{Step 4 - Fit the Selected Pattern(s)}

Once the best pattern(s) has been determined, we can now fit the components/elements from the candidate architecture to the capabilities described in the pattern(s). It is worth noting that, our architectural patterns preserve the flexibility for the concrete architecture; since whether two or more capabilities are combined and fit in one component; or one capability is fit in separate components can be based on the requirements and constraints. It is also an opportunity to refine/improve the candidate architecture during the mappings.

\subsection{Step 5 - Determine the Important Primitives and the Possible Alternatives for Non-functional Requirements}

The next step is to determine the important architectural primitives and the relevant alternatives (i.e., certain form of techniques) for the given problem context. Particularly, we should consider the non-functional requirements here and link them to the architectural primitives, including their attributes and the relevant techniques. Those primitives, their attributes and techniques that could influence the non-functional requirements would be extensively modeled and examined for justifying their benefits during next steps. In addition, this is also an opportunity to eliminate the primitives and techniques, which could be easily selected or are trivial to be considered. For example, if performance requirement is much more important than the other quality attributes and the environment conditions are dynamic, then the behavior primitives (e.g., the primitives send and handle) can be eliminated as it is almost certain that only multicast and parallel interactions are feasible here.

Previously, we have reported on a list of techniques and attributes for each architectural primitives. But not all of them are feasible depends on the problem context. The most important task is to determine the alternatives out of the candidate techniques and attributes. Here, the software engineer could also eliminate the techniques/primitives that could only affect functional requirements or that are useless due to some constraints of the environment. For instance, a supervised learning algorithm would not be useful in the case where only unlabeled data is available. In addition, more appropriate techniques that have not been reported could be added in to the consideration. The common practice of identifying alternative is to consider each technique as independent alternative. However, it is possible to create ensemble of techniques and attributes, therefore we consider such ensemble and  any of this combination as independent alternative.

From this step forward, our methodology would start focusing on the non-functional requirements because they are more difficult to be satisfied due to their compliances are often related to runtime uncertainty. In addition, the non-functional attributes are usually highly sensitive to the variants of patterns. The problem become even more complex when the non-functional requirements are conflict, e.g., accuracy vs overhead.

Up to this step, it is possible to go back to Step 2 if one feels that the chosen pattern(s) is inappropriate. This iterative process can continue till the final suitable pattern(s) is determined.

\subsection{Step 6 - Create Scenarios}

At this stage, we create the important scenarios that could influence each non-functional attribute and likely to occur at runtime. There is no restriction on the number or granularity of scenario per non-functional attribute, for instance, one could have a faulty scenario for a system running in stable state in order to examine the influence to availability attribute; in addition, he/she could also create a scenario under the unstable state (e.g., when new nodes join in or existing nodes leave). It is worth noting that the defined scenarios do not need to be exhaustive as using these scenarios do not imply the developed system is only able to cope with theses scenarios. Selecting pattern variants based on these scenarios could provide confidence about how the self-aware system would perform under the likely scenarios.

It is worth noting that there are cases where it is very difficult if not impossible to design scenarios, especially when these scenarios can not be foreseen and can only be dealt with in real time. In these case, the selection of pattern variants (step 6-8) can be skipped as it is impossible to assess and compare alternatives.  Instead, the proper variants can be determined based on the knowledge of domain experts, e.g., previous publications, experiments, implementations etc.

\subsection{Step 7 - Score the Alternative of Primitives Against each Non-functional Attribute using Analytical or Simulation Models}

In this step, we need to determine the score for each alternative of a primitive against each non-functional attribute under the considered scenarios. The purpose is to assess each alternative and justify their benefits with respect to the non-functional attributes. We assume that there is no or limited dependency between the architectural primitives in terms of how they affect the non-functional attributes, therefore we assess each primitive in isolation. In particular, we aim to gain relative score for every alternative in the context of each architectural primitive. Scoring an alternative can be achieved by using either analytical model or simulation: the former one refers to the empirical or statistical analysis of a particular alternative against a non-functional attribute, e.g., the complexity analysis. On the other hand, the later one uses the quantitative results from some simulations on an alternative under the considered scenarios. In particular, if the techniques of a primitive is difficult to be assessed in neither ways, then the software engineer could score the alternatives using empirical knowledge by assigning weights based on experience  \cite{Al-Naeem2005} . It is worth noting that in cases where the non-functional attributes can only be assessed using multiple primitives, then when scoring the alternatives of a primitive, it is important to ensure that the used alternatives for other primitives are equivalent. For example, when deciding the alternatives of stimulus-awareness in terms of system's overall adaptation quality, the chosen alternative for self-expression should be consistent. Precisely, the scoring process has three phases: 

Firstly, weight the relative importance of different non-functional attributes after negotiation amongst the
stakeholders. In particular, this can be achieved using the pair-wise comparison in AHP \cite{ahp}. This alternative can be
used to measured how much importance of attributes \textit{Q}{\textsubscript{a}} is over the attribute
\textit{Q}\textit{\textsubscript{b}} based on some scale.

Secondly, score each alternative against each non-functional attribute under every defined scenarios. Having this done, we then obtain a matrix of \emph{n} times \emph{m} for the \textit{hth} primitive under a given non-functional attribute \textit{k}, as shown below:

\begin{equation}
\genfrac{}{}{0pt}{0}{}{P_h=\begin{matrix}A_1\\...\\A_n\end{matrix}}\genfrac{}{}{0pt}{0}{\mathit{SC}_1,...,\mathit{SC}_m}{\left(\begin{matrix}S_{1, \ 1}^k&...&S_{\mathit{1, \ m}}^k\\...&...&...\\S_{\mathit{n, \ 1}}^k&...&S_{\mathit{n, \ m}}^k\end{matrix}\right)}
\end{equation}

\noindent where \textit{n} is the total number alternatives of the \textit{hth} primitive  $P_h$  and \textit{m} denotes the total
number of considered scenarios for non-functional attribute \textit{k}.  $\mathit{SC}_m$  denote the \textit{mth}
scenario;  $A_n$  mean the \textit{nth} alternative and  $S_{\mathit{n, \ m}}^k$  denote the score of the \textit{nth}
alternative under the \textit{mth} scenario. In cases where more than one scenario for a non-functional attribute, the
score of a alternative would the aggregative result of the scores under all scenarios. Therefore, we calculate the
total score of an alternative for the \textit{hth} primitive under non-functional attribute \textit{k} by aggregating
its score of all scenarios. The matrix can be then reduced to a vector as shown below:

\begin{equation}
P_h=\begin{matrix}A_1\\...\\A_n\end{matrix}\left(\begin{matrix}S_1^k\\...\\S_n^k\end{matrix}\right)\ \ \ \mathit{s.t.}\ \ S_n^k=\sum
_{x=1}^mS_{\mathit{n, \ x}}^k
\end{equation}

\noindent where  $S_n^k$  is the aggregative score for the \textit{nth} alternative of the \textit{hth} primitive under
non-functional attribute \textit{k}.

Finally, we normalize the raw score for the alternatives in each architectural primitive against a non-functional
attribute \textit{k}. To achieve this, we use the following formula:

\begin{equation}
\mathit{the} \ A_a \ \mathit{of} \ \mathit{primitive} \ P_h=\left\{\begin{matrix}\mathit{NS}_{\mathit{a, \ h}}^k \, \ \ \ \ \ \ \ \ \ \ \ \ \ \;\;\mathit{if} \ \mathit{max} \ Q_k\hfill\null
\\1-\mathit{NS}_{\mathit{a, \ h}}^k\ \ \ \ \ \ \ \ \ \mathit{if} \ \mathit{min} \ Q_k\hfill\null
\end{matrix}\right.\ \ \ \mathit{s.t.}\ \ \mathit{NS}_{\mathit{a, \ h}}^k=\frac{S_a^k}{\sum _{x=1}^pS_x^k}
\end{equation}

\noindent where is  $\mathit{NS}_{\mathit{a, \ h}}^k$  the normalized score for the \textit{ath} alternative of the \textit{hth}
primitive; \textit{p} denotes the total number of alternatives for the \textit{hth} primitive. By doing so, we can
normalize the score scaling from 0 to 1. In case where the non-functional attribute is to be minimized, the final normalized score
would be calculated as  $1-\mathit{NS}_{\mathit{a, \ h}}^k$ .

\subsection{Step 8 - Find the Best Alternatives for the Final Architecture View}

Once all the scores of alternatives have been obtained, the final task is to identify the best combination that
produces the highest total score. Specifically, we need to maximizing the formula below:

\begin{equation} \label{eq:solve}
\mathit{argmax}\sum _{k=1}^x(w_k\times \sum _{h=1}^y\mathit{NS}_{\mathit{s, \ h}}^k)
\end{equation}

\noindent where  $\mathit{NS}_{\mathit{s,\ h}}^k$  is the score of the \textit{sth} selected alternatives for the \textit{hth}
primitive;  $w_k$  is the weight for the \textit{kth} non-functional attribute. \textit{x} and \textit{y} are the total
number of non-functional attributes and architectural primitives respectively. It is easy to see that Equation \ref{eq:solve} can be solved by any optimization algorithms, and finally our
methodology provides a pattern based architecture with the best selected alternatives for all architectural primitives.

\section{Qualitative and Quantitative Evaluation}

We evaluate the methodology qualitatively using two scenarios: a cloud autoscaling case study and a smart camera networks case study. We also quantitatively assess the resulted system by comparing against a non-self-aware system for both case studies.

\subsection{Cloud Autoscaling Case Study}

\subsubsection{Introduction and Background}

\textstyleDefaultParagraphFont{\textcolor{black}{In cloud computing paradigm, the cloud-based services are deployed as
Software as-a-Service (SaaS) and are typically supported by the software stack in the Platform as-a-Service (PaaS)
layer \cite{paas}. They are also supported with Virtual Machines (VM) and hardware within the Infrastructure as-a-Service
(IaaS) layer \cite{iaas}. Under changing environmental conditions (e.g., workload, size of incoming job etc.), it is important
to manage and control the Quality of Service (QoS) of cloud-based services.
}}\textstyleDefaultParagraphFont{\textcolor{black}{By QoS, we refer to the non-functional attributes (e.g., throughput)
experienced by the end-users who use these services.}}\textstyleDefaultParagraphFont{\textcolor{black}{ In particular,
the QoS can be managed by various control knobs, which include software (e.g., threads) and hardware resources (e.g.,
CPU) in a shared infrastructure. Ho}}\textstyleDefaultParagraphFont{\foreignlanguage{english}{\textcolor{black}{however,
inappropriate use of software and hardware resources could result in large rental cost to the service. In this report,
we refer to these control knobs and environmental conditions in the cloud as primitives.}}}

\textstyleDefaultParagraphFont{With the context in mind, the term elasticity \cite{cloudelasiticity} in cloud refers to the ability to
adaptively scale control knobs to match the demand of cloud-based services at runtime. Given the uncertainty and
dynamics of QoS, there is an increasing demand on cloud where the realization of elasticity can be managed without
human intervention. In particular, for all cloud-based services, the cloud should dynamically and continuously select
an elastic strategy, which is the combinatorial decision of configurations for various control knobs; this process is
called autoscaling.}

As an important mechanism to realize elasticity in \ the cloud, autoscaling is an automatic and elastic process, typically running on each Physical Machine (PM), that adapts software configurations and hardware resources provisioning on-demand according to the changing environmental conditions. We argue that the autoscaling should be
cost and QoS optimized; more precisely, upon each provisioning and de-provisioning process, our aim is to design an
autoscaling mechanism that adaptively optimize QoS attributes and cost (i.e., cost of
\textstyleDefaultParagraphFont{\foreignlanguage{english}{\textcolor{black}{software and hardware resources}}}) for
\textbf{all} cloud-based applications and services at runtime by autoscaling to the best
\textstyleDefaultParagraphFont{\textcolor{black}{combinatorial values of control knobs. Due to the on-demand and
dynamic nature of cloud, human intervention and traditional analytical approaches are limited to achieve this goal.
Therefore, we intend to build a self-aware and self-expressive system to perform efficient and intelligent autoscaling.
To this end, this system should dynamically model QoS, which could use the primitives as inputs and predict the likely
QoS value as an output. These models can better express QoS sensitivity. By sensitivity, we are interested in
}}\textstyleDefaultParagraphFont{\textit{\textcolor{black}{which
}}}\textstyleDefaultParagraphFont{\textcolor{black}{(e.g., are throughput and CPU correlated?),
}}\textstyleDefaultParagraphFont{\textit{\textcolor{black}{when
}}}\textstyleDefaultParagraphFont{\textcolor{black}{(i.e., at which point in time they are correlated?) and
}}\textstyleDefaultParagraphFont{\textit{\textcolor{black}{how
}}}\textstyleDefaultParagraphFont{\textcolor{black}{(i.e., the magnitude of primitives in correlation) the primitives
correlate with QoS. In particular, the system must}}\textcolor{black}{ consider dynamics and uncertainty caused by
workload and QoS interference (both service-level and VM-level )}\textcolor{black}{.} By QoS interference, \ we refer
to scenarios where a service suffers wide disparity in its QoS depends on the fluctuated primitives of co-located
services on the same VM and co-hosted VMs on the same Physical Machine (PM). This is a typical consequence of resources
contention. In addition, the system shall also take objective-dependency into account as the objectives of a
cloud-based service could be either conflicted or harmonic with the objectives form the same service \ (intra-service
dependency) or the other co-located services (inter-service dependency).

\subsubsection{Terminology}

We advocate a fine-grained approach to the modeling and analysis of QoS. To achieve this, we decompose the notion of
primitives into two major categories: these are \textbf{Environmental Primitives (EP) }and \textbf{Control Primitives
(CP). }We posit that CP can be either software or hardware, which could be managed by cloud providers to support QoS
provisioning. In particular, software control primitives are software tactics and configurations; such as the number of
threads in thread pool and its life time, the number of connections in database connection pool, security and load
balancing policies etc. Whereas, hardware control primitives are computational resources provisioning, such as CPU,
memory and bandwidth. Software and hardware control primitives rely on the PaaS and IaaS layers respectively. In
particular, it is a non- trivial task to consider software control primitives when QoS modeling in the cloud as they
tend to influence QoSs significantly. On the other hand, we look at environmental primitives in the context of highly
dynamic scenarios, which reflect the cloud setting. The environmental primitives can significantly influence the QoS.
The providers often can not predict and fully control their behavior. Examples include unbounded workload and
unpredictable bound received data etc. If the provider would be able to predict and control the presence of these
scenarios, these can be then considered as control primitives.

We assume that cloud-based applications are composed of one or more services, each with its QoS requirements and can
experience different environmental changes (e.g., changes in workload). These services are deployed on a cloud software
stack, which can be setup using various configurations and tactics. In addition, they are hosted on the cloud
infrastructure, where resources are shared via VMs. As a result, the control knobs and environmental conditions could
significantly influences their QoSs. In distributed environment like cloud, each tier in a multi-tiers application,
composed of concrete services
\foreignlanguage{english}{\{}\foreignlanguage{english}{\textit{S}}\foreignlanguage{english}{\textit{\textsubscript{1}}}\foreignlanguage{english}{,
}\foreignlanguage{english}{\textit{S}}\foreignlanguage{english}{\textit{\textsubscript{2}}}\foreignlanguage{english}{,
…
}\foreignlanguage{english}{\textit{S}}\foreignlanguage{english}{\textit{\textsubscript{i}}}\foreignlanguage{english}{\}}
may have multiple replicas deployed on different VMs. The replica of a tier running on a VM is assumed to have the
replicas \ of its services running on the same VM. In this work, we refer to the replicas of concrete services as
service-instances: the \textit{j-th} service-instance of the \textit{i-th} concrete service is denoted by
\foreignlanguage{english}{\textit{S}}\foreignlanguage{english}{\textit{\textsubscript{ij}}}. Unlike existing work,
which focus on realizing elasticity at the application and VM level, we aim to adaptively optimize the QoS attributes
and rental cost of utilizing control knobs for each individual service-instance, considering the QoS interferences
caused by the co-located service-instances on a VM and the co-hosted VMs on a PM.

In addition, we do not consider global resources contention caused by shortage in cloud capacity; our architecture works
for cases where software and hardware resources tend to be available, which is normal in a cloud environment.
Henceforth, we assume that the maximum demand of software and hardware resources for all cloud service-instances (e.g.,
according to their budget) should be satisfied by the capability of the cloud provider. Under such assumption, we
eliminate extreme cases where the capacity of cloud provider reaches its limits causing likely global resources
contention. This is because the increasing demand of each service-instance would eventually be satisfied by scale
up/out as long as the cost does not exceed the budget. We believe this is a reasonable assumption as in realistic
scenarios, proper admission control can be applied to restrict the number of cloud-based service-instances. Moreover,
in case where the cloud provider actually encounters capacity shortage, the unsatisfied services can be switched to an
alternative provider via a cloud selection mechanism, which presumably hold our assumption. However, the design of
admission control and selection mechanism is outside the scope of this work.

\subsubsection{Objective and Models}

\textstyleDefaultParagraphFont{\textcolor{black}{We formulate an “online” QoS model, which captures both dynamic
sensitivity and interference with respect to the
}}\textstyleDefaultParagraphFont{\textcolor{black}{selected}}\textstyleDefaultParagraphFont{\textcolor{black}{
primitives over time. The model at given sampling interval t is formally expressed as:}}

\begin{equation}
\mathit{QoS}_k^{\mathit{ij}}(t)=f(\mathit{SP}_k^{\mathit{ij}}(t),\delta )
\end{equation}

\noindent \textstyleDefaultParagraphFont{\foreignlanguage{english}{\textcolor{black}{where }}} $\mathit{QoS}_k^{\mathit{ij}}(t)$
\textstyleDefaultParagraphFont{\foreignlanguage{english}{\textcolor{black}{ is the average value of
}}}\textstyleDefaultParagraphFont{\foreignlanguage{english}{\textit{\textcolor{black}{k-th }}}}\textstyleDefaultParagraphFont{\foreignlanguage{english}{\textcolor{black}{
QoS of
}}}\textstyleDefaultParagraphFont{\foreignlanguage{english}{\textit{\textcolor{black}{S}}}}\textstyleDefaultParagraphFont{\foreignlanguage{english}{\textit{\textcolor{black}{\textsubscript{ij }}}}}\textstyleDefaultParagraphFont{\foreignlanguage{english}{\textcolor{black}{
 at \ interval
}}}\textstyleDefaultParagraphFont{\foreignlanguage{english}{\textit{\textcolor{black}{t}}}}\textstyleDefaultParagraphFont{\foreignlanguage{english}{\textcolor{black}{.
}}}\textstyleDefaultParagraphFont{\foreignlanguage{english}{\textit{\textcolor{black}{f }}}}\textstyleDefaultParagraphFont{\foreignlanguage{english}{\textcolor{black}{
is the QoS function, which dynamically changes at runtime. }}} $\delta $
\textstyleDefaultParagraphFont{\foreignlanguage{english}{\textcolor{black}{refers to any other inputs that are required
by the algorithm to train \textit{f} apart from the primitives. Examples of other inputs may include historical QoS values and
tuning variables. To handle QoS interferences, we denote the input }}} $\mathit{SP}_k^{\mathit{ij}}(t)$
\textstyleDefaultParagraphFont{\foreignlanguage{english}{\textcolor{black}{of Eq. 3.5 as the
}}}\textstyleDefaultParagraphFont{\foreignlanguage{english}{\textcolor{black}{selected}}}\textstyleDefaultParagraphFont{\foreignlanguage{english}{\textcolor{black}{
primitives matrix of }}} $\mathit{QoS}_k^{\mathit{ij}}(t)$
\textstyleDefaultParagraphFont{\foreignlanguage{english}{\textcolor{black}{ at interval
}}}\textstyleDefaultParagraphFont{\foreignlanguage{english}{\textit{\textcolor{black}{t}}}}\textstyleDefaultParagraphFont{\foreignlanguage{english}{\textcolor{black}{.
}}}\textstyleDefaultParagraphFont{\foreignlanguage{english}{\textcolor{black}{This
matri}}}\textstyleDefaultParagraphFont{\foreignlanguage{english}{\textcolor{black}{x contains the
selected}}}\textstyleDefaultParagraphFont{\foreignlanguage{english}{\textcolor{black}{
}}}\textstyleDefaultParagraphFont{\foreignlanguage{english}{\textcolor{black}{primitive inputs of }}}
$\mathit{QoS}_k^{\mathit{ij}}(t)$ \textstyleDefaultParagraphFont{\foreignlanguage{english}{\textcolor{black}{ and it is
updated online. }}}

\textstyleDefaultParagraphFont{\foreignlanguage{english}{\textcolor{black}{In the context of cloud, utilizing control
primitives may be subject to certain monetary cost to the service owners, therefore the total costs model
f}}}\textstyleDefaultParagraphFont{\foreignlanguage{english}{\textcolor{black}{or
}}}\textstyleDefaultParagraphFont{\foreignlanguage{english}{\textit{\textcolor{black}{S}}}}\textstyleDefaultParagraphFont{\foreignlanguage{english}{\textit{\textcolor{black}{\textsubscript{ij }}}}}\textstyleDefaultParagraphFont{\foreignlanguage{english}{\textcolor{black}{
can be represented as:}}}

\begin{equation}
\mathit{Cost}^{\mathit{ij}}=\sum _{a=1}^ng(\mathit{CP}_a^{\mathit{ij}}(t),P_a)
\end{equation}

\noindent \textstyleDefaultParagraphFont{\textcolor{black}{where g is the fixed and unified cost function for each type of
}}\textstyleDefaultParagraphFont{\foreignlanguage{english}{\textcolor{black}{control
primitives}}}\textstyleDefaultParagraphFont{\textcolor{black}{, and n is the total number of
}}\textstyleDefaultParagraphFont{\foreignlanguage{english}{\textcolor{black}{control
primitive }}}\textstyleDefaultParagraphFont{\textcolor{black}{  type that used by service-instance
}}\textstyleDefaultParagraphFont{\textit{\textcolor{black}{S}}}\textstyleDefaultParagraphFont{\textit{\textcolor{black}{\textsubscript{ij }}}}\textstyleDefaultParagraphFont{\textcolor{black}{
to supports its QoS attributes. }} $\mathit{CP}_a^{\mathit{ij}}(t)$ \textstyleDefaultParagraphFont{\textcolor{black}{
is the amount of the
}}\textstyleDefaultParagraphFont{\textit{\textcolor{black}{a-th}}}\textstyleDefaultParagraphFont{\textcolor{black}{
}}\textstyleDefaultParagraphFont{\foreignlanguage{english}{\textcolor{black}{control
primitive}}}\textstyleDefaultParagraphFont{\textcolor{black}{ provision for
}}\textstyleDefaultParagraphFont{\textit{\textcolor{black}{S}}}\textstyleDefaultParagraphFont{\textit{\textcolor{black}{\textsubscript{ij }}}}\textstyleDefaultParagraphFont{\textcolor{black}{
at interval t. }} $P_a$ \textstyleDefaultParagraphFont{\textcolor{black}{denotes the corresponding price per unit of
the }}\textstyleDefaultParagraphFont{\textit{\textcolor{black}{a-th }}}\textstyleDefaultParagraphFont{\textcolor{black}{
}}\textstyleDefaultParagraphFont{\foreignlanguage{english}{\textcolor{black}{control
primitive}}}\textstyleDefaultParagraphFont{\textcolor{black}{. In this work, we assume that the price of each
}}\textstyleDefaultParagraphFont{\foreignlanguage{english}{\textcolor{black}{control
primitive}}}\textstyleDefaultParagraphFont{\textcolor{black}{ type is fixed for all service providers and their
service-instances. It is worth noting that the hardware
}}\textstyleDefaultParagraphFont{\foreignlanguage{english}{\textcolor{black}{control primitives
}}}\textstyleDefaultParagraphFont{\textcolor{black}{\ (e.g., CPU and memory) can be only provisioned for each VM
whereas the cost model is per-service based, thus the price of a hardware
}}\textstyleDefaultParagraphFont{\foreignlanguage{english}{\textcolor{black}{control
primitive}}}\textstyleDefaultParagraphFont{\textcolor{black}{ should be equally proportioned to each of the
service-instances deployed on the provisioned VM. }}

\textstyleDefaultParagraphFont{\foreignlanguage{english}{\textcolor{black}{To achieve globally-optimal QoS and cost in
elastic cloud via autoscaling, we aim at adaptively and dynamically determine and scale to the control primitive
configurations, which supports the best of all QoS attributes (Eq. 3.5) with minimal costs (Eq. 3.6) for all
service-instances in the cloud. In this work, we apply a linear weighted-sum aggregation to express the global result
for QoS attributes and costs of different service-instances in the cloud. Formally, at any given interval t, we aim to
optimize the global objective by maximizing the function in Eq. 3.7.}}}

\begin{equation}
\sum _{i=1}^n\sum _{j=1}^mw'_{\mathit{ij}}\cdot (\sum _a^lw_a\cdot \mathit{QoS}_a^{\mathit{ij}}(t)-\sum _b^rw_b\cdot
\mathit{QoS}_b^{\mathit{ij}}(t)-w_{(l+r+1)}\cdot \mathit{Cost}^{\mathit{ij}})
\end{equation}

\noindent \textstyleDefaultParagraphFont{\foreignlanguage{english}{\textcolor{black}{where
}}}\textstyleDefaultParagraphFont{\foreignlanguage{english}{\textit{\textcolor{black}{n}}}}\textstyleDefaultParagraphFont{\foreignlanguage{english}{\textcolor{black}{
and
}}}\textstyleDefaultParagraphFont{\foreignlanguage{english}{\textit{\textcolor{black}{m}}}}\textstyleDefaultParagraphFont{\foreignlanguage{english}{\textcolor{black}{
are the total number of services and their instances in the cloud;
}}}\textstyleDefaultParagraphFont{\foreignlanguage{english}{\textit{\textcolor{black}{w'}}}}\textstyleDefaultParagraphFont{\foreignlanguage{english}{\textit{\textcolor{black}{\textsubscript{ij}}}}}\textstyleDefaultParagraphFont{\foreignlanguage{english}{\textcolor{black}{\textsubscript{
}}}}\textstyleDefaultParagraphFont{\foreignlanguage{english}{\textcolor{black}{is the weight for each service-instance.
Because the global objective is to maximize Eq. 3.7, we need to carefully place the maximized QoS (e.g., throughput) and
the minimized ones (e.g., response time); thus
}}}\textstyleDefaultParagraphFont{\foreignlanguage{english}{\textit{\textcolor{black}{l}}}}\textstyleDefaultParagraphFont{\foreignlanguage{english}{\textcolor{black}{
and
}}}\textstyleDefaultParagraphFont{\foreignlanguage{english}{\textit{\textcolor{black}{r}}}}\textstyleDefaultParagraphFont{\foreignlanguage{english}{\textcolor{black}{
are the total number of the maximized and minimized QoS for
}}}\textstyleDefaultParagraphFont{\foreignlanguage{english}{\textit{\textcolor{black}{S}}}}\textstyleDefaultParagraphFont{\foreignlanguage{english}{\textit{\textcolor{black}{\textsubscript{ij }}}}}\textstyleDefaultParagraphFont{\foreignlanguage{english}{\textcolor{black}{
respectively;
}}}\textstyleDefaultParagraphFont{\foreignlanguage{english}{\textit{\textcolor{black}{w}}}}\textstyleDefaultParagraphFont{\foreignlanguage{english}{\textit{\textcolor{black}{\textsubscript{a}}}}}\textstyleDefaultParagraphFont{\foreignlanguage{english}{\textcolor{black}{,
}}}\textstyleDefaultParagraphFont{\foreignlanguage{english}{\textit{\textcolor{black}{w}}}}\textstyleDefaultParagraphFont{\foreignlanguage{english}{\textit{\textcolor{black}{\textsubscript{b}}}}}\textstyleDefaultParagraphFont{\foreignlanguage{english}{\textcolor{black}{\textsubscript{
\ }}}}\textstyleDefaultParagraphFont{\foreignlanguage{english}{\textcolor{black}{and
}}}\textstyleDefaultParagraphFont{\foreignlanguage{english}{\textit{\textcolor{black}{w}}}}\textstyleDefaultParagraphFont{\foreignlanguage{english}{\textit{\textcolor{black}{\textsubscript{(l+r+1)}}}}}\textstyleDefaultParagraphFont{\foreignlanguage{english}{\textcolor{black}{
are refer to the corresponding weight of the QoS and cost for
}}}\textstyleDefaultParagraphFont{\foreignlanguage{english}{\textit{\textcolor{black}{S}}}}\textstyleDefaultParagraphFont{\foreignlanguage{english}{\textit{\textcolor{black}{\textsubscript{ij}}}}}\textstyleDefaultParagraphFont{\foreignlanguage{english}{\textcolor{black}{.
In addition, the optimization of Eq.3.7 should be subject to the constraint of budget and SLA}}}

In the following, we qualitatively evaluate the proposed methodology by showing how it can be applied to engineer self-aware and self-expressive system in the cloud case study. We also show the experiments that compare the resulted system with a non-self-aware system. To simplify the explanation, we only consider the throughput and cost of cloud-based services as the objectives that need to be maintained.

\subsubsection{Step 1 - Collect Requirements and Constraints}

After negotiated with the stakeholders and analyzed the environment, the requirements and constraints are presented as the following Table:

\begin{table}[H]
\begin{adjustwidth}{-1in}{-1in} 
\begin{center}

\caption{The functional, non-functional requirements and constraints for the cloud case study.}

\tablefirsthead{}
\tablehead{}
\tabletail{}
\tablelasttail{}
\begin{supertabular}{|m{8.3cm}|m{8.3cm}|}
\hline
{\bfseries Functional Requirements} &
{\bfseries \ Explanation}\\\hline
The system must record historical data for analyzing the behaviors of cloud-based services. &
The data could be logs or real time measurement collected by the sensors.\\\hline
The system must able to control both software and hardware control primitives. &
Both types of control knobs could influence QoS significantly. \\\hline
The system must be aware of the changes in workload and deployment. &
This is the cause of dynamic and uncertainty in cloud\\\hline
The system should be able to aware of QoS interference. &
This could influence the autoscaling decisions.\\\hline
The system should support both vertical and horizontal scaling. &
Both actions could help to improve QoS.\\\hline
The system must cope with the conflicted objectives. &
This could influence the autoscaling decisions.\\\hline
The system must aware of the functional dependency between cloud-based services across nodes. &
This could influence the autoscaling decisions.\\\hline
The system should be able to cope with any given QoS attributes and cost objective of cloud-based services.  &
~
\\\hline
The system should be able to cope with any runtime changes of QoS and cost objectives made by the service providers. &
This is something could occur in the cloud.\\\hline
{\bfseries Non-functional Requirements} &
~
\\\hline
Accuracy: the accuracy of QoS modeling should be no less than 75\% to the actual QoS value. &
This could influence the effectiveness of autoscaling decisions.\\\hline
Adaptation Quality: the QoS of \ the cloud-based services that being managed should not be worse than 20\% of the
threshold in SLA for more than 5 mins. &
This is only applicable in case where the budget is allowed.\\\hline
Overhead: the overhead for making autoscaling decision should be less than 200 seconds. &
This could influence the effectiveness of autoscaling decisions.\\\hline
Reliability: to what extent the designer believe that the alternative would work at runtime when there is an emergent
scenario occurs? &
This can assess the confidence of simulation under unknown events at runtime (e.g., different workload) \ \\\hline
{\bfseries Constraints} &
~
\\\hline
VM can be added or removed &
Due to the dynamic cloud\\\hline
Service can be added or removed &
Due to the dynamic cloud\\\hline
Workload for each cloud-based service is fluctuated. &
Due to the dynamic cloud\\\hline
The cost of the cloud-based services that being managed should not exceed its defined budget. &
Should not let the cloud consumer pay more than they would like, even with the cost of worse QoS.\\\hline
QoS interference occurs once there are contentions, in which case the QoS could be negatively affected. &
Could caused by workload or improper configuration of control primitives.\\\hline
\end{supertabular}

\end{center}
 \end{adjustwidth}
\end{table}

\subsubsection{Step 2 - Propose Candidate Architecture}

Based on the aforementioned requirements and constraint, we scratch the an initial version of architectural view, as illustrated in the Figure below:

\begin{figure}[h!]
\centering
\includegraphics[width=5in]{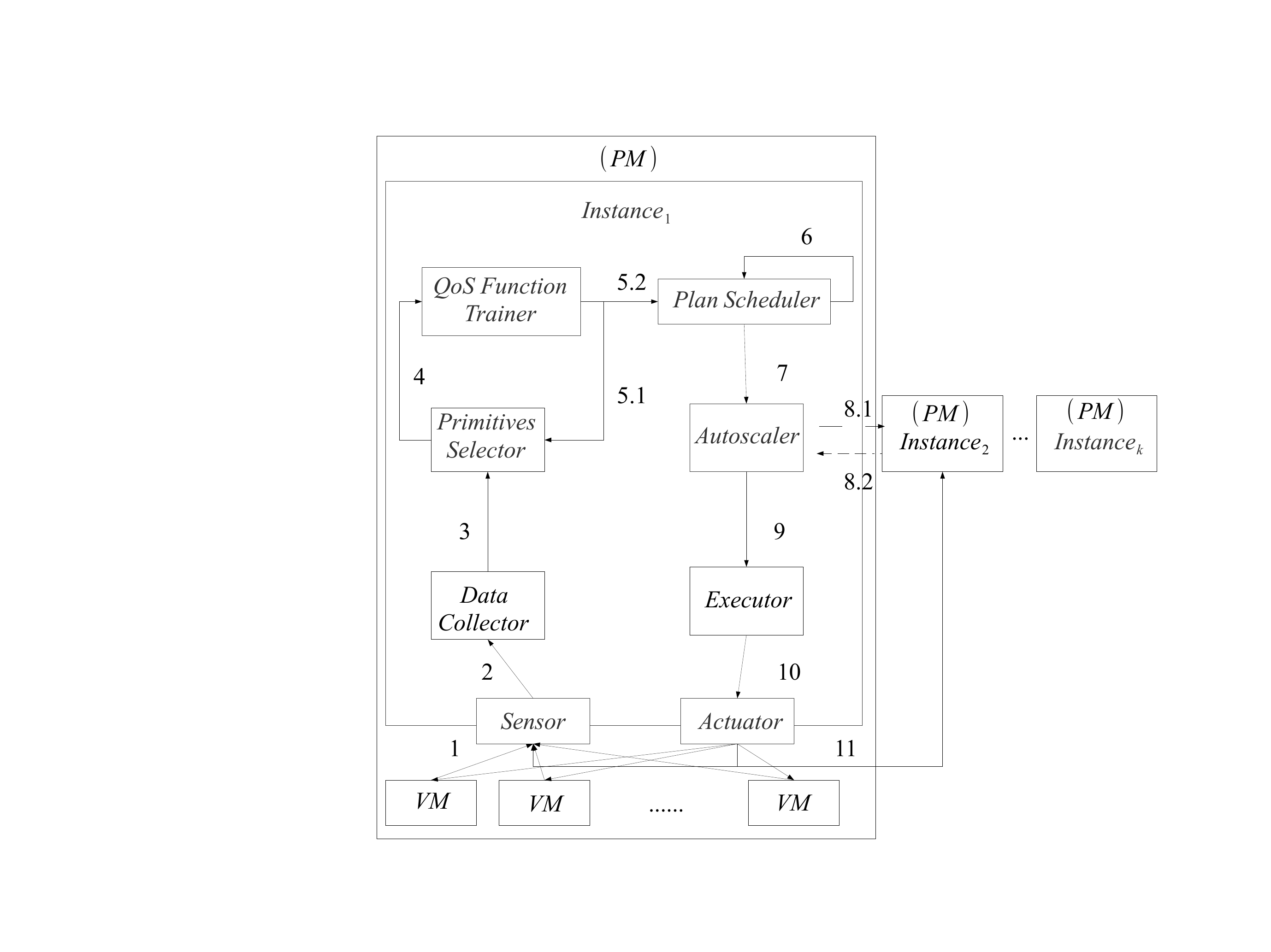}
\caption{The proposed architecture}
\label{fig:cloudproposed}
\end{figure}

\textstyleDefaultParagraphFont{\foreignlanguage{english}{\textcolor{black}{As we can see that the architecture is
deployed as distributed instances, each of which running on a separate VM (e.g.,
}}}\textstyleDefaultParagraphFont{\foreignlanguage{english}{\textit{\textcolor{black}{Dom0
}}}}\textstyleDefaultParagraphFont{\foreignlanguage{english}{\textcolor{black}{on Xen \cite{xen}) on every PM (node) in the
cloud. }}}The workflow of the proposed architecture has been shown in Figure ~\ref{fig:cloudproposed}. More precisely, the \textit{sensor }on
each PM collects the data (e.g., QoS values, CP usages and EP values) from the underlying VMs and service-instances;
and possibly from other PMs due to functional dependency (step 1). In addition, the sensor could sense deployment
changes and QoS sensitivity changes from other PMs. Next in step 2, the sensor passes raw information it received to
the \textit{Data Collector} for normalising the data. At step 3, the \textit{Primitives Selector} \textit{\ }receives
both current and historical data after normalisation, which would be used to determine the inputs of models. The
\textit{QoS Function Trainer }would be used to train the function (step 4-5.1). Once QoS models have been generated,
the propagation goes to step 5.2. In particular, the \ adaptation can be triggered if one or more of the following
symptoms is detected: 

\begin{itemize}
\item Symptoms 1: Proactively detect if the QoS of a service- instance is likely to violate SLA constraint for \textit{k} intervals by using the QoS models. 
\item Symptoms 2: Reactively detect if the QoS of a service- instance has violated its SLA constraint for \textit{k} intervals and/or if the utilisation of a CP has violated the constraint for \textit{k} intervals. 
\item Symptoms 3: occurrence of QoS sensitivity changes and deployment changes. 
\end{itemize}

All symptoms are handled by the\textit{ Plan Scheduler }component, which would be responsible for deciding whether to
trigger the Autoscaling decision making process or the area of effect caused by the violations (step 6). The
\textit{Autoscaler }component is designed to \ dynamically search the best adaptation strategies toward the optimal
result, using the QoS and cost models (step 7). In particular, the \textit{Autoscaler} of each node is triggered
independently and asynchronously. There are cases where the optimisation for a better autoscaling decision need to
communicate (for obtaining external QoS models) with other nodes because of the functional dependency between services.
In addition, it is critical to ensure the same objective is not optimised simultaneously on more than one nodes. These
processes are expressed as step 8.1 and 8.2. 

Once the elastic strategy is determined, the process proceeds to the \textit{Executor }via step 9. In particular, The
\textit{Executor}\textit{ }is responsible for determining which concrete actions (e.g., scale up/down, in/out and/or VM
migration and replication etc) need to be taken in order to fulfil the elastic strategy. In this work, we consider both
vertical and horizontal scaling and apply a simple solution to determine the actions, this is: we always try vertical
scaling (i.e., scale up/down) first before horizontal scaling (i.e., scale out/in). This is because horizontal scaling
is usually more expensive than vertical scaling. As for the VM migration/ replication decision, we always choose the
one that result in smaller overhead based on a predefined VM profiling pattern. Finally, the actions are taken by the
\textit{Actuator} via step 10 and 11. 

\subsubsection{Step 3 - Select the Best Pattern(s)}

We now select the pattern using the questions presented previously for each self-awareness and self-expression capability:

\begin{table}[H]
\begin{adjustwidth}{-1in}{-1in} 
\begin{center}

\caption{Questions and answers for deciding whether to include stimulus-awareness.}

\tablefirsthead{}
\tablehead{}
\tabletail{}
\tablelasttail{}
\begin{supertabular}{|m{4.005cm}|m{12.562cm}|}
\hline
\multicolumn{2}{|m{16.766998cm}|}{{\bfseries Stimulus-awareness}}\\\hline
What does the capability mean in your problem context. &
The ability to aware of newly-measured data (data for environmental primitives and control primitives ), violations of
QoS and utilisation thresholds, QoS sensitivity changes and deployment changes.\\\hline
What are the functional requirements that affected by this capability? &
1. The system must be aware of the changes in workload and deployment.

~

2. The system should be able to aware of QoS interference. 

~

3. The system must aware of the functional dependency between cloud-based services.\\\hline
What are the non-functional requirements that affected by this capability? &
1. Accuracy

2. Overhead

3. Adaptation Quality\\\hline
What are the constraints that could affect this capability? &
1. VM can be added or removed.

~

2. Service can be added or removed.

~

3. Workload for each cloud-based service is fluctuated.

~

4. QoS interference occurs once there are contentions, in which case the QoS could be negatively affected.\\\hline
Whether this capability is \ necessary or beneficial? &
Yes\\\hline
\end{supertabular}

\end{center}
 \end{adjustwidth}
\end{table}

\begin{table}[H]
\begin{adjustwidth}{-1in}{-1in} 
\begin{center}

\caption{Questions and answers for deciding whether to include time-awareness.}

\tablefirsthead{}
\tablehead{}
\tabletail{}
\tablelasttail{}
\begin{supertabular}{|m{4.005cm}|m{12.562cm}|}
\hline
\multicolumn{2}{|m{16.766998cm}|}{{\bfseries Time-awareness}}\\\hline
What does the capability mean in your problem context. &
The ability to aware of historical behaviors of cloud-based services, the continuous consequences of autoscaling
decisions and the emergent events that occurred in the past.\\\hline
What are the functional requirements that affected by this capability? &
1. The system must record historical data for analyzing the behaviors of cloud-based services.\\\hline
What are the non-functional requirements that affected by this capability? &
1. Accuracy

2. Adaptation Quality

3. Overhead\\\hline
What are the constraints that could affect this capability? &
~
\\\hline
Whether this capability is \ necessary or beneficial? &
Yes\\\hline
\end{supertabular}

\end{center}
 \end{adjustwidth}
\end{table}

\begin{table}[H]
\begin{adjustwidth}{-1in}{-1in} 
\begin{center}

\caption{Questions and answers for deciding whether to include interaction-awareness.}

\tablefirsthead{}
\tablehead{}
\tabletail{}
\tablelasttail{}
\begin{supertabular}{|m{4.005cm}|m{12.562cm}|}
\hline
\multicolumn{2}{|m{16.766998cm}|}{{\bfseries Interaction-awareness}}\\\hline
What does the capability mean in your problem context. &
The ability to aware of the state (e.g., QoS models) of other nodes because of functional dependency; and also the
possible internal interactions of local services.\\\hline
What are the functional requirements that affected by this capability? &
1. The system must aware of the functional dependency between cloud-based services across nodes.\\\hline
What are the non-functional requirements that affected by this capability? &
1. Accuracy

2. Adaptation Quality

3. Overhead\\\hline
What are the constraints that could affect this capability? &
~
\\\hline
Whether this capability is \ necessary or beneficial? &
Yes\\\hline
\end{supertabular}

\end{center}
 \end{adjustwidth}
\end{table}

\begin{table}[H]
\begin{adjustwidth}{-1in}{-1in} 
\begin{center}

\caption{Questions and answers for deciding whether to include goal-awareness.}

\tablefirsthead{}
\tablehead{}
\tabletail{}
\tablelasttail{}
\begin{supertabular}{|m{4.005cm}|m{12.562cm}|}
\hline
\multicolumn{2}{|m{16.766998cm}|}{{\bfseries Goal-awareness}}\\\hline
What does the capability mean in your problem context. &
The ability to aware of the QoS and cost objectives of cloud-based services. In addition, it should also aware of the
changes.\\\hline
What are the functional requirements that affected by this capability? &
1. The system must aware of the functional dependency between cloud-based services.

~

2. The system should be able to cope with any given QoS attributes and cost objective of cloud-based services. 

~

3. The system should be able to cope with any runtime changes of QoS and cost objectives made by the service
providers.\\\hline
What are the non-functional requirements that affected by this capability? &
1. Adaptation Quality

2. Overhead\\\hline
What are the constraints that could affect this capability? &
1. The cost of the cloud-based services that being managed should not exceed its defined budget.\\\hline
Whether this capability is \ necessary or beneficial? &
Yes\\\hline
\end{supertabular}

\end{center}
 \end{adjustwidth}
\end{table}

\begin{table}[H]
\begin{adjustwidth}{-1in}{-1in} 
\begin{center}

\caption{Questions and answers for deciding whether to include self-expression.}

\tablefirsthead{}
\tablehead{}
\tabletail{}
\tablelasttail{}
\begin{supertabular}{|m{4.005cm}|m{12.562cm}|}
\hline
\multicolumn{2}{|m{16.766998cm}|}{{\bfseries Self-expression}}\\\hline
What does the capability mean in your problem context. &
The ability to change the configurations, tactics and resource provisioning of a node.\\\hline
What are the functional requirements that affected by this capability? &
1. The system must able to control both software and hardware control primitives.

~

2. The system should support both vertical and horizontal scaling.

~
\\\hline
What are the non-functional requirements that affected by this capability? &
1. Adaptation Quality

2. Overhead\\\hline
What are the constraints that could affect this capability? &
1. VM can be added or removed.

~

2. Service can be added or removed.\\\hline
Whether this capability is \ necessary or beneficial? &
Yes\\\hline
\end{supertabular}

\end{center}
 \end{adjustwidth}
\end{table}

\begin{table}[H]
\begin{adjustwidth}{-1in}{-1in} 
\begin{center}

\caption{Questions and answers for deciding whether to include met-self-awareness.}

\tablefirsthead{}
\tablehead{}
\tabletail{}
\tablelasttail{}
\begin{supertabular}{|m{4.005cm}|m{12.562cm}|}
\hline
\multicolumn{2}{|m{16.766998cm}|}{{\bfseries Meta-self-awareness}}\\\hline
What does the capability mean in your problem context. &
The ability to improve the goal-awareness by either dynamically finding the best technique or using ensemble
methods.\\\hline
What are the functional requirements that affected by this capability? &
1. The system must be aware of the changes in workload and deployment.

~

2. The system should be able to aware of QoS interference. 

~

3. The system must aware of the functional dependency between cloud-based services.

~
\\\hline
What are the non-functional requirements that affected by this capability? &
1. Accuracy

2. Reliability

3. Adaptation Quality

4. Overhead\\\hline
What are the constraints that could affect this capability? &
~
\\\hline
Whether this capability is \ necessary or beneficial? &
Yes\\\hline
\end{supertabular}

\end{center}
 \end{adjustwidth}
\end{table}

According to the above analysis and the aforementioned Table 3.1, the selected pattern for this cloud case study is the \textit{Goal Sharing with time-awareness capability Pattern}.

\subsubsection{Step 4 - Fit the Selected Pattern(s)}

We now fit the proposed architecture to the selected pattern, as shown in the Figure~\ref{fig:cloudfit} below.

\begin{figure}[h!]
\centering
\includegraphics[width=5in]{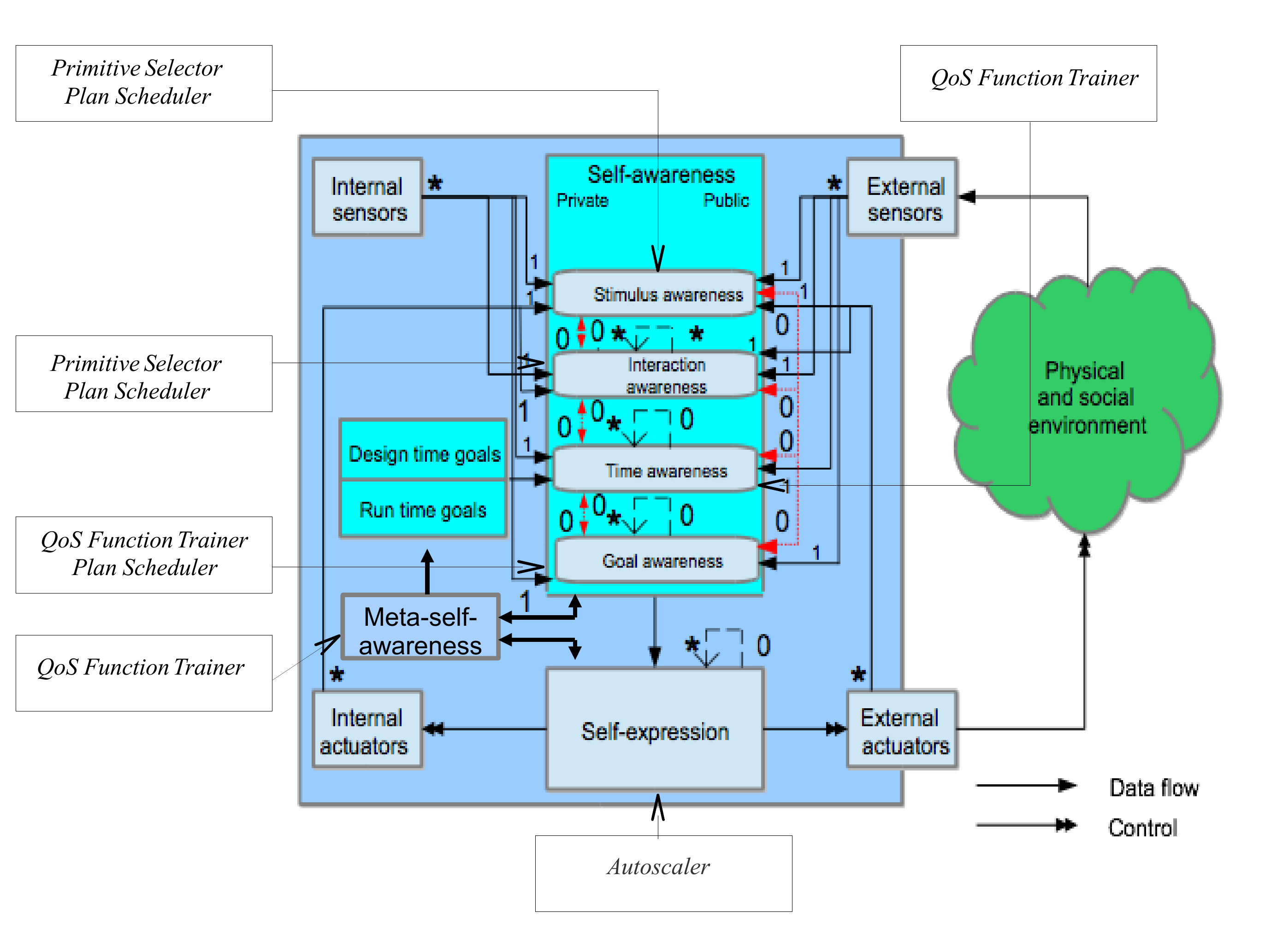}
\caption{Fit the proposed architecture to Goal Sharing with time-awareness capability Pattern}
\label{fig:cloudfit}
\end{figure}

\subsubsection{Step 5 - Determine the Important Primitives and the Possible Alternatives  for Non-functional Requirements}

\textstyleDefaultParagraphFont{\foreignlanguage{english}{\textcolor{black}{It is worth noting that certain primitives
are eliminated form consideration as they are trivial in this problem context, these are:
}}}\textstyleDefaultParagraphFont{\foreignlanguage{english}{\textit{\textcolor{black}{transit, link
}}}}\textstyleDefaultParagraphFont{\foreignlanguage{english}{\textcolor{black}{and}}}\textstyleDefaultParagraphFont{\foreignlanguage{english}{\textit{\textcolor{black}{
structure}}}}\textstyleDefaultParagraphFont{\foreignlanguage{english}{\textcolor{black}{.
}}}\textstyleDefaultParagraphFont{\foreignlanguage{english}{\textit{\textcolor{black}{Transit}}}}\textstyleDefaultParagraphFont{\foreignlanguage{english}{\textcolor{black}{
is eliminated because all the capabilities are associated with at least one functional requirements, therefore we do
not need to consider the possibility of switch on/off certain capabilities at runtime in this case. As for
}}}\textstyleDefaultParagraphFont{\foreignlanguage{english}{\textit{\textcolor{black}{link}}}}\textstyleDefaultParagraphFont{\foreignlanguage{english}{\textcolor{black}{,
we consider that the topology of components is constrained by the environment and functional requirements, thus it does
not significantly influence the non-functional requirements. Similarity, the
}}}\textstyleDefaultParagraphFont{\foreignlanguage{english}{\textit{\textcolor{black}{structure
}}}}\textstyleDefaultParagraphFont{\foreignlanguage{english}{\textcolor{black}{primitive is eliminated because how the
capabilities are distributed into components is not significantly associated with the non-functional requirements in
our case.}}}

In addition, we eliminated some techniques as they are fundamentally not applicable in our case due to the constraints and non-functional requirements. Precisely, the Table below list the rest architectural primitives and the possible alternatives for this case study:

\begin{table}[H]
\begin{adjustwidth}{-1in}{-1in} 
\begin{center}

\caption{The chosen architectural
primitives and their alternatives for the cloud case study.}

\tablefirsthead{}
\tablehead{}
\tabletail{}
\tablelasttail{}
\begin{supertabular}{|m{3.654cm}|m{12.946cm}|}
\hline
\centering{\bfseries Architectural Primitives} &
\centering\arraybslash{\bfseries \ Alternatives}\\\hline
stimulus-awareness &
Symmetric Uncertainty Measurement, Queuing Network, Queuing Network +Simple Update Function, Sensitivity and
Region-based Partitioning+Symmetric Uncertainty Measurement, Symmetric Uncertainty Measurement+ Conditions-Actions
Rule, \ Symmetric Uncertainty Measurement+ Conditions-Actions Rule+ Sensitivity and Region-based Partitioning\\\hline
interaction-awareness &
Threshold-based Algorithm, Symmetric Uncertainty Measurement, Sensitivity and Region-based Partitioning+Symmetric
Uncertainty Measurement\\\hline
time-awareness &
Linear ARMAX, Neural Network, Regression Tree, Linear ARMAX+Neural Network+ Regression Tree (need
meta-self-awareness)\\\hline
goal-awareness &
Linear ARMAX, Neural Network, Regression Tree, Sensitivity and Region-based Partitioning+Linear ARMAX, Sensitivity and
Region-based Partitioning+Neural Network, Sensitivity and Region-based Partitioning+Regression Tree, Sensitivity and
Region-based Partitioning+Linear ARMAX+Neural Network+ Regression Tree (need meta-self-awareness)\\\hline
self-expression &
Random Optimisation, Static Mapping, Brute Force Optimization,\\\hline
meta-self-awareness &
Bucket of Models, Ensemble Method\\\hline
send &
synchronous function call+asynchronous multicast, asynchronous function call+asynchronous multicast \\\hline
handle &
First-Come-First-Serve (FCFS), multi-threading, First-Come-First-Serve (FCFS)+multi-threading\\\hline
state &
proactive Goal-awareness, reactive Goal-awareness, proactive+reactive Goal-awareness\\\hline
existence &
exist, non-exist\\\hline
\end{supertabular}

\end{center}
 \end{adjustwidth}
\end{table}

\subsubsection{Step 6 - Create Scenarios}

We now defined some scenarios for each non-functional attribute. In particular, the accuracy analysis uses 3 scenarios represents the most common facts in the cloud:

\begin{itemize}
\item The cloud-based services are under different level of burst workload.
\item QoS interference occurs due to contention.
\item There are VM migration/replication taken place due to actuations.
\end{itemize}

For adaptation quality analysis, we use 3  scenarios to assess how the self-aware and self-expressive system behaves.

\begin{itemize}
\item There are some amount of conflicted and harmonic objectives of different cloud-based services.
\item There are some amount of conflicted and harmonic objectives of different cloud-based services.
\item The cloud-based services are under different level of burst workload.
\end{itemize}

For overhead analysis, we assume 2  scenarios, representing the anticipated way in which the system could suffer from overhead. 

\begin{itemize}
\item More than one cloud-based services located on each VM.
\item More than one VMs hosted on each node.
\end{itemize}

For the reliability analysis, we only use one scenario. In particular, this attribute is measured by empirical method [4] instead of simulation models as it involved unknown workload.

\begin{itemize}
\item The cloud-based services are under unknown workload and/or events
\end{itemize}

\subsubsection{Step 7 - Score the Alternative of Primitives Against each Non-functional Attribute using Analytical or Simulation Models}

According to the steps mentioned previously, we first weight the relative importance of different non-functional attributes after negotiation amongst the stackholders. The results are shown as below:

\begin{table}[H]
\begin{adjustwidth}{-1in}{-1in} 
\begin{center}

\caption{The weights of different
non-functional attributes for the cloud case study.}

\tablefirsthead{}
\tablehead{}
\tabletail{}
\tablelasttail{}
\begin{supertabular}{|m{8.3cm}|m{8.3cm}|}
\hline
{\bfseries Attribute} &
{\bfseries Weight}\\\hline
Accuracy &
0.1\\\hline
Adaptation Quality &
0.75\\\hline
Overhead &
0.05\\\hline
Reliability at runtime &
0.1\\\hline
\end{supertabular}

\end{center}
 \end{adjustwidth}
\end{table}

Secondly, we score each alternative of all primitives against each non-functional attribute under every defined scenarios. In particular, we run simulation for each alternative under the aforementioned scenarios. The results are shown as below (the score of 0 means they have no or limited impact to the non-functional attribute ):

\tablefirsthead{}
\tablehead{}
\tabletail{}
\tablelasttail{}
\setlength\LTleft{-1in}

\begin{longtable}{|m{1.634cm}|m{9.037001cm}|m{1.7789999cm}|m{1.7789999cm}|m{1.77cm}|}
\caption{The scores of different alternatives for accuracy. }\\

\hline
\multicolumn{5}{|m{16.799cm}|}{{\bfseries Accuracy (\%)}}\\\hline
\multicolumn{2}{|m{10.871cm}|}{{\bfseries Alternative}} &
{\bfseries Scenario 1} &
{\bfseries Scenario 2} &
{\bfseries Scenario 3}\\\hline
stimulus-awareness &
Symmetric Uncertainty Measurement &
84.1 &
84.5 &
85.7\\\hline
~
 &
Threshold-based Algorithm &
54.3 &
39.5 &
55.4\\\hline
~
 &
Simple Update Function &
56.4 &
40.7 &
56.2\\\hline
~
 &
Sensitivity and Region-based Partitioning+Symmetric Uncertainty Measurement &
84.1 &
84.5 &
85.7\\\hline
~
 &
Symmetric Uncertainty Measurement+ Conditions-Actions Rule &
84.1 &
84.5 &
85.7\\\hline
~
 &
Symmetric Uncertainty Measurement+ Conditions-Actions Rule+ Sensitivity and Region-based Partitioning &
84.1 &
84.5 &
85.7\\\hline
interaction-awareness &
Conditions-Actions Rule &
0 &
0 &
0\\\hline
~
 &
Symmetric Uncertainty Measurement &
84.1 &
84.5 &
85.7\\\hline
~
 &
Sensitivity and Region-based Partitioning+Symmetric Uncertainty Measurement &
84.1 &
84.5 &
85.7\\\hline
time-awareness &
Linear ARMAX &
83.3 &
82.1 &
80.2\\\hline
~
 &
Neural Network &
85.4 &
85 &
86.1\\\hline
~
 &
Regression Tree &
75.4 &
70.2 &
73.3\\\hline
~
 &
Linear ARMAX+Neural Network+ Regression Tree (need meta-self-awareness) &
84.1 &
84.5 &
85.7\\\hline
goal-awareness &
Linear ARMAX &
83.3 &
82.1 &
80.2\\\hline
~
 &
Neural Network &
85.4 &
85 &
86.1\\\hline
~
 &
Regression Tree &
75.4 &
70.2 &
73.3\\\hline
~
 &
Sensitivity and Region-based Partitioning+Linear ARMAX &
83.3 &
82.1 &
80.2\\\hline
~
 &
Sensitivity and Region-based Partitioning+Neural Network &
85.4 &
85 &
86.1\\\hline
~
 &
Sensitivity and Region-based Partitioning+Regression Tree &
75.4 &
70.2 &
73.3\\\hline
~
 &
Sensitivity and Region-based Partitioning+Linear ARMAX+Neural Network+ Regression Tree (need meta-self-awareness) &
84.1 &
84.5 &
85.7\\\hline
self-expression &
Random Optimization &
0 &
0 &
0\\\hline
~
 &
Static Mapping &
0 &
0 &
0\\\hline
~
 &
Brute Force Optimization &
0 &
0 &
0\\\hline
meta-self-awareness &
Bucket of Models &
84.1 &
84.5 &
85.7\\\hline
~
 &
Ensemble Method &
78.6 &
81.1 &
80.5\\\hline
send &
synchronous function call+asynchronous multicast &
0 &
0 &
0\\\hline
~
 &
asynchronous function call+asynchronous multicast  &
0 &
0 &
0\\\hline
handle &
First-Come-First-Serve (FCFS) &
0 &
0 &
0\\\hline
~
 &
multi-threading &
0 &
0 &
0\\\hline
~
 &
First-Come-First-Serve (FCFS)+multi-threading &
0 &
0 &
0\\\hline
state &
proactive Goal-awareness &
0 &
0 &
0\\\hline
~
 &
reactive Goal-awareness &
0 &
0 &
0\\\hline
~
 &
proactive+reactive Goal-awareness &
0 &
0 &
0\\\hline
existence &
exist &
84.1 &
84.5 &
85.7\\\hline
~
 &
non-exist &
80.3 &
76.1 &
79.2\\\hline

\end{longtable}

\tablefirsthead{}
\tablehead{}
\tabletail{}
\tablelasttail{}
\setlength\LTleft{-1in}
\begin{longtable}{|m{1.634cm}|m{9.037001cm}|m{1.7789999cm}|m{1.7789999cm}|m{1.77cm}|}
\caption{The scores of different alternatives for adaptation quality.  }\\
\hline
\multicolumn{5}{|m{16.799cm}|}{{\bfseries Adaptation Quality (calculated by Eq 7.)}}\\\hline
\multicolumn{2}{|m{10.871cm}|}{{\bfseries Alternative}} &
{\bfseries Scenario 1} &
{\bfseries Scenario 2} &
{\bfseries Scenario 3}\\\hline
stimulus-awareness &
Symmetric Uncertainty Measurement &
5.7 &
4.2 &
4.7\\\hline
~
 &
Threshold-based Algorithm &
4.8 &
4.1 &
4.5\\\hline
~
 &
Simple Update Function &
5.2 &
4.7 &
3.9\\\hline
~
 &
Sensitivity and Region-based Partitioning+Symmetric Uncertainty Measurement &
5.8 &
4.4 &
4.8\\\hline
~
 &
Symmetric Uncertainty Measurement+ Conditions-Actions Rule &
5.8 &
4.5 &
4.7\\\hline
~
 &
Symmetric Uncertainty Measurement+ Conditions-Actions Rule+ Sensitivity and Region-based Partitioning &
6.2 &
5.6 &
4.9\\\hline
interaction-awareness &
Conditions-Actions Rule &
4.3 &
4 &
3.5\\\hline
~
 &
Symmetric Uncertainty Measurement &
5.7 &
4.2 &
4.7\\\hline
~
 &
Sensitivity and Region-based Partitioning+Symmetric Uncertainty Measurement &
5.9 &
5.2 &
4.9\\\hline
time-awareness &
Linear ARMAX &
4.3 &
4.5 &
3.9\\\hline
~
 &
Neural Network &
6 &
5.1 &
5.2\\\hline
~
 &
Regression Tree &
4 &
4.4 &
4.5\\\hline
~
 &
Linear ARMAX+Neural Network+ Regression Tree (need meta-self-awareness) &
5.8 &
4.8 &
5.1\\\hline
goal-awareness &
Linear ARMAX &
4.3 &
4.5 &
3.9\\\hline
~
 &
Neural Network &
6.1 &
5.2 &
5\\\hline
~
 &
Regression Tree &
4 &
4.4 &
4.5\\\hline
~
 &
Sensitivity and Region-based Partitioning+Linear ARMAX &
4.4 &
4.8 &
4.1\\\hline
~
 &
Sensitivity and Region-based Partitioning+Neural Network &
5.9 &
5.3 &
5.2\\\hline
~
 &
Sensitivity and Region-based Partitioning+Regression Tree &
4.5 &
4.8 &
4.6\\\hline
~
 &
Sensitivity and Region-based Partitioning+Linear ARMAX+Neural Network+ Regression Tree (need meta-self-awareness) &
5.8 &
4.8 &
5.1\\\hline
self-expression &
Random Optimization &
4.8 &
4.9 &
4.8\\\hline
~
 &
Static Mapping &
4.7 &
4.6 &
4.9\\\hline
~
 &
Brute Force Optimization &
5 &
5.2 &
5.1\\\hline
meta-self-awareness &
Bucket of Models &
5.5 &
5.4 &
5.3\\\hline
~
 &
Ensemble Method &
5.9 &
5.1 &
4.9\\\hline
send &
synchronous function call+asynchronous multicast &
0 &
0 &
0\\\hline
~
 &
asynchronous function call+asynchronous multicast  &
0 &
0 &
0\\\hline
handle &
First-Come-First-Serve (FCFS) &
0 &
0 &
0\\\hline
~
 &
multi-threading &
0 &
0 &
0\\\hline
~
 &
First-Come-First-Serve (FCFS)+multi-threading &
0 &
0 &
0\\\hline
state &
proactive Goal-awareness &
4.6 &
4.5 &
4.1\\\hline
~
 &
reactive Goal-awareness &
4.8 &
4.9 &
3.9\\\hline
~
 &
proactive+reactive Goal-awareness &
5 &
5.1 &
5\\\hline
existence &
exist &
5.8 &
5.3 &
5.1\\\hline
~
 &
non-exist &
4.1 &
4.5 &
4.1\\\hline
\end{longtable}

\tablefirsthead{}
\tablehead{}
\tabletail{}
\tablelasttail{}
\setlength\LTleft{-1in}
\begin{longtable}{|m{1.634cm}|m{9.037001cm}|m{1.7789999cm}|m{3.7489998cm}|}
\caption{The scores of different alternatives for overhead.  }\\
\hline
\multicolumn{4}{|m{16.799cm}|}{{\bfseries Overhead (s) (instead of measure the overhead of the whole system, we measure
the overhead of each alternative)}}\\\hline
\multicolumn{2}{|m{10.871cm}|}{{\bfseries Alternative}} &
{\bfseries Scenario 1} &
{\bfseries Scenario 2}\\\hline
stimulus-awareness &
Symmetric Uncertainty Measurement &
1.2 &
1.3\\\hline
~
 &
Threshold-based Algorithm &
0.7 &
0.6\\\hline
~
 &
Simple Update Function &
0.7 &
0.6\\\hline
~
 &
Sensitivity and Region-based Partitioning+Symmetric Uncertainty Measurement &
1.7 &
1.5\\\hline
~
 &
Symmetric Uncertainty Measurement+ Conditions-Actions Rule &
1.2 &
1.3\\\hline
~
 &
Symmetric Uncertainty Measurement+ Conditions-Actions Rule+ Sensitivity and Region-based Partitioning &
1.7 &
1.5\\\hline
interaction-awareness &
Conditions-Actions Rule &
0.8

~
 &
0.8

~
\\\hline
~
 &
Symmetric Uncertainty Measurement &
1.2 &
1.3\\\hline
~
 &
Sensitivity and Region-based Partitioning+Symmetric Uncertainty Measurement &
1.7 &
1.4\\\hline
time-awareness &
Linear ARMAX &
1.2 &
1.2\\\hline
~
 &
Neural Network &
35.4 &
27.9\\\hline
~
 &
Regression Tree &
5.1 &
3.6\\\hline
~
 &
Linear ARMAX+Neural Network+ Regression Tree (need meta-self-awareness) &
41.2 &
33.3\\\hline
goal-awareness &
Linear ARMAX &
1.2 &
1.2\\\hline
~
 &
Neural Network &
35.4 &
27.9\\\hline
~
 &
Regression Tree &
5.1 &
3.6\\\hline
~
 &
Sensitivity and Region-based Partitioning+Linear ARMAX &
2.7 &
2.3\\\hline
~
 &
Sensitivity and Region-based Partitioning+Neural Network &
36.9 &
29\\\hline
~
 &
Sensitivity and Region-based Partitioning+Regression Tree &
6.6 &
4.7\\\hline
~
 &
Sensitivity and Region-based Partitioning+Linear ARMAX+Neural Network+ Regression Tree (need meta-self-awareness) &
42.7 &
34.4\\\hline
self-expression &
Random Optimization &
55.4 &
49.3\\\hline
~
 &
Static Mapping &
135.3 &
118.2\\\hline
~
 &
Brute Force Optimization &
154.3 &
144.2\\\hline
meta-self-awareness &
Bucket of Models &
5.5 &
5\\\hline
~
 &
Ensemble Method &
7.8 &
6.3\\\hline
send &
synchronous function call+asynchronous multicast &
37.3 &
35.3\\\hline
~
 &
asynchronous function call+asynchronous multicast  &
11.4 &
13.2\\\hline
handle &
First-Come-First-Serve (FCFS) &
35.5 &
37\\\hline
~
 &
multi-threading &
12.3 &
10\\\hline
~
 &
First-Come-First-Serve (FCFS)+multi-threading &
18.8 &
23.7\\\hline
state &
proactive Goal-awareness &
0 &
0\\\hline
~
 &
reactive Goal-awareness &
0 &
0\\\hline
~
 &
proactive+reactive Goal-awareness &
0 &
0\\\hline
existence &
exist &
41.2 &
33.3\\\hline
~
 &
non-exist &
35.4 &
27.9\\\hline
\end{longtable}

\tablefirsthead{}
\tablehead{}
\tabletail{}
\tablelasttail{}
\setlength\LTleft{-1in}
\begin{longtable}{|m{1.634cm}|m{11.043cm}|m{3.723cm}|}
\caption{The scores of different alternatives for reliability.  }\\
\hline
\multicolumn{3}{|m{16.8cm}|}{{\bfseries Reliability (relative weights) }}\\\hline
\multicolumn{2}{|m{12.877cm}|}{{\bfseries Alternative}} &
{\bfseries Scenario 1}\\\hline
stimulus-awareness &
Symmetric Uncertainty Measurement &
5\\\hline
~
 &
Threshold-based Algorithm &
1\\\hline
~
 &
Simple Update Function &
1\\\hline
~
 &
Sensitivity and Region-based Partitioning+Symmetric Uncertainty Measurement &
5\\\hline
~
 &
Symmetric Uncertainty Measurement+ Conditions-Actions Rule &
7\\\hline
~
 &
Symmetric Uncertainty Measurement+ Conditions-Actions Rule+ Sensitivity and Region-based Partitioning &
7\\\hline
interaction-awareness &
Conditions-Actions Rule &
1\\\hline
~
 &
Symmetric Uncertainty Measurement &
7\\\hline
~
 &
Sensitivity and Region-based Partitioning+Symmetric Uncertainty Measurement &
7\\\hline
time-awareness &
Linear ARMAX &
1\\\hline
~
 &
Neural Network &
1\\\hline
~
 &
Regression Tree &
1\\\hline
~
 &
Linear ARMAX+Neural Network+ Regression Tree (need meta-self-awareness) &
7\\\hline
goal-awareness &
Linear ARMAX &
1\\\hline
~
 &
Neural Network &
1\\\hline
~
 &
Regression Tree &
1\\\hline
~
 &
Sensitivity and Region-based Partitioning+Linear ARMAX &
1\\\hline
~
 &
Sensitivity and Region-based Partitioning+Neural Network &
1\\\hline
~
 &
Sensitivity and Region-based Partitioning+Regression Tree &
1\\\hline
~
 &
Sensitivity and Region-based Partitioning+Linear ARMAX+Neural Network+ Regression Tree (need meta-self-awareness) &
7\\\hline
self-expression &
Random Optimization &
5\\\hline
~
 &
Static Mapping &
1\\\hline
~
 &
Brute Force Optimization &
5\\\hline
meta-self-awareness &
Bucket of Models &
1\\\hline
~
 &
Ensemble Method &
1\\\hline
send &
synchronous function call+asynchronous multicast &
1\\\hline
~
 &
asynchronous function call+asynchronous multicast  &
1\\\hline
handle &
First-Come-First-Serve (FCFS) &
1\\\hline
~
 &
multi-threading &
1\\\hline
~
 &
First-Come-First-Serve (FCFS)+multi-threading &
1\\\hline
state &
proactive Goal-awareness &
1\\\hline
~
 &
reactive Goal-awareness &
1\\\hline
~
 &
proactive+reactive Goal-awareness &
1\\\hline
existence &
exist &
1\\\hline
~
 &
non-exist &
1\\\hline
\end{longtable}

Once we obtain all the scores and calculate the total score for all scenarios, we then normalised the scores using Eq. 3.3, the results are shown in Table 3.15. 

\begin{table}[H]
\begin{adjustwidth}{-1in}{-1in} 
\begin{center}

\caption{The normalised scores of different alternatives for all non-functional attributes. }

\tablefirsthead{}
\tablehead{}
\tabletail{}
\tablelasttail{}
\begin{supertabular}{|m{1.6129999cm}|m{7.282cm}|m{1.643cm}|m{1.813cm}|m{1.671cm}|m{1.7579999cm}|}
\hline
\multicolumn{2}{|m{9.095cm}|}{{\bfseries Alternative}} &
{\bfseries Accuracy} &
{\bfseries Adaptation Quality} &
{\bfseries Overhead} &
{\bfseries Reliability}\\\hline
stimulus-awareness &
Symmetric Uncertainty Measurement &
0.192 &
0.165 &
0.821 &
0.19\\\hline
~
 &
Threshold-based Algorithm &
0.112 &
0.151 &
0.907 &
0.04\\\hline
~
 &
Simple Update Function &
0.116 &
0.156 &
0.907 &
0.04\\\hline
~
 &
Sensitivity and Region-based Partitioning+Symmetric Uncertainty Measurement &
0.192 &
0.169 &
0.771 &
0.19\\\hline
~
 &
Symmetric Uncertainty Measurement+ Conditions-Actions Rule &
0.192 &
0.169 &
0.821 &
0.27\\\hline
~
 &
Symmetric Uncertainty Measurement+ Conditions-Actions Rule+ Sensitivity and Region-based Partitioning &
0.2 &
0.189 &
0.771 &
0.27\\\hline
interaction-awareness &
Conditions-Actions Rule &
0 &
0.278 &
0.777 &
0.07\\\hline
~
 &
Symmetric Uncertainty Measurement &
0.5 &
0.344 &
0.653 &
0.47\\\hline
~
 &
Sensitivity and Region-based Partitioning+Symmetric Uncertainty Measurement &
0.5 &
0.377 &
0.569 &
0.47\\\hline
time-awareness &
Linear ARMAX &
0.252 &
0.22 &
0.984 &
0.1\\\hline
~
 &
Neural Network &
0.263 &
0.283 &
0.575 &
0.1\\\hline
~
 &
Regression Tree &
0.224 &
0.224 &
0.942 &
0.1\\\hline
~
 &
Linear ARMAX+Neural Network+ Regression Tree (need meta-self-awareness) &
0.261 &
0.273 &
0.5 &
0.7\\\hline
goal-awareness &
Linear ARMAX &
0.145 &
0.125 &
0.99 &
0.08\\\hline
~
 &
Neural Network &
0.151 &
0.161 &
0.729 &
0.08\\\hline
~
 &
Regression Tree &
0.129 &
0.127 &
0.963 &
0.08\\\hline
~
 &
Sensitivity and Region-based Partitioning+Linear ARMAX &
0.145 &
0.131 &
0.979 &
0.08\\\hline
~
 &
Sensitivity and Region-based Partitioning+Neural Network &
0.151 &
0.162 &
0.718 &
0.08\\\hline
~
 &
Sensitivity and Region-based Partitioning+Regression Tree &
0.129 &
0.137 &
0.952 &
0.08\\\hline
~
 &
Sensitivity and Region-based Partitioning+Linear ARMAX+Neural Network+ Regression Tree (need meta-self-awareness) &
0.15 &
0.155 &
0.670 &
0.54\\\hline
self-expression &
Random Optimization &
0 &
0.33 &
0.841 &
0.45\\\hline
~
 &
Static Mapping &
0 &
0.323 &
0.614 &
0.09\\\hline
~
 &
Brute Force Optimization &
0 &
0.348 &
0.545 &
0.45\\\hline
meta-self-awareness &
Bucket of Models &
0.514 &
0.509 &
0.573 &
0.5\\\hline
~
 &
Ensemble Method &
0.486 &
0.491 &
0.427 &
0.5\\\hline
send &
synchronous function call+asynchronous multicast &
0 &
0 &
0.253 &
0.5\\\hline
~
 &
asynchronous function call+asynchronous multicast  &
0 &
0 &
0.747 &
0.5\\\hline
handle &
First-Come-First-Serve (FCFS) &
0 &
0 &
0.472 &
0.33\\\hline
~
 &
multi-threading &
0 &
0 &
0.838 &
0.33\\\hline
~
 &
First-Come-First-Serve (FCFS)+multi-threading &
0 &
0 &
0.690 &
0.33\\\hline
state &
proactive Goal-awareness &
0 &
0.315 &
0 &
0.33\\\hline
~
 &
reactive Goal-awareness &
0 &
0.324 &
0 &
0.33\\\hline
~
 &
proactive+reactive Goal-awareness &
0 &
0.360 &
0 &
0.33\\\hline
existence &
exist &
0.519 &
0.541 &
0.46 &
0.5\\\hline
~
 &
non-exist &
0.481 &
0.459 &
0.54 &
0.5\\\hline
\end{supertabular}

\end{center}
 \end{adjustwidth}
\end{table}

\subsubsection{Step 8 - Find the Best Alternatives for the Final Architecture View}

Finally, we search for the alternative that resulting the highest value using Eq. 3.4. The final output of the selected alternatives is list as in Table 3.16. 

\begin{table}[H]
\begin{adjustwidth}{-1in}{-1in} 
\begin{center}

\caption{The normalised scores of
selected alternatives for all non-functional attributes. }

\tablefirsthead{}
\tablehead{}
\tabletail{}
\tablelasttail{}
\begin{supertabular}{|m{1.634cm}|m{7.695cm}|m{1.4319999cm}|m{1.8869998cm}|m{1.569cm}|m{1.583cm}|}
\hline
\multicolumn{2}{|m{9.529cm}|}{{\bfseries Selected Alternative}} &
{\bfseries Accuracy} &
{\bfseries Adaptation Quality} &
{\bfseries Overhead} &
{\bfseries Reliability}\\\hline
stimulus-awareness &
Symmetric Uncertainty Measurement+ Conditions-Actions Rule+ Sensitivity and Region-based Partitioning &
0.2 &
0.189 &
0.771 &
0.27\\\hline
interaction-awareness &
Sensitivity and Region-based Partitioning+Symmetric Uncertainty Measurement &
0.5 &
0.377 &
0.569 &
0.47\\\hline
time-awareness &
Linear ARMAX+Neural Network+ Regression Tree (need meta-self-awareness) &
0.261 &
0.273 &
0.5 &
0.7\\\hline
goal-awareness &
Sensitivity and Region-based Partitioning+Linear ARMAX+Neural Network+ Regression Tree (need meta-self-awareness) &
0.15 &
0.155 &
0.670 &
0.54\\\hline
self-expression &
Random Optimization &
0 &
0.33 &
0.841 &
0.45\\\hline
meta-self-awareness &
Bucket of Models &
0.514 &
0.509 &
0.573 &
0.5\\\hline
send &
asynchronous function call+asynchronous multicast  &
0 &
0 &
0.747 &
0.5\\\hline
handle &
multi-threading &
0 &
0 &
0.838 &
0.33\\\hline
state &
proactive+reactive Goal-awareness &
0 &
0.360 &
0 &
0.33\\\hline
existence &
exist &
0.519 &
0.541 &
0.46 &
0.5\\\hline
\end{supertabular}

\end{center}
 \end{adjustwidth}
\end{table}

This selection gives us the highest score of 15.437 according to Eq 4. Once we combine the results with those primitives, which were eliminated at the beginning of this step, the detailed variation of our architectural instance based on the Goal Sharing Pattern with time-awareness is shown in the Table 3.17.

\begin{table}[H]
\begin{adjustwidth}{-1in}{-1in} 
\begin{center}

\caption{The  selected alternatives for the cloud case study. }

\tablefirsthead{}
\tablehead{}
\tabletail{}
\tablelasttail{}
\begin{supertabular}{|m{1.5cm}|m{1.5cm}|m{1.5cm}|m{1.5cm}|m{1.5cm}|m{1.5cm}|m{1.5cm}|m{1.5cm}|m{1.5cm}|m{1.5cm}|}
\hline
\multicolumn{2}{|m{3.2cm}|}{~

~
} &
\centering stimulus-awareness &
\centering interaction-awareness &
\centering time-awareness &
\centering goal-awareness &
\centering self-expression &
\centering meta-self-awareness &
\centering sensor &
\centering\arraybslash actuator\\\hline
\multicolumn{2}{|m{3.2cm}|}{\centering Selected alternative(s)} &
Symmetric Uncertainty Measurement+ Conditio-ns-Actions Rule+ Sensitivity and Region-based Partitioning &
Sensitivity and Region-based Partit-

ioning+S-ymmetric Uncertainty Measurement &
Linear ARMAX+N

{}-eural Network+ Regression Tree  &
Sensitivity and Region-based Partitioning+Linear ARMAX+N

{}-eural Network+ Regression Tree  &
Random Optimization &
Bucket of Models &
~
 &
~
\\\hline
\centering send &
\centering synchr-onous &
\centering N/A &
\centering N/A &
\centering N/A &
\centering N/A &
\centering N/A &
\centering N/A &
\centering N/A &
\centering\arraybslash N/A\\\hline
~
 &
\centering asynch-ronous &
~

\centering function call &
~

\centering function call &
~

\centering function call &
~

\centering function call &
~

\centering function call &
~

\centering function call &
~

\centering function call &
~

\centering\arraybslash multicast\\\hline
\centering handle &
\centering sequential &
\centering N/A &
\centering N/A &
\centering N/A &
\centering N/A &
\centering N/A &
\centering N/A &
\centering N/A &
\centering\arraybslash N/A\\\hline
~
 &
\centering parallel &
\centering multi-threading &
\centering multi-threading &
\centering multi-threading &
\centering multi-threading &
\centering multi-threading &
\centering multi-threading &
\centering multi-threading &
\centering\arraybslash multi-threading\\\hline
\centering state &
~
 &
\centering reactive &
\centering reactive &
\centering reactive &
\centering proactive and reactive &
\centering proactive &
\centering proactive &
~
 &
~
\\\hline
\centering transit &
~
 &
~
 &
~
 &
~
 &
~
 &
~
 &
\centering N/A &
~
 &
~
\\\hline
\centering link &
~
 &
\centering one-to-one &
\centering many-to-many, one-to-one &
\centering one-to-one &
\centering one-to-one &
\centering one-to-many &
\centering one-to-many &
~
 &
~
\\\hline
\centering structure &
~
 &
{\centering combine+\par}

\centering separate &
{\centering combine+\par}

\centering separate &
\centering combine &
{\centering combine+\par}

\centering separate &
\centering compact &
\centering combine &
\centering compact &
\centering\arraybslash compact\\\hline
\centering existence &
~
 &
~
 &
~
 &
~
 &
~
 &
~
 &
\centering exist &
~
 &
~
\\\hline
\end{supertabular}

\end{center}
 \end{adjustwidth}
\end{table}

\subsubsection{Quantitative Experiments}

In this section, we conduct quantitative evaluation by experimenting our self-aware and self-expressive system against a non self-aware system, which adapts simple rule-based policies. We primary assess the adaptation quality for cloud-based services under the management of these two systems. The observed adaptation quality is measured by score, which is the average result calculated by Eq. 3.7 for the interval after a previous elasticity decision point and before the next one. Each of these intervals is referred to as effect point. 

To evaluate global benefit of the elastic strategies produced by our architecture and the overhead for reaching these strategies, we have conducted an experimental evaluation. In particular, we have implemented the architecture prototype using Java JDK1.6, and we assessed the elastic scaling of 8 hypothetical cloud-based service-instances under the control of our prototype. In the experiment setup, each service-instance was deployed on software stack including Apache, Tomcat and MySQL. We simulate a synthetical workload to each service-instance. The workload has been designed in a way that the intensity was sufficient for causing QoS interference on the co-located services and co-hosted VMs. The testbed is a private cloud, where PMs are connected by Gigabit Ethernet and a switch. Xen \cite{xen} is used as the underlying hypervisor. The initial deployment and the considered CP/EP of our experiments are shown on Table 3.18. The scale of each CP and their corresponding prices are specified in Table 3.19. 

For simplicity, we assume that the service-instances and their QoS/cost are equivalently important and thus all weights in the global objective function (Eq. 3.7) are set to 1. In addition, we consider both vertical and horizontal scaling; and apply a simple solution to determine the actions, this is: we always try vertical scaling (i.e., scale up/down) first before horizontal scaling (i.e., scale out/in). This is because horizontal scaling is usually more expensive than vertical scaling. 

\begin{table}[H]
\begin{adjustwidth}{-1in}{-1in} 
\begin{center}

\caption{Initial deployments and the examined objectives/primitives }

\tablefirsthead{}
\tablehead{}
\tabletail{}
\tablelasttail{}
\begin{supertabular}{|m{1.074cm}|m{1.2129999cm}|m{2.483cm}|m{3.031cm}|m{2.486cm}|m{2.986cm}|m{2.328cm}|}
\hline
\centering{\bfseries PM} &
\centering{\bfseries VM} &
\centering{\bfseries Service-instance} &
\centering{\bfseries Objectives} &
\centering{\bfseries Software CP} &
\centering{\bfseries Hardware CP} &
\centering\arraybslash{\bfseries EP}\\\hline
\centering{\bfseries\color{black} \textmd{PM1}} &
\centering VM &
\centering{\selectlanguage{english}\color{black} \textit{S}\textit{\textsubscript{11}}} &
\centering Throughput and cost &
\centering The max threads &
\centering CPU and Memory &
\centering\arraybslash workload\\\hline
~
 &
~
 &
\centering{\selectlanguage{english}\color{black} \textit{S}\textit{\textsubscript{21}}} &
\centering{\color{black} Throughput and cost} &
\centering{\color{black} The max threads} &
~
 &
\centering\arraybslash{\color{black} workload}\\\hline
~
 &
\centering VM &
\centering{\selectlanguage{english}\color{black} \textit{S}\textit{\textsubscript{31}}} &
\centering Throughput and cost &
\centering The max threads &
\centering CPU and Memory &
\centering\arraybslash workload\\\hline
~
 &
~
 &
\centering{\selectlanguage{english}\color{black} \textit{S}\textit{\textsubscript{41}}} &
\centering Throughput and cost &
\centering The max threads &
~
 &
\centering\arraybslash workload\\\hline
\centering{\color{black} PM2} &
\centering VM &
\centering{\selectlanguage{english}\color{black} \textit{S}\textit{\textsubscript{12}}} &
\centering{\color{black} Throughput and cost} &
\centering The max threads &
\centering CPU and Memory &
\centering\arraybslash workload\\\hline
~
 &
~
 &
\centering{\selectlanguage{english}\color{black} \textit{S}\textit{\textsubscript{51}}} &
\centering Throughput and cost &
\centering The max threads &
~
 &
\centering\arraybslash workload\\\hline
\centering{\color{black} PM3} &
\centering VM &
\centering{\selectlanguage{english}\color{black} \textit{S}\textit{\textsubscript{32}}} &
\centering{\color{black} Throughput and cost} &
\centering The max threads &
\centering CPU and Memory &
\centering\arraybslash workload\\\hline
~
 &
~
 &
\centering{\selectlanguage{english}\color{black} \textit{S}\textit{\textsubscript{61}}} &
\centering Throughput and cost &
\centering The max threads &
~
 &
\centering\arraybslash workload\\\hline
\end{supertabular}
\end{center}
 \end{adjustwidth}
\end{table}

\begin{table}[H]
\begin{adjustwidth}{-1in}{-1in} 
\begin{center}

\caption{Scaling options and price of control primitives}

\tablefirsthead{}
\tablehead{}
\tabletail{}
\tablelasttail{}
\begin{supertabular}{|m{2.093cm}|m{7.9020004cm}|m{2.181cm}|m{4.025cm}|}
\hline
\centering{\bfseries CP} &
\centering{\bfseries Optional Values} &
\centering{\bfseries Unit} &
\centering\arraybslash{\bfseries Price}\\\hline
\centering{\color{black} Max Threads} &
\centering{\color{black} 5,10,15,20,25,30,35,40,45,50} &
\centering{\color{black} Thread count} &
\centering\arraybslash{\color{black} \$0.8 for each 5 unit per hr}\\\hline
\centering{\color{black} CPU} &
\centering{\color{black} 1, 2,3, 4,5,6, 7, 8} &
\centering{\color{black} Compute Unit} &
\centering\arraybslash{\color{black} \$2.5 for each 1 unit per hr}\\\hline
\centering{\color{black} Memory} &
{\centering\color{black} 0.1,0.2,0.3,0.4,0.5,0.6,0.7,0.8,0.9,1,1.1,1.2,1.3,1.4,\par}

\centering{\color{black} 1.5,1.6,1.7,1.8,1.9,2} &
\centering{\color{black} GB} &
\centering\arraybslash{\color{black} \$1.5 for each 0.1 unit per hr}\\\hline
\end{supertabular}

\end{center}
 \end{adjustwidth}
\end{table}

\begin{figure}[h!]
\centering
\includegraphics[width=5in]{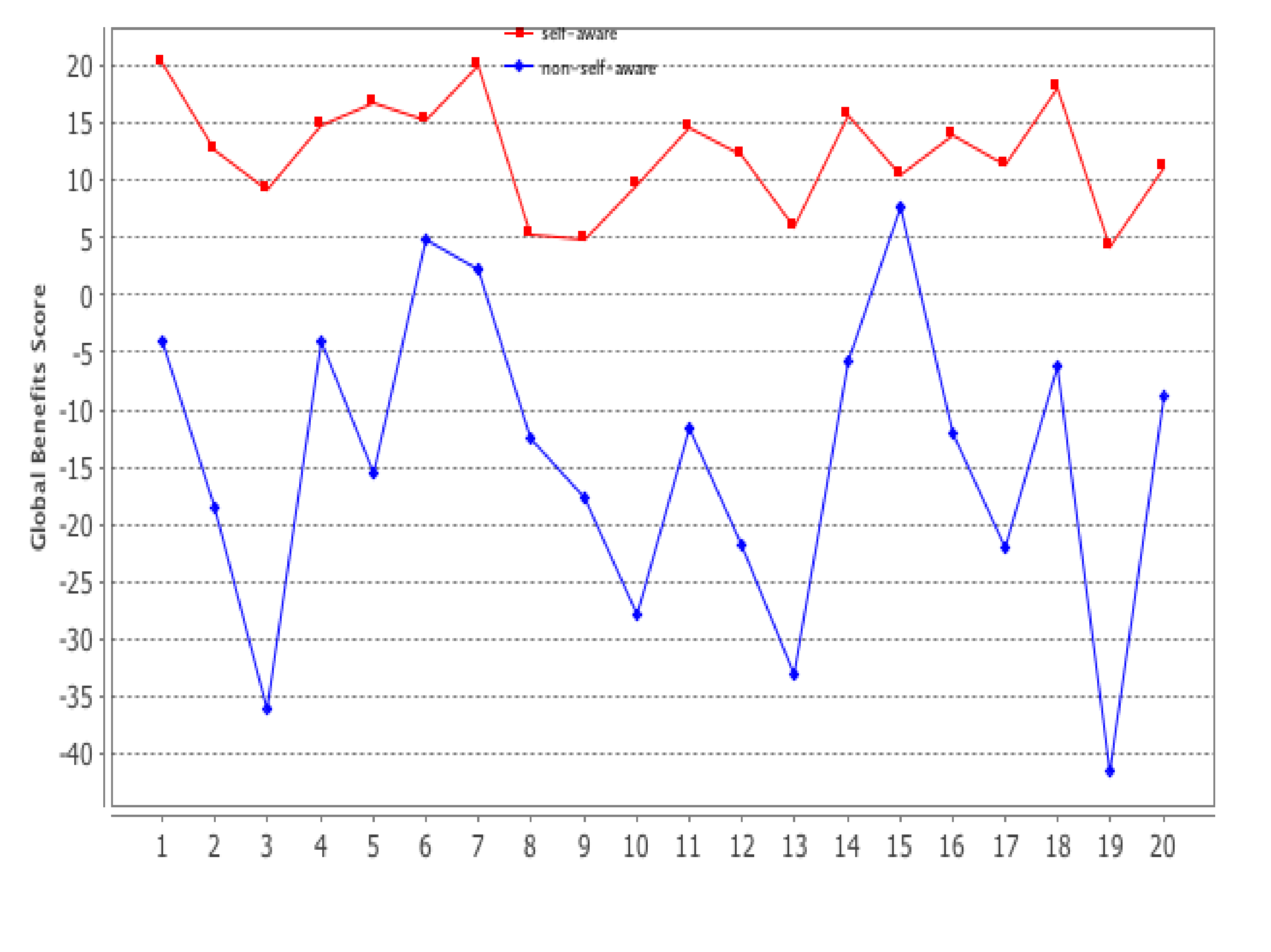}
\caption{The global adaptation quality with respect to effect points}
\label{fig:cloudresult}
\end{figure}

Figure~\ref{fig:cloudresult} illustrate the results of the score (y- axis) in relation to each effect point (x-axis).  we can clearly see that the self-aware system perform much better than the non-self-aware one along the entire time series. This is due to the fact that the non-self-aware system ignores the sensitivity caused by QoS interferences on co-located services and co-hosted VMs, which are significant in our experiments. 

\subsection{Smart Camera Networks Case Study}

In the following, we qualitatively evaluate the proposed methodology by showing how it can be applied in the smart camera networks case study. We also show the experiments that compare the resulted system with a non-self-aware system.

\subsubsection{Step 1 - Collect Requirements and Constraints}

The requirements and constraints of the smart camera networks context as shown in the Table below:

\begin{table}[H]
\begin{adjustwidth}{-1in}{-1in} 
\begin{center}

\caption{The functional, non-functional requirements and constraints for the smart camera networks case study. }

\tablefirsthead{}
\tablehead{}
\tabletail{}
\tablelasttail{}
\begin{supertabular}{|m{16.8cm}|}
\hline
{\bfseries Functional Requirements}\\\hline
\textcolor{black}{The system should continuously track objects while they are visible to at least one camera of the
network.}\\\hline
\textcolor{black}{The system has to coordinate tracking of objects within a network of smart cameras via
handover.}\\\hline
\textcolor{black}{Each camera of the system has to be able to track objects autonomously within its own FOV.}\\\hline
\textcolor{black}{The system has to be able to re-identify objects reliably within various cameras with different
viewpoints.}\\\hline
\textcolor{black}{Each camera has to be able to record information about its local handover behaviour. }\\\hline
\textcolor{black}{The system should notice disappeared objects.}\\\hline
\textcolor{black}{The system should be robust to node failures.}\\\hline
\textcolor{black}{The system should be extensible (add new cameras during runtime). }\\\hline
\textcolor{black}{The system should minimise communication effort while maximising tracking responsibility. }\\\hline
{\bfseries Non-functional Requirements}\\\hline
\textcolor{black}{Maximise tracking utility.}\\\hline
\textcolor{black}{Minimise the number of exchange messages in the network.}\\\hline
{\bfseries Constraints}\\\hline
\textcolor{black}{Cameras can be added or removed during runtime.}\\\hline
\textcolor{black}{Tracker can fail during execution.}\\\hline
\textcolor{black}{Resources on each camera may not be exceeded.}\\\hline
\textcolor{black}{the observed area has to be illuminated.}\\\hline
\end{supertabular}

\end{center}
 \end{adjustwidth}
\end{table}

\subsubsection{Step 2 - Propose Candidate Architecture}

\begin{figure}[h!]
\centering
\includegraphics[width=.95\linewidth]{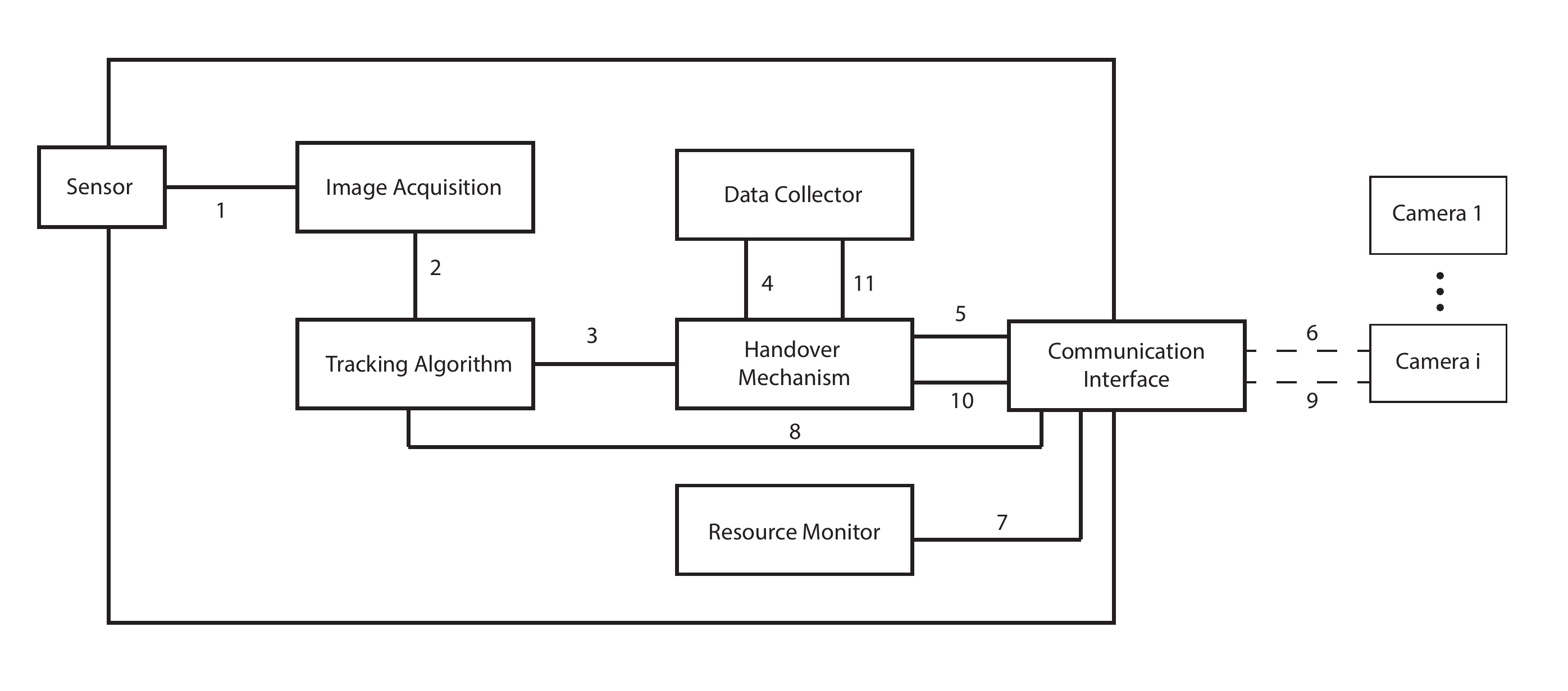}
\caption{The proposed architecture}
\label{fig:cameraproposed}
\end{figure}

The presented architecture is implemented on distributed smart cameras. The workflow of the proposed architecture has been shown in Figure~\ref{fig:cameraproposed} . The sensor on each camera senses the current environment. The image acquisition collects the image data (step 1) from the sensor and transmits it to the tracking algorithm (step 2).
The tracking algorithm detects and identifies the objects of interest. In case the object moves out of the scope of the current camera, the tracking algorithm initializes the handover mechanism (step 3).
The handover mechanism requests available cameras to continue tracking from the 
data collector (step 4). Afterwards, the handover mechanism notifies those 
cameras and requests their tracking capabilities via the communication 
interface (step 5 and 6).
Upon receiving such a request, the camera analyzes its available resources (step 7) and tries to detect the object of interest (step 8). If the camera is able to track the object, it notifies the initial camera (step 9).
When the initial camera received a reply from all contacted cameras, the next camera can be selected via the handover mechanism (step 10). Information about the cameras able to track the object are stored in the data collector and serve as a reference for future coordination (step 11).

\subsubsection{Step 3 - Select the Best Pattern(s)}
We now select the pattern using the questions presented previously for each 
self-awareness and self-expression capability:
%
%In the EPiCS smart camera demonstrator, we require two architectures to 
%reflect on the problems. As a result, we need to select two patterns and for 
%each of these patterns, we follow the same procedure by anwsering the 
%questions 
%for each capability, as described in Step 3 of the methodology. The procedure 
%is shown as below:

\begin{table}[H]
\begin{adjustwidth}{-1in}{-1in} 
\begin{center}

\caption{Questions and answers for deciding whether to include 
stimulus-awareness. }

\tablefirsthead{}
\tablehead{}
\tabletail{}
\tablelasttail{}
\begin{supertabular}{|m{4.005cm}|m{12.562cm}|}
\hline
\multicolumn{2}{|m{16.766998cm}|}{{\bfseries Stimulus-awareness}}\\\hline
What does the capability mean in your problem context. &
\textcolor{black}{Locate and value objects within own field of view of a camera.}\\\hline
What are the functional requirements that affected by this capability? &
\textcolor{black}{1. The system should continuously track objects while they are visible to at least one camera of the
network.}

~

\textcolor{black}{2. The system has to coordinate tracking of objects within a network of smart cameras via handover.}

~

\textcolor{black}{3. Each camera of the system has to be able to track objects autonomously within its own FOV.}

~

\textcolor{black}{4. The system has to be able to re-identify objects reliably within various cameras with different
viewpoints.}

~

\textcolor{black}{5. The system should notice disappeared objects.}\\\hline
What are the non-functional requirements that affected by this capability? &
\textcolor{black}{1. Maximize tracking utility.}\\\hline
What are the constraints that could affect this capability? &
\textcolor{black}{1. Cameras can be added or removed during runtime.}

~

\textcolor{black}{2. Tracker can fail during execution.}

~

\textcolor{black}{3. Resources on each camera may not be exceeded.}

~

\textcolor{black}{4. The observed area has to be illuminated.}\\\hline
Whether this capability is \ necessary or beneficial? &
\textcolor{black}{Yes}\\\hline
\end{supertabular}

\end{center}
 \end{adjustwidth}
\end{table}

\begin{table}[H]
\begin{adjustwidth}{-1in}{-1in} 
\begin{center}

\caption{Questions and answers for deciding whether to include time-awareness. }

\tablefirsthead{}
\tablehead{}
\tabletail{}
\tablelasttail{}
\begin{supertabular}{|m{4.005cm}|m{12.562cm}|}
\hline
\multicolumn{2}{|m{16.766998cm}|}{{\bfseries Time-awareness}}\\\hline
What does the capability mean in your problem context. &
{\color{black} Unlearn previously learnt information in case something changes. 
Explore and exploit behavioral
strategies.}

~
\\\hline
What are the functional requirements that affected by this capability? &
1. \textcolor{black}{The system has to coordinate tracking of objects within a 
network of smart cameras via handover.}

~

\textcolor{black}{2. Each camera has to be able to record information about its 
local handover behavior. }

~

\textcolor{black}{3. The system should minimize communication effort while 
maximizing tracking responsibility. }\\\hline
What are the non-functional requirements that affected by this capability? &
1. Maximize tracking utility.

~

2. Minimize the number of exchange messages in the network.\\\hline
What are the constraints that could affect this capability? &
~
\\\hline
Whether this capability is \ necessary or beneficial? &
Yes\\\hline
\end{supertabular}

\end{center}
 \end{adjustwidth}
\end{table}

%
%\tablefirsthead{}
%\tablehead{}
%\tabletail{}
%\tablelasttail{}
%\begin{supertabular}{|m{4.005cm}|m{12.562cm}|}
%\hline
%\multicolumn{2}{|m{16.766998cm}|}{{\bfseries Time-awareness}}\\\hline
%What does the capability mean in your problem context. &
%~
%\\\hline
%What are the functional requirements that affected by this capability? &
%~
%\\\hline
%What are the non-functional requirements that affected by this capability? &
%~
%\\\hline
%What are the constraints that could affect this capability? &
%~
%\\\hline
%Whether this capability is \ necessary or beneficial? &
%No\\\hline
%\end{supertabular}
%
%\end{center}
% \end{adjustwidth}
%\end{table}

\begin{table}[H]
\begin{adjustwidth}{-1in}{-1in} 
\begin{center}

\caption{Questions and answers for deciding whether to include 
interaction-awareness. }

\tablefirsthead{}
\tablehead{}
\tabletail{}
\tablelasttail{}
\begin{supertabular}{|m{4.005cm}|m{12.562cm}|}
\hline
\multicolumn{2}{|m{16.766998cm}|}{{\bfseries Interaction-awareness}}\\\hline
What does the capability mean in your problem context. &
{\color{black} Reaction to auctions, bids and handover. Definition of neighborhood based on auctions.}

~
\\\hline
What are the functional requirements that affected by this capability? &
1. \textcolor{black}{The system has to coordinate tracking of objects within a network of smart cameras via handover.}

~

\textcolor{black}{2. The system has to be able to re-identify objects reliably within various cameras with different
viewpoints.}

~

\textcolor{black}{3. The system should be robust to node failures.}

~

\textcolor{black}{4. The system should be extensible (add new cameras during runtime). }\\\hline
What are the non-functional requirements that affected by this capability? &
1. Minimize the number of exchange messages in the network.\\\hline
What are the constraints that could affect this capability? &
\textcolor{black}{1. Cameras can be added or removed during runtime.}

~

\textcolor{black}{2. Tracker can fail during execution.}

~

\textcolor{black}{3. Resources on each camera may not be exceeded.}

~

\textcolor{black}{4. The observed area has to be illuminated.}\\\hline
Whether this capability is \ necessary or beneficial? &
Yes\\\hline
\end{supertabular}

\end{center}
 \end{adjustwidth}
\end{table}

\begin{table}[H]
\begin{adjustwidth}{-1in}{-1in} 
\begin{center}

\caption{Questions and answers for deciding whether to include goal-awareness. }

\tablefirsthead{}
\tablehead{}
\tabletail{}
\tablelasttail{}
\begin{supertabular}{|m{4.005cm}|m{12.562cm}|}
\hline
\multicolumn{2}{|m{16.766998cm}|}{{\bfseries Goal-awareness}}\\\hline
What does the capability mean in your problem context. &
{\color{black} Utility function for objects to be tracked. Performance measurement of different strategies.}

~
\\\hline
What are the functional requirements that affected by this capability? &
1. \textcolor{black}{Each camera of the system has to be able to track objects autonomously within its own FOV.}

~

2. \textcolor{black}{The system has to be able to re-identify objects reliably within various cameras with different
viewpoints.} 

~

3. \textcolor{black}{The system should minimize communication effort while maximizing tracking responsibility. }\\\hline
What are the non-functional requirements that affected by this capability? &
1. Maximize tracking utility.

~

2. Minimize the number of exchange messages in the network.\\\hline
What are the constraints that could affect this capability? &
~
\\\hline
Whether this capability is \ necessary or beneficial? &
Yes\\\hline
\end{supertabular}

\end{center}
 \end{adjustwidth}
\end{table}

\begin{table}[H]
\begin{adjustwidth}{-1in}{-1in} 
\begin{center}

\caption{Questions and answers for deciding whether to include self-expression. 
}

\tablefirsthead{}
\tablehead{}
\tabletail{}
\tablelasttail{}
\begin{supertabular}{|m{4.005cm}|m{12.562cm}|}
\hline
\multicolumn{2}{|m{16.766998cm}|}{{\bfseries Self-expression}}\\\hline
What does the capability mean in your problem context. &
{\color{black} Sending out auction invitations and bids. }

~
\\\hline
What are the functional requirements that affected by this capability? &
1. \textcolor{black}{The system has to coordinate tracking of objects within a network of smart cameras via handover.}

~

2. \textcolor{black}{Each camera has to be able to record information about its local handover behavior. }

~

3. \textcolor{black}{The system should be robust to node failures.}

~

\textcolor{black}{4. The system should be extensible (add new cameras during runtime). }\\\hline
What are the non-functional requirements that affected by this capability? &
1. Minimize the number of exchange messages in the network.\\\hline
What are the constraints that could affect this capability? &
\textcolor{black}{1. Cameras can be added or removed during runtime.}

~

\textcolor{black}{2. Tracker can fail during execution.}

~

\textcolor{black}{3. Resources on each camera may not be exceeded.}\\\hline
Whether this capability is \ necessary or beneficial? &
Yes\\\hline
\end{supertabular}

\end{center}
 \end{adjustwidth}
\end{table}

\begin{table}[H]
\begin{adjustwidth}{-1in}{-1in} 
\begin{center}

\caption{Questions and answers for deciding whether to include 
meta-self-awareness. }

\tablefirsthead{}
\tablehead{}
\tabletail{}
\tablelasttail{}
\begin{supertabular}{|m{4.005cm}|m{12.562cm}|}
\hline
\multicolumn{2}{|m{16.766998cm}|}{{\bfseries Meta-self-awareness}}\\\hline
What does the capability mean in your problem context. &
\textcolor{black}{Bandit solvers.}\\\hline
What are the functional requirements that affected by this capability? &
1. \textcolor{black}{The system should notice disappeared objects.}

~
\\\hline
What are the non-functional requirements that affected by this capability? &
1. Maximize tracking utility.

~

2. Minimize the number of exchange messages in the network.\\\hline
What are the constraints that could affect this capability? &
~
\\\hline
Whether this capability is \ necessary or beneficial? &
Yes\\\hline
\end{supertabular}

\end{center}
 \end{adjustwidth}
\end{table}

In summary, we again select the \textit{Goal Sharing with time-awareness 
capability pattern including meta-self-awareness capabilities}. This selection 
is based on the aforementioned Table~\ref{tab:pattern_select}.

%the selected patterns are the \textit{Goal Sharing Pattern} and the 
%\textit{Temproal Goal Aware Pattern}.

\subsubsection{Step 4 - Fit the Selected Pattern(s)}

We now fit the proposed architecture to the selected pattern, as shown in the 
Figure~\ref{fig:camfit} below.

\begin{figure}[h!]
\centering
\includegraphics[width=.95\linewidth]{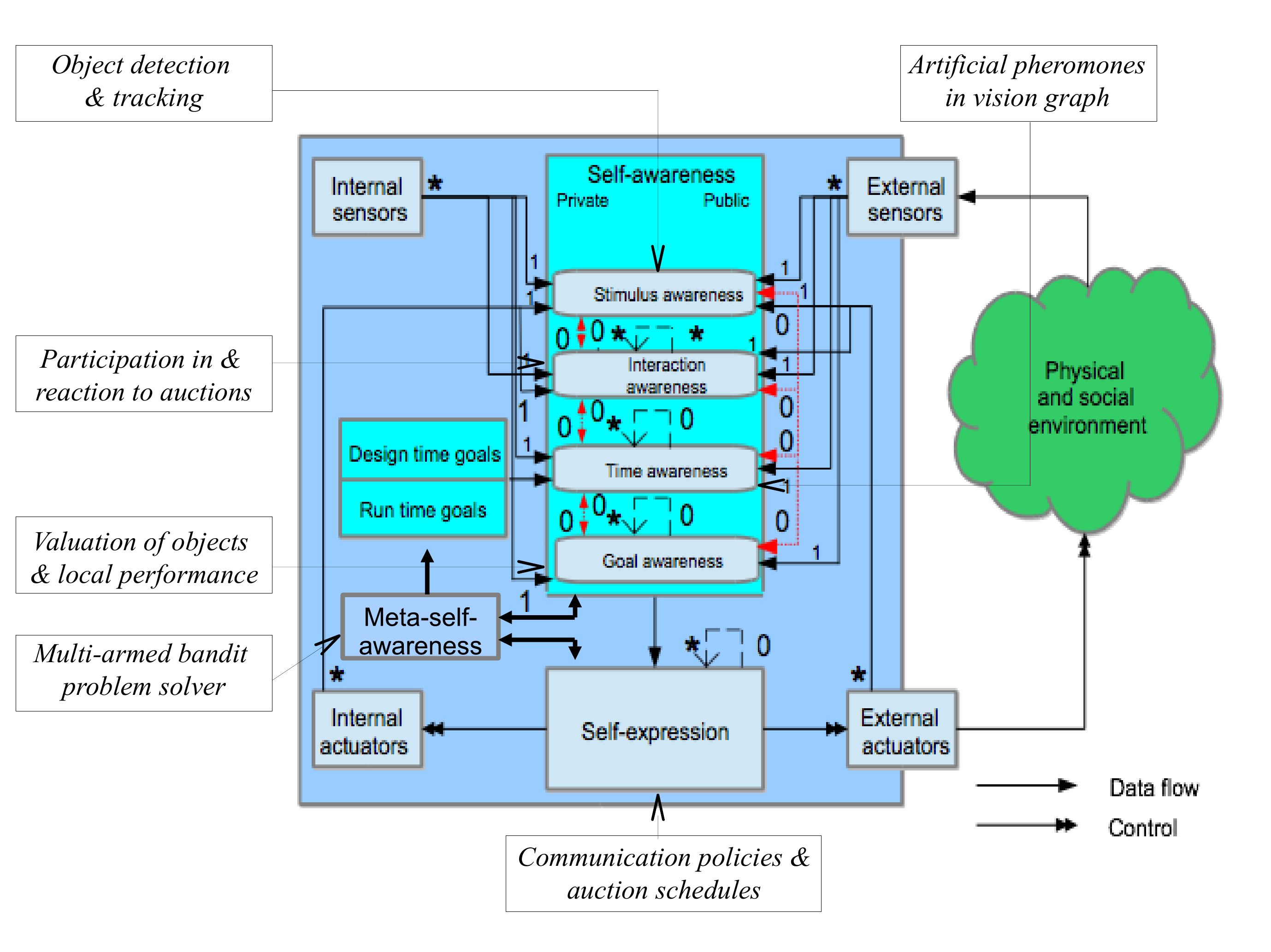}
\caption{Fit the proposed architecture to Goal Sharing with time-awareness 
capability Pattern including meta-self-awareness capabilities.}
\label{fig:camfit}
\end{figure}

\subsubsection{Step 6 to Step 8}

In smart camera systems, any alternative mechanisms apply either a central component (server) or introduce a priori knowledge about the scenario to coordinate tracking responsibilities.
In the EPiCS smart camera demonstrator we do not have both assumptions and 
deploy our system without any knowledge of the scenario and without any central 
coordination. This allows a quick deployment of such a system in a highly 
dynamic environment. Due to the lack of applicable alternatives, we are not 
able to compare our approach directly. However, the benefits of our 
socio-economic approach in comparison to a prior knowledge and fixed 
communication partners is presented in the quantitative experiments.

The details of the final architecture are shown in the 
Table~\ref{tab:cam_selection} below:

 \begin{table}
 \begin{adjustwidth}{-1in}{-1in} 
 \begin{center}
 
 \caption{The  selected alternatives for the smart-camera case study. }
 \label{tab:cam_selection}
 \tablefirsthead{}
 \tablehead{}
 \tabletail{}
 \tablelasttail{}
 \begin{supertabular}{|m{1.5cm}|m{1.5cm}|m{1.5cm}|m{1.5cm}|m{1.5cm}|m{1.5cm}|m{1.5cm}|m{1.5cm}|m{1.5cm}|m{1.5cm}|}
 \hline
 \multicolumn{2}{|m{3.2cm}|}{~
 
 ~
 } &
 \centering stimulus-awareness &
 \centering interaction-awareness &
 \centering time-awareness &
 \centering goal-awareness &
 \centering self-expression &
 \centering meta-self-awareness &
 \centering sensor &
 \centering\arraybslash actuator\\\hline
 \multicolumn{2}{|m{3.2cm}|}{\centering Selected alternative(s)} &
 Object detection and tracking&
 Reaction to auctions, bids and handover &
 Artificial pheromones in vision graph  &
 Utility function and local performance measurement &
 Communication policies and auction schedules  &
 Multi-armed bandit problem solvers &
 ~
  &
 ~
 \\\hline
 \centering send &
 \centering synchr-onous &
 \centering N/A &
 \centering N/A &
 \centering N/A &
 \centering N/A &
 \centering N/A &
 \centering N/A &
 \centering N/A &
 \centering\arraybslash N/A\\\hline
 ~
  &
 \centering asynch-ronous &
 ~
 
 \centering function call &
 ~
 
 \centering function call &
 ~
 
 \centering function call &
 ~
 
 \centering function call &
 ~
 
 \centering function call &
 ~
 
 \centering function call &
 ~
 
 \centering function call &
 ~
 
 \centering\arraybslash multicast\\\hline
 \centering handle &
 \centering sequential &
 \centering N/A &
 \centering N/A &
 \centering N/A &
 \centering N/A &
 \centering N/A &
 \centering N/A &
 \centering N/A &
 \centering\arraybslash N/A\\\hline
 ~
  &
 \centering parallel &
 \centering multi-threading &
 \centering multi-threading &
 \centering multi-threading &
 \centering multi-threading &
 \centering multi-threading &
 \centering multi-threading &
 \centering multi-threading &
 \centering\arraybslash multi-threading\\\hline
 \centering state &
 ~
  &
 \centering reactive &
 \centering proactive and reactive &
 \centering proactive and reactive &
 \centering reactive &
 \centering proactive and reactive &
 \centering proactive and reactive &
 ~
  &
 ~
 \\\hline
 \centering transit &
 ~
  &
 ~
  &
 ~
  &
 ~
  &
 ~
  &
 ~
  &
 \centering N/A &
 ~
  &
 ~
 \\\hline
 \centering link &
 ~
  &
 \centering one-to-one &
 \centering many-to-many, one-to-one &
 \centering one-to-one &
 \centering one-to-one &
 \centering one-to-many &
 \centering one-to-many &
 ~
  &
 ~
 \\\hline
 \centering structure &
  &
 \centering compact &
 \centering combine+ separate &  
 \centering compact &
 \centering combine+ separate &
 \centering combine+ separate &
 \centering compact &
 \centering compact &
 \centering\arraybslash compact\\\hline
 \centering existence &
 ~
  &
 ~
  &
 ~
  &
 ~
  &
 ~
  &
 ~
  &
 \centering exist &
 ~
  &
 ~
 \\\hline
 \end{supertabular}
 
 \end{center}
  \end{adjustwidth}
 \end{table}

\subsubsection{Quantitative Experiments}

We conduct experiments with our smart camera case study using our self-aware, socio-economic approach and compare the results with a non-self-aware approach where each camera only communicates with its direct neighbors. In the self-aware approach, these neighborhood relations are learnt online while in the non-self-aware approach the neighborhood relationships are defined a priori and are not adapted during runtime.
We simulate different scenarios and change the network of cameras during runtime. These so-called uncertainties affect the camera network only in the form of adding new cameras, remove existing cameras for a certain time or change the location and/or orientation of a camera. We measure the generated utility during runtime and show the accumulated utility for the entire network over time for all self-aware as well as the non-self-aware approach.

\begin{figure}
    \centering
		\subfloat[Scenario 
		1]{\label{fig:c1s1a2c1f}\includegraphics[width=0.45\columnwidth]{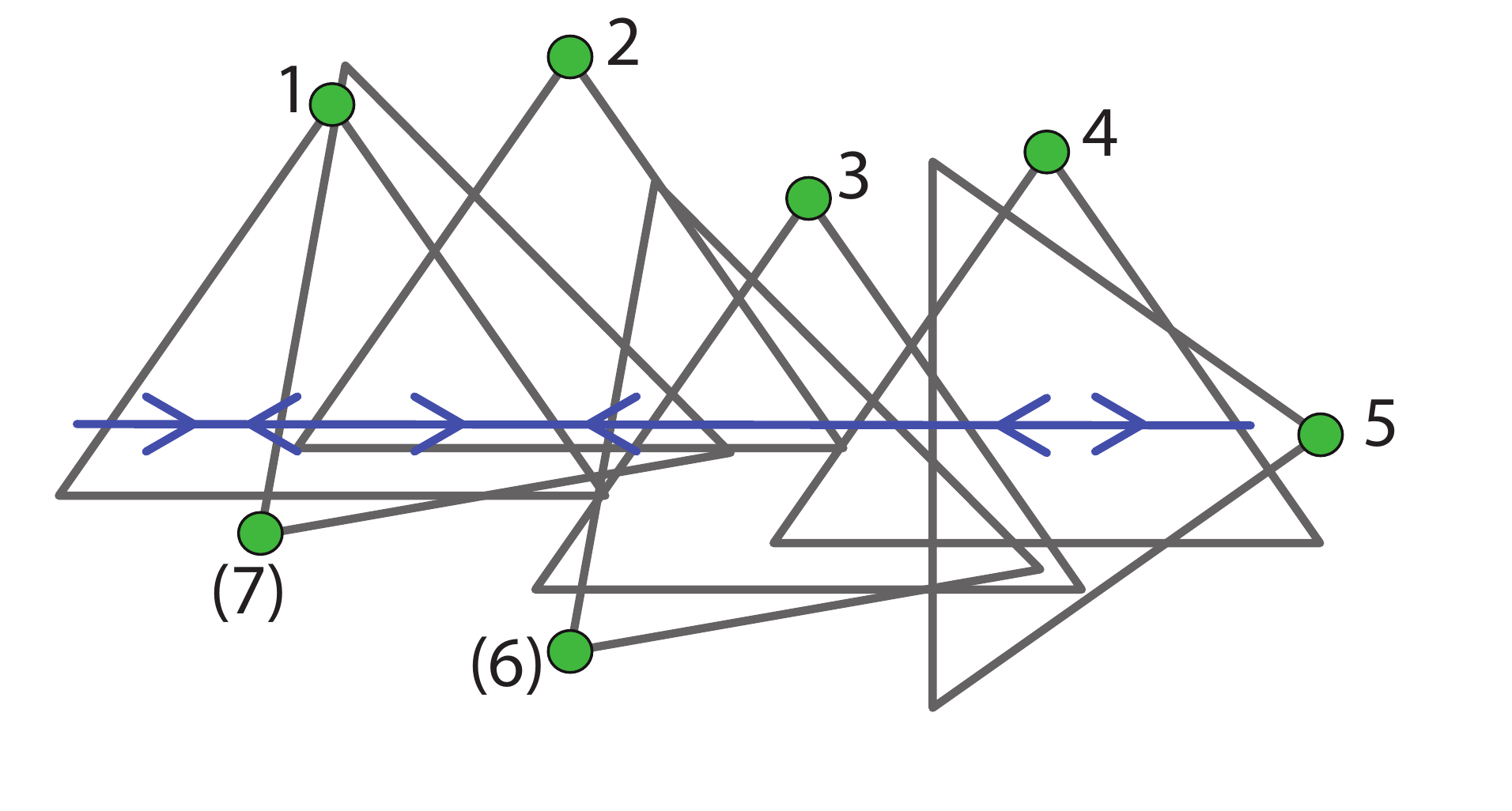}}
		\hspace{.25cm}
		\subfloat[Scenario 
		2]{\label{fig:c1s1a2c2f}\includegraphics[width=0.45\columnwidth]{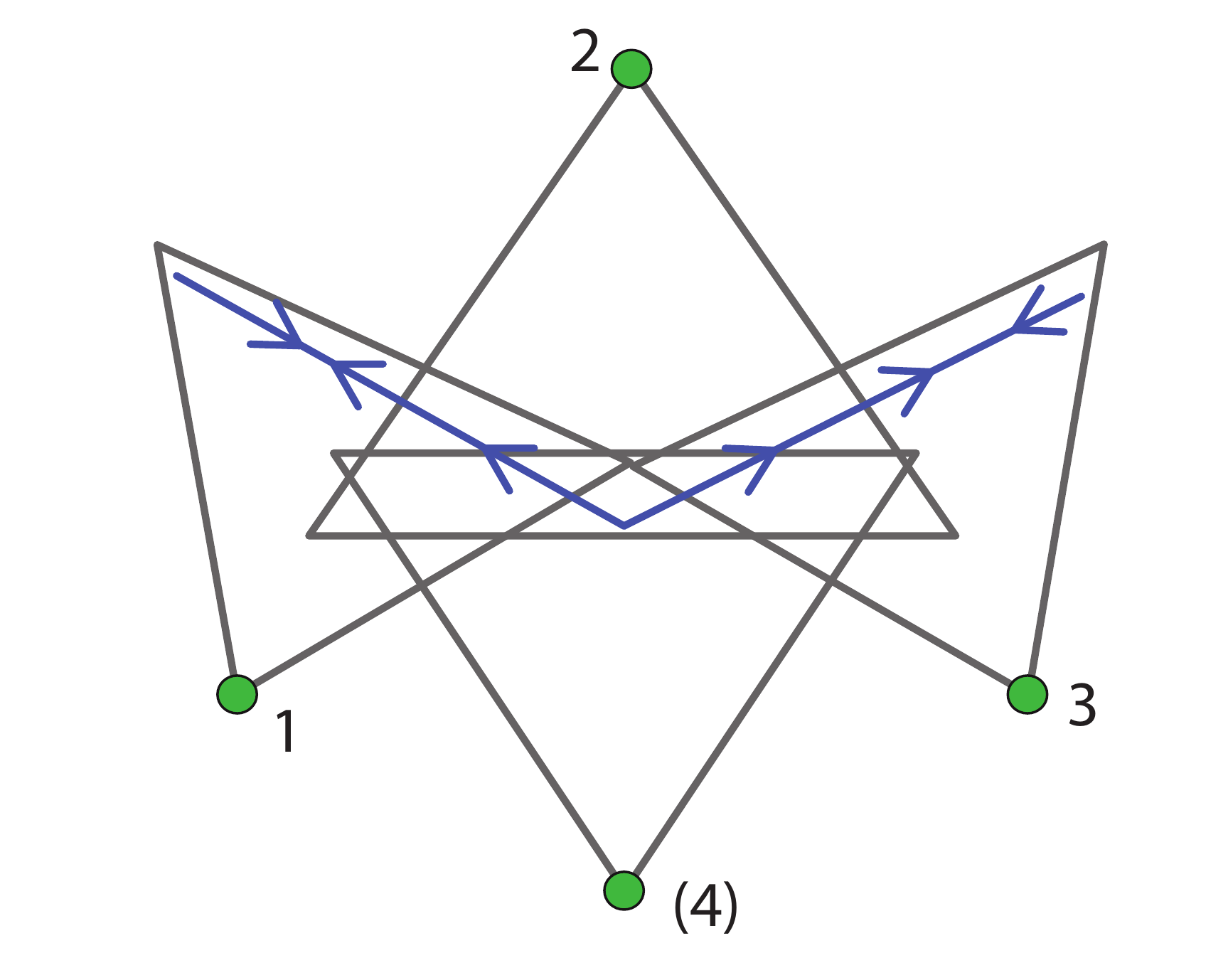}}
		\hspace{.25cm}
		\subfloat[Scenario 
		6]{\label{fig:c6}\includegraphics[width=0.7\columnwidth]{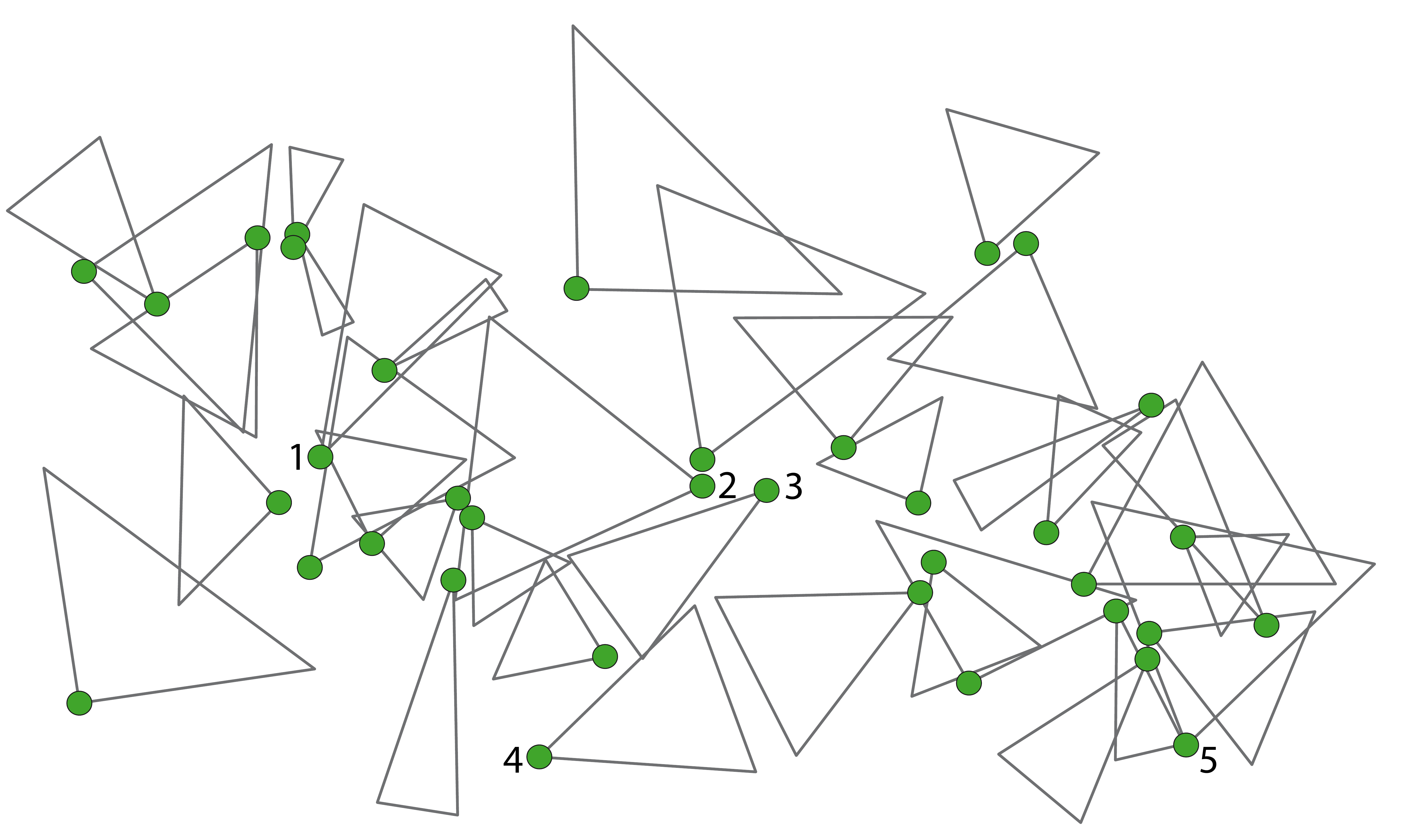}}
	    \caption[Evaluated scenarios]{Three qualitative different scenarios 
	    with various uncertainties. Green dots represent cameras while grey 
	    triangles indicate the corresponding field of view}
	    \label{fig:scen_smartcams}
\end{figure}

Since we are interested in performing repeatable experiments to investigate adaptivity and robustness issues, we used the simulation environment CamSim~\cite{2013_Esterle_SASOW} with different scripted experimental setups of smart-camera networks. In the following subsection the different experimental scenarios will be described. For our experiments we considered three general scenarios
and executed these scenarios with a variety of objects, paths and events. The 
different scenarios are illustrated in Figure~\ref{fig:scen_smartcams}.

For the first and second scenarios we defined paths for the object to traverse along. These paths are illustrated as blue
lines. For scenario three, the objects move in a straight line in a random direction. For each scenario we defined different
experiments using our events. For our three distinctive scenarios we conducted experiments where we added a camera during runtime, removed an camera from the test environment and changed the extrinsic parameters of a camera. An overview is given in the following Table:

\begin{table}[H]
\begin{adjustwidth}{-1in}{-1in} 
\begin{center}

\caption{The experiments configurations. }

\tablefirsthead{}
\tablehead{}
\tabletail{}
\tablelasttail{}
\begin{supertabular}{|m{5.467cm}|m{5.467cm}|m{5.467cm}|}
\hline
\centering{\bfseries Experiment No.} &
\centering{\bfseries Scenario} &
\centering\arraybslash{\bfseries Action}\\\hline
\centering 1 &
\centering Scenario 1 &
\centering\arraybslash Add Camera (6)\\\hline
\centering 2 &
\centering Scenario 1 &
\centering\arraybslash Remove Camera 3\\\hline
\centering 3 &
\centering Scenario 1 &
\centering\arraybslash Change Position 3 to (7)\\\hline
\centering 4 &
\centering Scenario 2 &
\centering\arraybslash Add Camera (4)\\\hline
\centering 5 &
\centering Scenario 2 &
\centering\arraybslash Remove Camera 2\\\hline
\centering 6 &
\centering Scenario 2 &
\centering\arraybslash Change Orientation of Camera 2 by -55 degree\\\hline
\centering 7 &
\centering Scenario 3 &
\centering\arraybslash Remove Cameras 1, 2, 3, 4, 5\\\hline
\end{supertabular}

\end{center}
 \end{adjustwidth}
\end{table}

For the non-self-aware approach applied in the first and second scenario, neighborhood relations are defined between cameras only when they have overlapping FOVs. For scenario three, neighborhood relations are defined whenever an object can traverse from the FOV of one camera to another in a straight line without appearing in the FOV of any other camera.

\begin{figure}[h]
	\centering
	\includegraphics[angle=270,width=0.85\linewidth]{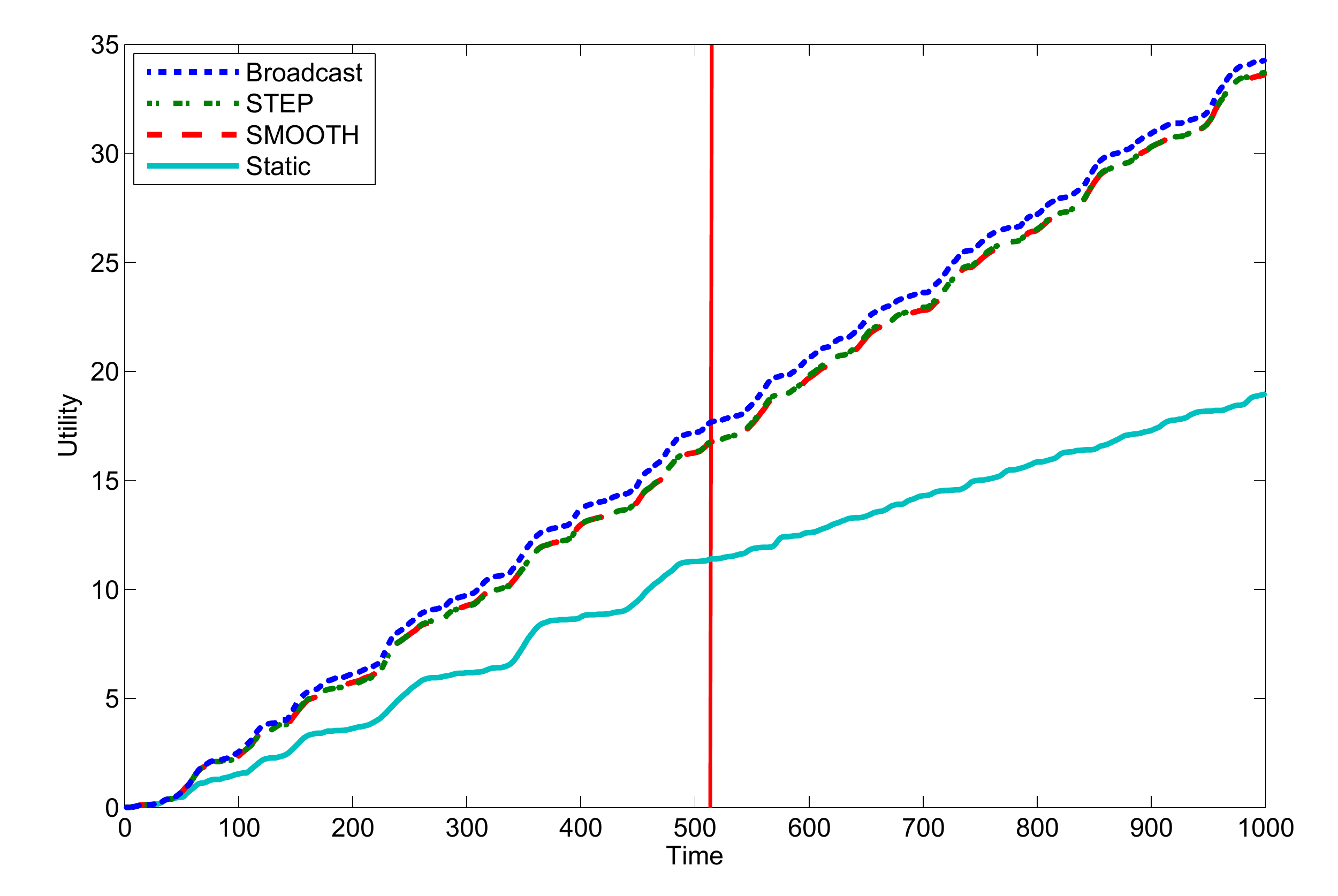}
	\caption{Cumulative sum of the entire network utility over time for a 
	typical	simulation run of experiment 3 (Scenario 1 with change event) and 
	using our passive approaches. The vertical line indicates the timestep when 
	the event
	occurred. The simulation ran for 1000 timesteps. We changed the position of
	a single camera within the environment to show the ability of our approach
	to deal with changes of the extrinsic parameters of cameras.}
	\label{fig:res_2}
\end{figure}

Figure \ref{fig:res_2} shows results for experiment number 3 employing our 
active approach. We changed the position of a single camera within the 
environment to show the ability of our approach to deal with changes of the 
extrinsic parameters of cameras. The vertical line shows the time at which the 
event happened. The drop in utility gain for the static approach after the 
event occurred is apparent, demonstrating its inability to adapt to the change. 
While the static approach loses overall utility, the SMOOTH and STEP policies 
are able to keep a high utility after the event, indicating their robustness to 
change. 

\begin{figure}[h]
	\centering
	\includegraphics[angle=270,width=0.85\linewidth]{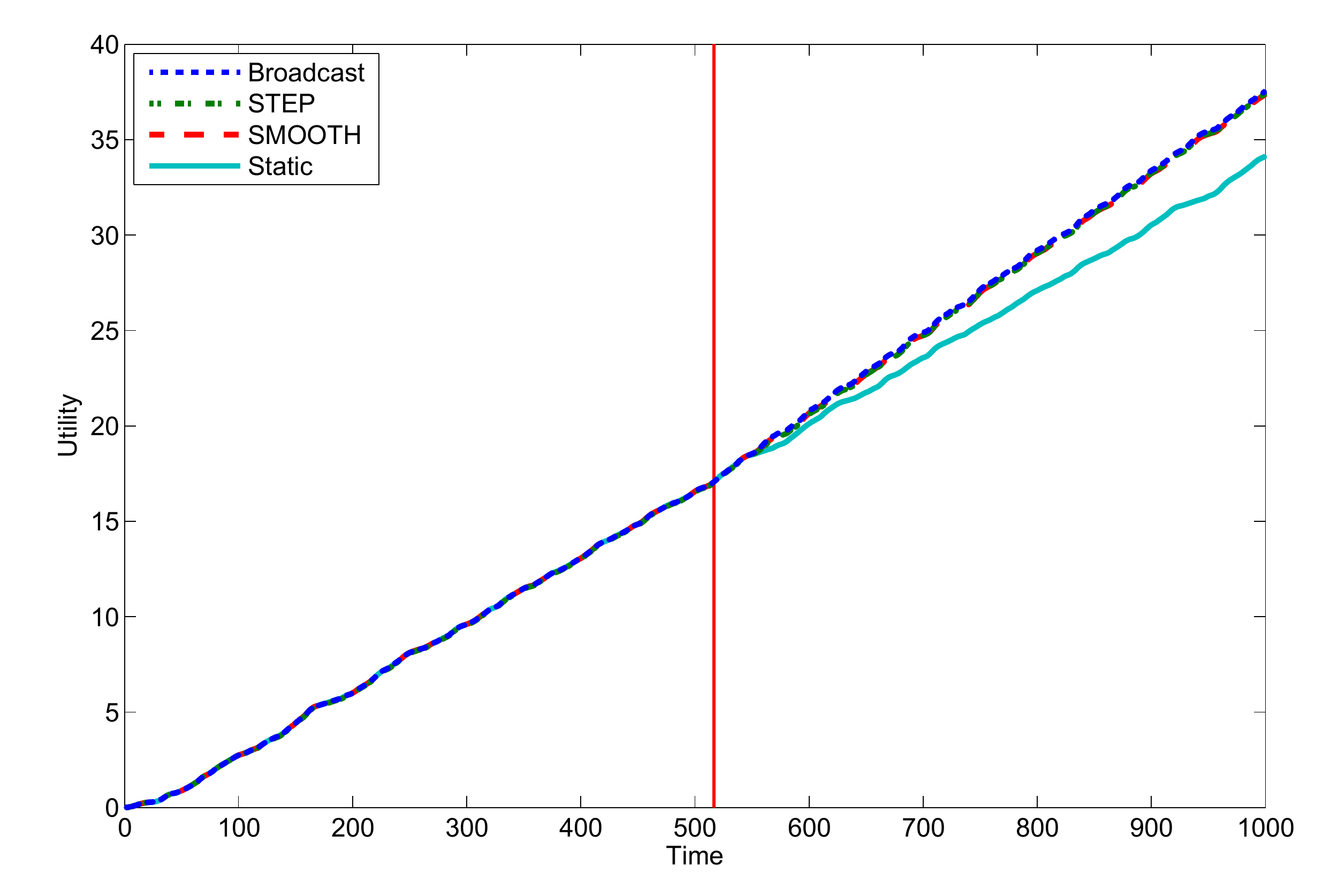}
	\caption{Cumulative sum of the entire network utility over time for a
	typical simulation run of experiment 4 comparing our active socio-economic
	approaches with a static handover. The red vertical line indicates the 
	timestep when the event occurred. The simulation ran for 1000 timesteps.}
	\label{fig:res_1}
\end{figure}

The results of experiment 4 are shown in Figure \ref{fig:res_1} where we added 
a new camera during runtime. The occurrence of the event is indicated with a 
red vertical line again at time step 518. The increased accumulated utility 
using the active SMOOTH and STEP approach is apparent. Since the camera was 
placed at a location which was already covered by a different camera, the 
improvement was rather small. 

\begin{figure}[h]
	\centering
	\includegraphics[angle=270,width=0.85\linewidth]{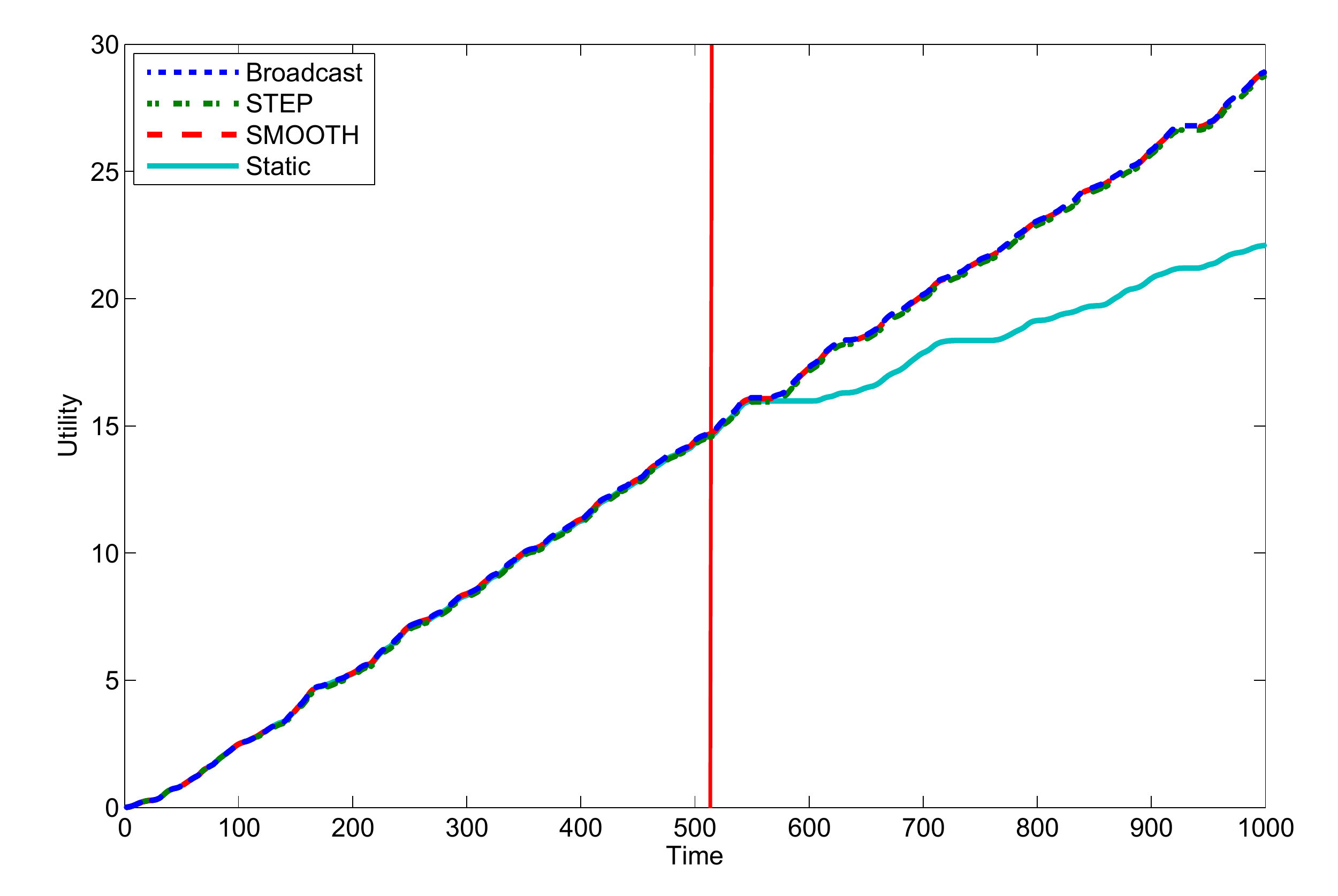}
	\caption{Cumulative sum of the entire network utility over time for a 
	typical	simulation run of Scenario 2 with an error event (experiment 5) 
	and using our active approaches. The red vertical line indicates the 
	timestep when the event occurred. The simulation lasted for 1000 timesteps.}
	\label{fig:res_3}
\end{figure}

Figure \ref{fig:res_3} illustrates the results of scenario 2 with a 
camera failure event (experiment 5), when passive approaches were used. Here 
the drop of the accumulated utility is obvious for the static approach, while 
the socio-economic approaches are able to relearn the vision graph online and 
continue tracking the object within the entire network.

\bibliography{ref}{}
\bibliographystyle{plain}

\end{document}